\documentclass[12pt,a4wide]{article}
\pdfoutput=1
\usepackage{jheppub}
\usepackage{amsfonts}
\usepackage{amscd}
\usepackage{amssymb}
\usepackage{amsmath,bbm}
\usepackage{epsfig}
\usepackage{latexsym}
\usepackage{mathtools}
\usepackage{hyperref}
\usepackage[vcentermath]{youngtab}

\def \be  {\begin{equation}}
\def \ee  {\end{equation}}
\def \ba  {\begin{eqnarray}}
\def \ea  {\end{eqnarray}}
\def \bea {\begin{equation}\begin{aligned}}
\def \eea {\end{aligned}\end{equation}}
\def \bb  {\begin{thebibliography}}
\def \eb  {\end{thebibliography}}
\def \lab #1 {\label{#1}}

\newcommand\cF{\mathcal{F}}
\newcommand\cG{\mathcal{G}}

\newcommand\cI{\mathcal{I}}

\newcommand\cK{\mathcal{K}}

\newcommand\cN{\mathcal{N}}
\newcommand\cO{\mathcal{O}}

\newcommand\cQ{\mathcal{Q}}

\newcommand\cS{\mathcal{S}}
\newcommand\cT{\mathcal{T}}

\newcommand\cZ{\mathcal{Z}}
\newcommand\al{\alpha}

\newcommand\rd{\mathrm{d}}

\newcommand\lb{\lambda}

\newcommand\la{\langle}
\newcommand\ra{\rangle}
\newcommand\del{\partial}

\newcommand\tr{\mathrm{Tr}}

\title{The superconformal index and an elliptic algebra of surface defects}
\author[1]{Mathew~Bullimore,}
\author[2]{Martin~Fluder,}
\author[2]{Lotte~Hollands,}
\author[2]{Paul~Richmond\,}
\affiliation[1]{Perimeter Institute for Theoretical Physics, \\
  Waterloo, Ontario, N2L 2Y5, Canada.} 
\affiliation[2]{Mathematical Institute, University of Oxford,\\ Andrew Wiles Building, Radcliffe Observatory Quarter, \\ Woodstock Road, Oxford, OX2 6GG, UK.\\}

\emailAdd{mbullimore@perimeterinstitute.ca}
\emailAdd{fluder@maths.ox.ac.uk}
\emailAdd{hollands@maths.ox.ac.uk}
\emailAdd{richmond@maths.ox.ac.uk}

\abstract{In this paper we continue the study of the superconformal index of four-dimensional $\cN=2$ theories of class $\cS$ in the presence of surface defects. Our main result is the construction of an algebra of difference operators, whose elements are labeled by irreducible representations of $A_{N-1}$. For the fully antisymmetric tensor representations these difference operators are the Hamiltonians of the elliptic Ruijsenaars-Schneider system. The structure constants of the algebra are elliptic generalizations of the Littlewood-Richardson coefficients. In the Macdonald limit, we identify the difference operators with local operators in the two-dimensional TQFT interpretation of the superconformal index. We also study the dimensional reduction to difference operators acting on the three-sphere partition function, where they characterize supersymmetric defects supported on a circle, and show that they are transformed to supersymmetric Wilson loops under mirror symmetry. Finally, we compare to the difference operators that create 't Hooft loops in the four-dimensional $\cN=2^*$ theory on a four-sphere by embedding the three-dimensional theory as an S-duality domain wall. 
 
}

\begin{document}
\maketitle

\bibliographystyle{JHEP}
\section{Introduction}

Surface defects are an interesting class of non-local observables in four-dimensional gauge theories~\cite{Gukov:2006jk}. In this paper, we consider surface defects in four-dimensional $\cN=2$ superconformal field theories of class $\cS$, which are obtained by compactifying the partially twisted six-dimensional $(2,0)$ theory on a decorated Riemann surface $C$~\cite{Gaiotto:2009we,Gaiotto:2009hg}. The six-dimensional $(2,0)$ theory is characterized by a Lie algebra $\mathfrak{g}$ of ADE type. In this paper we focus on the case of $A_{N-1}$.
In this case, the six-dimensional $(2,0)$ theory arises as the infrared limit of the worldvolume theory on a stack of $N$ coincident M5-branes. 
Surface defects in four-dimensional theories of class $\cS$ can be formed from both codimension-two and codimension-four defects in the six-dimensional parent theory. This is summarized in Table~\ref{table}.

\begin{table}[h]
\begin{center}
\begin{tabular}{c|c|c|c}
  & $X$ & $C$ & Name \\
\hline
(i) & 4 & 0 & flavor puncture \\ 
(ii) & 2 & 2  & surface defect \\
(iii) & 2 & 0  & surface defect \\
\end{tabular}
\end{center}
\caption{Summary of the defects in the six-dimensional $(2,0)$ theory on $X \times C$. $X$ is the four dimensional space-time and $C$ is a decorated Riemann surface. (i) and (ii) show configurations of codimension-two defects while (iii) shows the configuration of a codimension-four defect.  \label{table}}
\end{table}

Let us first discuss codimension-two defects of the $(2,0)$ theory in six-dimensions, which are labeled by embeddings $\rho: su(2) \to \mathfrak{g}$. These defects play an important role in the construction of theories of class $\cS$: a codimension-two defect that is inserted at a point on the Riemann surface $C$ and spans all four space-time dimensions corresponds to a flavor puncture in the construction of~\cite{Gaiotto:2009we,Gaiotto:2009hg} - see (i) of Table~\ref{table}. Alternatively, wrapping the same codimension-two defect on the whole Riemann surface $C$ leads to a surface defect in the four-dimensional theory - see (ii) of Table~\ref{table}. This class of surface defects has been studied, for example, in~\cite{Alday:2010vg,Kanno:2011fw}. 

On the other hand, there are codimension-four defects in the $(2,0)$ theory in six-dimensions, which are expected to be labeled by an irreducible representation of $\mathfrak{g}$, see for example~\cite{Moore} and references therein. Inserting a codimension-four defect at a point on the Riemann surface $C$ engineers another class of surface defects in the four-dimensional theory - see (iii) of Table~\ref{table}. In this paper, we study this second class of surface defects in four-dimensional $\cN=2$ theories of class $\cS$.

Important evidence for the classification of codimension-four defects in terms of irreducible representations of $\mathfrak{g}$ comes from the correspondence between four-sphere partition functions of $\cN=2$ theories of class $\cS$ and correlation functions in Liouville or Toda conformal field theory on $C$~\cite{Alday:2009aq,Wyllard:2009hg}. In this correspondence, flavor punctures are represented by vertex operators labeled by non-degenerate and semi-degenerate representations of the Virasoro or $W_N$-algebra. There are also completely degenerate representations labeled by two dominant integral weights of $\mathfrak{g}$, or equivalently, by two irreducible representations $R_1$ and $R_2$ of $\mathfrak{g}$. Correlation functions with additional insertions of completely degenerate vertex operators compute the four-sphere partition function in the presence of surface defects~\cite{Alday:2009fs}. In particular, the labels $R_1$ and $R_2$ characterize the surface defects supported on orthogonal two-spheres. 

\begin{figure}[t]
\centering
\includegraphics[width=0.9\textwidth]{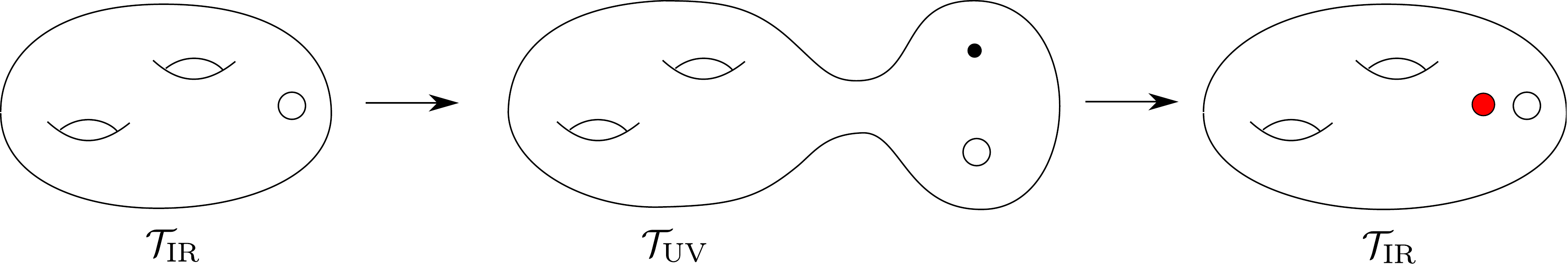}
\caption{Schematic illustration of the renormalization group flow $\cT_{UV} \to \cT_{IR}$ that can be used to introduce surface defects. The white dots represent full punctures with $SU(N)$ symmetry while the black dot is a simple puncture with $U(1)$ symmetry. The red dot represents a codimension-four defect engineering a surface defect in four dimensions.} 
\label{fig:GRRconstruction}
\end{figure}

Inspired by the connection to degenerate vertex operators and the analytic structure of Virasoro/$W_N$-algebra conformal blocks, the authors of reference~\cite{Gaiotto:2012xa} introduced a renormalization group flow that can be used to construct the surface defects from vortex configurations in a larger theory. Let us consider the simplest example of this procedure illustrated in Figure~\ref{fig:GRRconstruction}. 

The starting point is a theory $\cT_{IR}$ with a full puncture encoding an $SU(N)$ flavor symmetry. We then form the larger theory $\cT_{UV}$ by adding a simple puncture nearby with $U(1)$ flavor symmetry. This corresponds to adding an additional hypermultiplet in the bifundamental of $SU(N)\times SU(N)$ by gauging the diagonal $SU(N)$. The extra $U(1)$ symmetry corresponds to the baryonic symmetry of the bifundamental hypermultiplet and the position of the simple puncture controls the gauge coupling of the gauged $SU(N)$. 

The theories $\cT_{IR}$ and $\cT_{UV}$ are connected by a renormalization group flow that is initiated by turning on a constant vacuum expectation value for the hypermultiplet scalar. By turning on a position-dependent vacuum expectation value corresponding to a half-BPS vortex configuration in $\cT_{UV}$, the endpoint of the renormalization group flow is a surface defect in the original theory $\cT_{IR}$. These surface defects are labeled by a pair of positive integers $(r_1,r_2)$ corresponding to the vortex numbers in orthogonal two-planes. This construction is analogous to the Toda construction of codimension-four surface operators \cite{Alday:2009fs}. Hence our working conjecture is that they give a representation of codimensions-four surface defects labelled by a pair of symmetric tensor representations of $\mathfrak{g}$.

A concrete prescription was given in~\cite{Gaiotto:2012xa} to implement this renormalization group flow at the level of the superconformal index. The superconformal index is a trace over states of a superconformal field theory in radial quantization~\cite{Kinney:2005ej}. It is a much simpler observable than the four-sphere partition function because it does not depend on the marginal couplings of the theory. For previous work on the superconformal index of theories of class $\cS$ see~\cite{Gadde:2009kb,Gadde:2010te,Gadde:2011ik,Gadde:2011uv,Gaiotto:2012uq}. In full generality, the $\cN=2$ superconformal index depends on three parameters denoted by $\{p,q,t\}$ that are associated to combinations of bosonic conserved charges commuting with a chosen supercharge. It also depends on flavor parameters $\{a_1,\ldots,a_N\}$, such that $\prod_j a_j=1$, for each global $SU(N)$ symmetry and an additional parameter $b$ for each $U(1)$ symmetry. The superconformal index is thus denoted by
\be
\cI(p,q,t,a_j,b,\ldots) \, .
\ee

The superconformal index of the theory $\cT_{IR}$ with surface defects is obtained by computing a residue of the superconformal index of the theory $\cT_{UV}$ in the additional fugacity $b$ associated to the additional $U(1)$ symmetry. The result is a difference operator $G_{r_1,r_2}$ that acts on the superconformal index of the original theory $\cT_{IR}$ by shifting the fugacities of the $SU(N)$ flavor symmetry. Schematically, the difference operator is defined by
\be
G_{r_1,r_2} \cdot \cI_{IR}(a_j,\ldots) \sim \underset{b =  t^{\frac{1}{2}}p^{r_1/N}q^{r_2/N} }{\mathrm{Res}} \, \Big[ \, \frac{1}{b} \, \cI_{UV}(a_j,b,\ldots) \, \Big],
\ee
where the proportionality constant is discussed in \S\ref{section:index}. The difference operator $G_{r_1,r_2}$ corresponds to inserting a surface defect in the original theory $\cT_{IR}$ that is labeled by the pair $(r_1,r_2)$. 

In what follows we concentrate on the case $r_1=0$ and simply label the difference operators by $G_{r}$, where $r\in\mathbb{Z}_{\geq 0}$. The label $r$ can be thought of as denoting a symmetric tensor representation of rank $r$. The resulting expression for $G_r$ is
\be
G_r \cdot \cI(a_j) = \sum_{\sum_{k=1}^N m_k = r}  \;
\prod_{j,k=1}^N \left[ \,  \prod_{m=0}^{m_k-1} \frac{\theta\left( q^{m+m_k-m_j} t a_j / a_k ; p\right)}{\theta\left( q^{m-m_k} a_k/a_j ; p \right)} \right]  \; \cI\left(a_j \mapsto q^{\frac{r}{N}-m_j} \, a_j \right) \, ,
\ee
where the theta-function $\theta(z,p)$ is defined in section~\S\ref{section:index}.

Following our arguments above, we expect that there exist difference operators $G_{R}$ corresponding to surface defects labeled by all irreducible representations $R$ of $\mathfrak{g}$. In principle, they could be constructed by starting from a theory $\cT_{UV}$ with an additional puncture with a larger flavor symmetry. However, this would involve non-Lagrangian ingredients and, although the index can be bootstrapped as in \cite{Gaiotto:2012xa}, the analytic structure needed for this approach is not manifest. 

Instead we follow the line of reasoning introduced in~\cite{Alday:2013kda} and complete the algebra of difference operators. For the difference operator associated to the representation $R$ we make an ansatz
\begin{equation}
G_R \cdot \cI(a_j) = \sum_{\lambda} C_{R,\lambda}
(p,q,t,a_j) \,  \cI(q^{- (\lambda, h_j)}a_j), 
\end{equation}
where the sum is over the weights $\lambda$ of the representation $R$, $(\, , \,)$ is the standard inner product on the Cartan subalgebra of $\mathfrak{g}$, and $h_j$ are the weights of the fundamental representation. This ansatz is compatible with what we already know about difference operators $G_{r}$ for symmetric tensor representations $R=(r)$. 

The coefficients $C_{R,\lambda}(p,q,t,a_j)$ are then determined by imposing that the full set of difference operators $G_R$ is closed under composition
\begin{equation}\label{eqn:algebra}
G_{R_1} \circ G_{R_2} = \sum_{R_3} {\cN_{R_1, R_2}}^{R_3}(p,q,t) \, G_{R_3}\, ,
\end{equation}
and forms a commutative algebra. Since the symmetric tensor representations form an over-complete basis, there are many compatibility conditions for the system~(\ref{eqn:algebra})  to be solved consistently. It is thus non-trivial that a solution exists. Nevertheless, we can find a solution using the following method. 

First, we notice that all irreducible representations in the case $\mathfrak{g}=su(2)$ are symmetric tensor representations, so that there are no additional difference operators. Even though it is not obvious and requires numerous functional identities for theta-functions, the system~(\ref{eqn:algebra}) can be solved uniquely in this case. The structure coefficients ${\cN_{R_1, R_2}}^{R_3}(p,q,t)$ turn out to be an elliptic generalization of the $(q,t)$-deformed Littlewood-Richardson coefficients. In \S{\ref{subsubsec:SU2} we give a recipe to obtain the elliptic coefficients ${\cN_{R_1, R_2}}^{R_3}(p,q,t)$ uniquely from the $(q,t)$-deformed ones. 

If we then assume that for any rank of the gauge group the structure coefficients ${\cN_{R_1, R_2}}^{R_3}(p,q,t)$ are given by this elliptic generalization of the Littlewood-Richardson coefficients, the system~(\ref{eqn:algebra}) can be solved consistently and uniquely for all of the difference operators $G_R$. The coefficients $C_{R,\lambda}$ are in general sums of products of ratios of theta-functions. 
Let us stress once more that the fact that we can find a consistent solution to the system~(\ref{eqn:algebra}) is highly non-trivial and involves numerous identities for theta-functions. We see this as strong evidence that  a class of surface defects labeled by general irreducible representations $R$ of $\mathfrak{g}$ exists. 

In particular, we find that the difference operators $G_{(1^r)}$ labeled by the rank $r$ antisymmetric tensor representations, can be conjugated to the Hamiltonians of the $N$-body elliptic Ruijsenaars-Schneider integrable system. This is an extension of the fact, noted in~\cite{Gaiotto:2012xa}, that the fundamental operator in the case of $A_1$ can be conjugated to the Hamiltonian of the two-body elliptic Ruijsenaars-Schneider integrable system.

A microscopic definition of a large class of surface defects can be given by coupling the four-dimensional theory to two-dimensional $\cN=(2,2)$ degrees of freedom supported on the surface~\cite{Alday:2009fs,Dimofte:2010tz,Gaiotto:2009fs,Gaiotto:2013sma}. The superconformal
index in the presence of such surface defects has been constructed
recently in \cite{Gadde:2013dda}. Thus it is natural to ask whether the surface defects introduced by the operators $G_R$ can be understood in this approach. For the rank $r$ symmetric tensor representation, it was already noted in~\cite{Gadde:2013dda} that the two-dimensional degrees of freedom consist of an $\cN=(2,2)$ gauge theory with gauge group $U(r)$, coupled to $N$ fundamental 
and $N$ anti-fundamental chiral fields and an additional chiral field
in the adjoint representation of $U(r)$. Using the same techniques, we find that the relevant two-dimensional degrees of freedom for the rank $r$ antisymmetric tensor representation are the same as above, but without the adjoint chiral field. For other representations, it is not clear to us whether the surface defect can be constructed by coupling to an $\cN=(2,2)$ supersymmetric gauge theory. We make a few additional remarks about this in the discussion in \S\ref{section_Conclusions}.

The superconformal index of $\cN=2$ theories of class $\cS$ has a dual description in terms of a two-dimensional topological quantum field theory on the surface $C$~\cite{Gadde:2011ik,Gadde:2011uv}. We continue in this paper by showing that the difference operators $G_R$ are natural objects in this two-dimensional TQFT.  When we focus on the Macdonald slice $
\{p=0,q,t\}$, the TQFT is given as an analytic continuation of refined Chern-Simons theory on $S^1 \times C$ \cite{Aganagic:2011sg}. 

In the Macdonald limit, the operators $G_{(1^r)}$, labeled by antisymmetric tensor representations, can be conjugated to the so-called Macdonald operators, whose eigenfunctions are the Macdonald polynomials $P_S(a,q,t)$ labeled by an irreducible representation $S$. 
We find that the eigenvalue of a general, conjugated, difference operator $G^c_R$ in the Macdonald limit is given by
\be
G^c_{R} \cdot P_S(a_j,q,t) = \frac{S_{R,S}}{S_{0,S}} P_{S} (a_j,q,t) \, ,
\ee
where $S_{R,S}$ is an analytic continuation of the modular S-matrix of refined Chern-Simons theory, which depends on $q$ and $t$. A consequence is that the surface defect introduced by the operator $G^c_R$ is equivalent to a Wilson loop wrapping around the $S^1$ of the three-manifold $S^1 \times C$. 

In the Macdonald limit, the structure constants ${\cN_{R_1, R_2}}^{R_3}(q,t)$ become the $(q,t)$-deformed Littlewood-Richardson coefficients and the algebra of difference operators $G_R$ is identified with the Verlinde algebra. We expect that this Verlinde algebra has a natural interpretation in the (analytically continued) chiral boundary theory on the two-torus boundary near a puncture of $C$.  

We find further confirmation of the physical relevance of the difference operators $G_R$ by reducing the 
superconformal index to the three-sphere partition function, following~\cite{Gadde:2011ia,Dolan:2011rp,Imamura:2011uw}. In particular, we consider the dimensional reduction of the four-dimensional $\cT_N$ theory, which is obtained by compactifying the six-dimensional $(2,0)$ theory on a three-punctured sphere
with three full punctures. The dimensionally reduced $\cT_N$ theory has a Lagrangian mirror description as a
star-shaped quiver theory~\cite{Benini:2010uu}. This is illustrated in Figure~\ref{fig:3d-dualities}. 
\begin{figure}[t]
\centering
\includegraphics[width=0.65\textwidth]{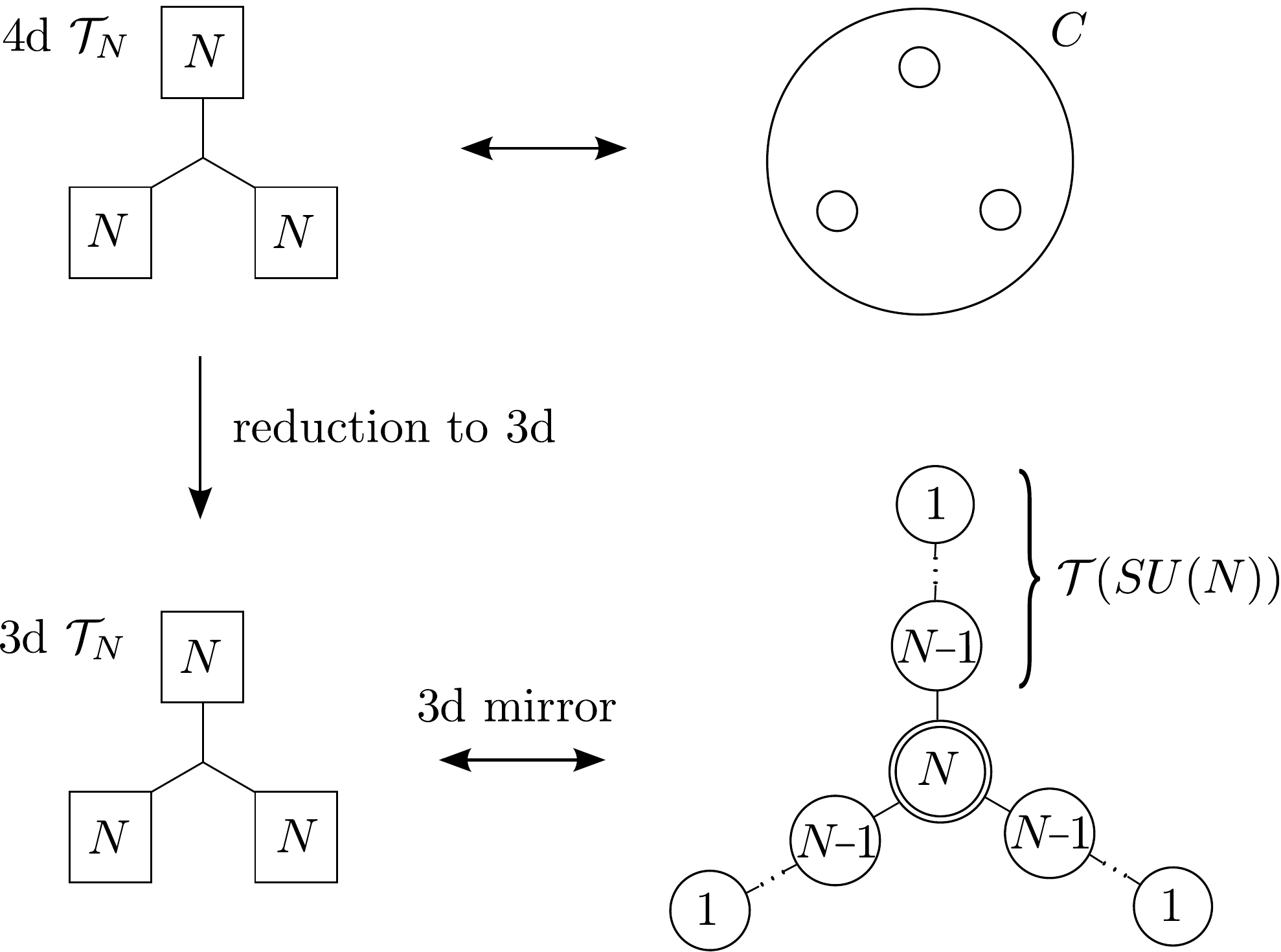}
\caption{Sequence of dualities that maps the four-dimensional $\cT_N$
  theory (upper-left) to the three-dimensional star-shaped quiver
  theory (lower-right).} 
\label{fig:3d-dualities}
\end{figure}
In
particular, each full puncture of the three-punctured sphere is represented
by a three-dimensional linear quiver theory called $\cT(SU(N))$. 

It is expected that the surface defects introduced by the dimensional reduction of the operators $G_R$ correspond to supersymmetric Wilson loops in the representation $R$ for the central node of the star-shaped quiver. This is in fact equivalent to the statement that the partition function of the $\cT(SU(N))$ theory is an eigenfunction of the dimensionally reduced operators $G^{(\mathrm{3d})}_{R}$. 
The partition function $\cZ(x,y)$ of the $\cT(SU(N))$ theory depends on two mass parameters $x$ and $y$ associated to the Higgs branch and the Coulomb branch respectively, and is symmetric under $x\leftrightarrow y$. For the case of a round four-sphere, we show indeed that
\begin{align}\label{eqn:intertwine}
G^{(\mathrm{3d})}_{(1^r)}(y) \cdot \cZ(x,y) =  W_{(1^r)}(x) \, \cZ(x,y),  
\end{align}
where $W_{(1^r)}(x)$ is a supersymmetric Wilson loop in the $r$-th antisymmetric tensor representation. 

For other (non-minuscule) representations we find that this is not quite correct. In particular, the Wilson loops obey the algebra
\begin{align}
W_{R_1} \cdot W_{R_2} =
\sum_{R_3} {N_{R_1, R_2}}^{R_3} \, W_{R_3} \, ,
\end{align}
where ${N_{R_1, R_2}}^{R_3}$ are the ordinary
Littlewood-Richardson coefficients, whereas the algebra of the three-dimensional operators $G^{(\mathrm{3d})}_{R}$ is not of this form. Instead, we find that when the representation $R$ is non-minuscule, the dimensionally reduced operators $G^{(\mathrm{3d})}_R$ are linear combinations of operators
$\tilde{G}^{(\mathrm{3d})}_S$, with
$|S| \leq |R|$, that \emph{are} dual to Wilson loop operators\footnote{We defined the partial ordering of representation by $|R_1| < |R_2|$ iff the dimension of the representation $R_1$ is less than the dimension of $R_2$.}. This gives a simple invertible linear transformation on the algebra of difference operators. 

Finally, by embedding the three-dimensional $\cT(SU(N))$ theory as an S-duality domain wall in the four-dimensional $\cN=2^*$ theory, we interpret the dimensionally reduced difference operators $G ^{(\mathrm{3d})}_R$ as operators that introduce 't Hooft defects, labeled by irreducible representations $R$, into the four-sphere partition function of the $\cN=2^*$ theory. Again, when the representation $R$ is an antisymmetric tensor representation, we find perfect agreement with both localization~\cite{Gomis:2011pf} and (in the case of the fundamental representation) computations of Verlinde operators in Liouville/Toda conformal field theory~\cite{Alday:2009fs,Drukker:2009id,Drukker:2010jp,Gomis:2010kv}, while for other representations we once more find an invertible linear transformation on the algebra of operators.

\medskip

The outline of this paper is as follows. In \S\ref{section:index} we
construct the difference operators $G_R$ by completing the algebra generated by the difference operators $G_r$, which are labeled by symmetric tensor representations, and we interpret the operators $G_R$  as computing the $\cN=2$ superconformal index in the presence of surface defects. In \S\ref{sec:2dTQFT} we
interpret the difference operators $G_R$ in the limit $p=0$ as Wilson loops wrapping the $S^1$ in an analytic continuation of refined Chern-Simons theory on $S^1 \times C$. In \S\ref{section_Reduction_to_3d} we reduce
the difference operators $G_R$ to three dimensions, and interpret
them as operators that describe line defects when added to the three-sphere partition function. In \S\ref{sec:thooft} we relate the
dimensionally reduced operators $G^{(\mathrm{3d})}_R$ to operators that introduce 't Hooft loops into the four-sphere partition function of the four-dimensional $\cN=2^*$ theory. We finish in \S\ref{section_Conclusions} with a discussion of our findings. Some longer calculations are presented in Appendices \ref{MacPolys}, \ref{appendix:sdualitykernel} and \ref{appendix:special}.


\section{Elliptic algebra of four-dimensional surface defects}
\label{section:index}


\subsection{The superconformal index}

The superconformal index is a trace over the states of a superconformal field theory in radial quantization, or equivalently, a twisted partition function on $S^1 \times S^3$. The most general superconformal index of four-dimensional $\cN=2$ theories is
\begin{align}
\cI = \mathrm{Tr} (-1)^F p^{j_z-r}
q^{j_w-R} t^{r+R} \prod_j a_j^{f_j} \, ,
\label{indexdef}
\end{align}
where the trace is taken over states of the theory in radial quantization annihilated by a single supercharge $\tilde{Q}_{1,\dot{-}}$. Here, we are parametrizing $S^3$ by two complex coordinates $(z,w)$ obeying $|z|^2+|w|^2=1$, and the generators $j_z$ and $j_w$ are rotations in the orthogonal $z$ and $w$-planes respectively. The symbol $r$ denotes the generator of the superconformal $U(1)_r$ and $R$ the generator of the Cartan subalgebra of $SU(2)_R$. The $f_j$ are generators of the Cartan subalgebra of the flavor symmetry group. 

The combinations of generators appearing in the powers of $(p,q,t,a_j)$ in equation~\eqref{indexdef} are those combinations that commute with the supercharge $\tilde{Q}_{1,\dot{-}}$. The letters $p$, $q$, $t$ and $a_i$ are fugacities for these symmetries and obey 
\be
|p|, |q|, |t|, |pq/t|<1, \qquad |a_j|=1 \, ,
\label{eq:fugacityconditions}
\ee 
which ensure that the index is well-defined. 

If there exists a weakly coupled Lagrangian, the superconformal index can be computed from single-letter indices by the plethystic exponential. The basic ingredients are the single letter indices of a half-hypermultiplet and vectormultiplet,
\bea
i_H & = \frac{\sqrt{t} - \frac{pq}{\sqrt{t}} }{(1-p)(1-q)} \, , \\
i_V & = -\frac{p}{1-p} - \frac{q}{1-q} + \frac{\frac{pq}{t}-t}{(1-p)(1-q) } \, .
\eea
For example, the superconformal index of a free hypermultiplet in the bifundamental representation of $SU(N)\times SU(N)$ is
\bea
\cI(a_j,b_j,c) & = \mathrm{PE} \left[ i_H  \sum_{i,j=1}^N \left( a_i b_j c +\frac{1}{a_i b_j c} \right)  \right] \\
& = \prod_{i,j=1}^N \Gamma\left( \sqrt{t} (a_i b_j c)^{\pm};p,q \right),  
\eea
where PE stands for the plethystic exponential. The parameters $\{a_i\}$ and $\{b_j\}$ are fugacities for the $SU(N)\times SU(N)$ symmetry and $c$ is the fugacity for the overall $U(1)$ symmetry. The elliptic gamma function $\Gamma(z;p,q)$ is defined as
\bea
\Gamma(z;p,q) = \prod_{i,j=0}^{\infty} \frac{(1-z^{-1} p^{i+1} q^{j+1})}{(1-z p^{i} q^{j})} \, . 
\eea

An important operation on the superconformal index is that of gauging a global symmetry. Given the superconformal index $\cI(a)$ of a theory with $SU(N)$ flavor symmetry, the superconformal index of the theory where this symmetry has been gauged is
\be
\oint \Delta(a) \,  \cI_V(a) \, \cI(a)\, ,
\ee
where
\be
\cI_V(a) = \mathrm{PE} \Big[ \, i_V \Big( \sum_{i,j=1}^N \frac{a_i }{ a_j} -1 \Big) \,  \Big]
\ee
is the superconformal index of an $SU(N)$ vectormultiplet and
\be
\Delta(a) = \left[ \prod_{j=1}^{N-1}  \frac{da_j}{2\pi i a_j} \, \right] \frac{1}{N!} \prod_{i\neq j}^N\Big(1-\frac{a_i}{a_j} \Big)
\ee
is the Haar measure on the maximal torus of $SU(N)$. 


\subsection{Surface defects from vortices}

In this section, we review the construction of the superconformal
index in the presence of a certain class of surface defects, which
arise as the infinite tension limit of background vortex
configurations~\cite{Gaiotto:2012xa}. They are labeled by a nonnegative integer $r$, the
vortex number, which may be interpreted as the magnetic flux through
the vortex core. 
  
The starting point is any superconformal 
field theory  $\cT_{IR}$ with a global flavor symmetry 
$SU(N)$. By gauging this flavor symmetry, the theory may be
coupled to a hypermultiplet in the bifundamental representation of
$SU(N) \times SU(N)$.  The resulting superconformal field theory $\cT_{UV}$ has an
additional baryonic $U(1)$ symmetry acting on the bifundamental
hypermultiplet. 

The two theories $\cT_{IR}$ and $\cT_{UV}$ are related by a
renormalization flow initiated by turning on a Higgs branch
vacuum expectation value for the bifundamental scalar field $Q$. When
this expectation value is a constant, the RG flow brings us back
to the theory $\cT_{IR}$. When the expectation value is taken to be
coordinate-dependent, the theory $\cT_{IR}$ is modified along a
surface and in the low energy limit we recover the theory $\cT_{IR}$
in the presence of a surface defect. 

More precisely, we can introduce a vacuum expectation value for the baryon operator $B = \det Q$ of the form
\be
B(z) = \prod_{i=1}^r (z-z_i) \, ,
\ee
where $z$ is a complex coordinate in a two-plane, the degree $r$ corresponds to the
vortex number, and the parameters $z_i$ are the positions of the vortex strings. Taking the $z_i=0$, we have $r$ coincident vortices. This construction then leads to surface defects labeled by $r\in \mathbb{Z}_{\geq 0}$. For $\cN=2$ superconformal field theories of class $\cS$, this construction has an elegant interpretation in terms of the curve $C$ - see Figure~\ref{fig:UVIR1}.
 
\begin{figure}[t]
\centering
\includegraphics[width=0.7\textwidth]{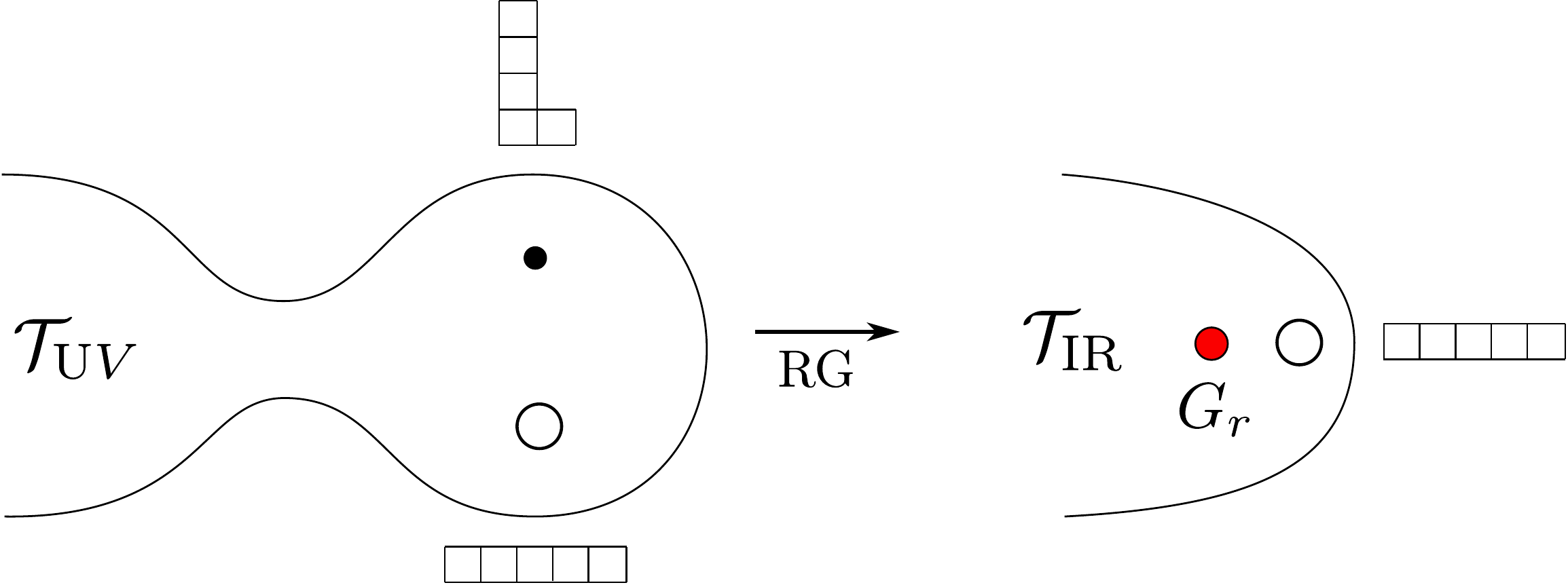}
\caption{The left picture illustrates
    the Riemann surface $C$ corresponding to 
  a theory $\cT_{UV}$, which is obtained by coupling the theory
  $\cT_{IR}$ to a bifundamental field. An RG flow, that is
  initiated by turning on a Higgs vev for the 
bifundamental scalar, relates the theory $\cT_{UV}$ to the original
theory $\cT_{IR}$ with a surface defect $G_r$. This is illustrated on
the right.} 
\label{fig:UVIR1}
\end{figure}

This field theoretic construction of surface defects can be implemented concretely in the superconformal index for surface defects supported on the $S^1\times S^1$ defined by the locus $\{ z=0\}$. Denoting the superconformal index of $\cT_{IR}$ by $\cI_{IR}(a_j,\ldots)$, then the superconformal index of $\cT_{UV}$ is
\be
\cI_{UV}(b_j,c,\ldots) = \oint  \Delta(a_i) \, \cI_{V}(a_i) \, \cI_H(a_i,b_j,c) \, \cI_{\rm IR}(a_i^{-1}, \ldots) \, .
\ee
This has simple poles that originate from simple poles in the
integrand pinching the contour. We consider the simple poles of the integrand coming from the bifundamental hypermultiplet index at
\be
\label{eqn:poleai}
a_i = t^{\frac{1}{2}}q^{m_i} \frac{1}{b_{\sigma(i)} c} \, ,
\ee 
where $\sigma$ is a permutation of $\{1, \ldots, N\}$ and $\sum_{i}m_i = r$ where $r\in \mathbb{Z}_{\geq0}$. They correspond to the chiral ring generated by derivatives of components of the bifundamental scalar field, $(\partial_w)^{m_i}Q_i^{\sigma(i)}$. For each permutation $\sigma$, these poles pinch the contour when
\be
\label{eqn:polec}
c =  t^{\frac{1}{2}} q^{\frac{r }{ N }} \, ,
\ee
leading to a simple pole in the integral at this point. This pole then corresponds to the chiral ring generated by derivatives of the baryon operator $(\partial_w)^r B$ where $B = \det Q$, which is charged only under the $U(1)$. The residue at this pole corresponds to
the index of $\cT_{IR}$ in the presence of a surface defect obtained by giving an expectation value $B = z^r$ to the baryon operator of $\cT_{UV}$ and flowing to the IR.

As demonstrated in~\cite{Gaiotto:2012xa}, the residue takes the form of a difference operator $G_r$ acting on the superconformal index of $\cT_{IR}$. There is one term in the operator for each distinct set of integers $\{m_1,\ldots,m_N\}$ such that $\sum_i m_i =r$. The precise prescription defining the difference operator is
\be
G_r \cdot \cI_{IR}(b_i,\ldots) = N \, \cI_V(b_i) \, \underset{c =  t^{\frac{1}{2}}q^{\frac{r}{N}} }{\mathrm{Res}} 
\, \left[ \, \frac{1}{c}\, \cI_{\rm UV}(c,b_i,\ldots) \, \right].
\ee
The result of the computation is
\be
\label{eqn:symmelldiffop}
G_r \cdot \cI(b_i) = \sum_{\sum_{j=1}^N m_j = r}  \;
\prod_{i,j=1}^N \left[ \,  \prod_{m=0}^{m_j-1} \frac{\theta\left( q^{m+m_j-m_i} t b_i / b_j ; p\right)}{\theta\left( q^{m-m_j} b_j/b_i ; p \right)} \right]  \; \cI\left(b_i \mapsto q^{\frac{r}{N}-m_i} \, b_i \right) \, ,
\ee
where the theta-function is defined as
\bea
\theta(z;p) = \prod_{i=0}^{\infty} \left(1-zp^i\right)\left(1-\frac{p^{i+1}}{z}\right).
\eea

The difference operators $G_r$ constructed by this method are formally self-adjoint with respect to the measure $\Delta(a) \, \cI_V(a)$ used for gauging. They are labeled by a
nonnegative integer $r \in \mathbb{Z}_{\geq 0}$. Furthermore each
term in the operator can be identified with a weight of the $r$-th
symmetric tensor representation of $su(N)$. In particular, the numbers
$\{m_1,m_2,\ldots,m_N\}$ denote the number of times the integers $\{1,\ldots,N\}$
appear in the corresponding Young tableau. Based on this observation,
we associate these operators to surface defects labeled by the
symmetric tensor representations of $su(N)$. 

It is, however, expected
that there exist surface defects labeled by \emph{arbitrary}
irreducible representations of $su(N)$. The necessity of such defects
becomes apparent when the difference operators are composed.


\subsection{Composition of difference operators}

Let us now consider the composition of two difference operators, $G_{r_1} \circ G_{r_2}$. This can be given a physical interpretation by coupling the theory $\cT_{IR}$ to a
single hypermultiplet $Q_{1}$ in the bifundamental representation of $SU(N)
\times SU(N)$ and then to another bifundamental
hypermultiplet $Q_2$. The resulting theory $\mathcal{T}'_{UV}$ is
illustrated in Figure~\ref{fig:UVIR2}. It has two additional flavor symmetries
$U(1)_{1}$ and $U(1)_{2}$ that act on the two bifundamental
hypermultiplets $Q_1$ and $Q_2$ respectively.
  
The original theory $\cT_{IR}$ is reached by turning on constant vacuum expectation values for both baryon operators $B_1 = \det Q_1$ and $B_2 = \det Q_2$ charged under the additional flavor symmetries $U(1)_{1}$ and $U(1)_{2}$. In the superconformal index, this corresponds to the residues of $\cI_{UV}$ at the simple poles $c_1 = t^{1/2}$ and $c_2 = t^{1/2}$ in the fugacities associated to $U(1)_{f,1}$ and $U(1)_{f,2}$ respectively. Turning on position dependent vacuum expectation values $B_1 = z^{r_1}$ and $B_2 = z^{r_2}$ corresponds to computing the residues at simple poles $c_1 = t^{1/2}q^{r_1}$ and $c_2 = t^{1/2} q^{r_2}$. The order in which the residues are computed is irrelevant and  the result
\be
G_{r_1} \cdot \left( G_{r_2} \cdot \cI_{IR} \right) = G_{r_2} \cdot \left( G_{r_1} \cdot \cI_{IR} \right)\, ,
\ee
defines the (commutative) composition $G_{r_1} \circ G_{r_2}$. This construction again has an interpretation in terms of the curve $C$ for theories of class $\cS$, shown in Figure~\ref{fig:UVIR2}.

\begin{figure}[t]
\centering
\includegraphics[width=0.8\textwidth]{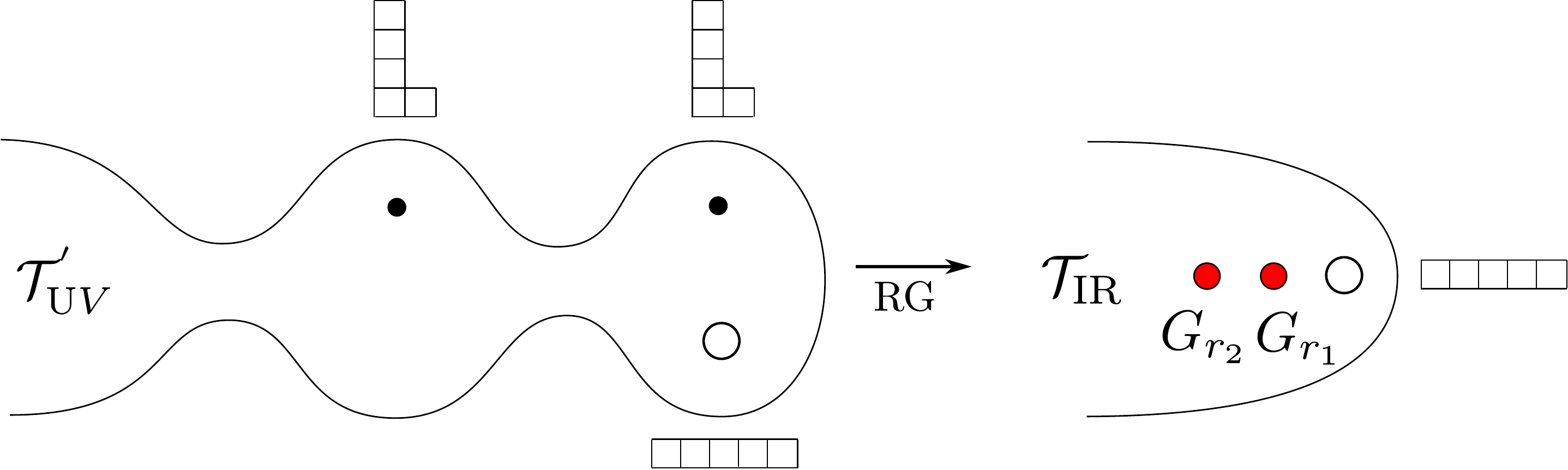}
\caption{The left picture illustrates the Riemann surface $C$ corresponding to
  the theory $\cT_{UV}^{'}$, which is obtained by coupling the theory
  $\cT_{IR}$ to two bifundamental fields. An RG flow, that is
  initiated by turning on Higgs vevs for both 
bifundamental scalars, relates the theory $\cT_{\rm UV}^{'}$ to the original
theory $\cT_{\rm IR}$ with two surface defects $G_{r_1}$ and $G_{r_2}$. This is illustrated on
the right.}
\label{fig:UVIR2}
\end{figure}


\subsection{The algebra of surface defects}

The operators $G_{r}$ constructed above do not form a closed algebra 
under composition and addition. More precisely, except for $su(2)$, the composition $G_{r_1} \circ
G_{r_2}$ cannot be decomposed as a sum of other operators
$G_{r_3}$ with coefficients that are independent of the flavor
fugacities $\{a_j\}$ acted on by the operators. 

In order to close the algebra, we need to enlarge the set of difference operators
$G_r$. Having identified the label $r$ with the $r$-fold symmetric
tensor representation of $su(N)$, it is natural to introduce operators
$G_R$ for any irreducible representation $R$ of $su(N)$ and to force them to
obey the algebra
\be
G_{R_1} \circ G_{R_2} = \sum_{R_3} {\cN_{R_1,R_2\,}}^{R_3}~G_{R_3} \, ,
\ee
where the coefficient ${\cN_{R_1,R_2\,}}^{R_3}$ is non-zero only when
the representation $R_3$ appears in the direct sum decomposition of the tensor product $R_1 \otimes
R_2$. Indeed, it turns out that this determines the operators  $G_{R}$ and the algebra coefficients ${\cN_{R_1,R_2\,}}^{R_3}$ essentially
uniquely, in a sense we explain in detail below. The closure of the algebra is a highly non-trivial  
statement, however, depending on intricate theta-function
identities.  

Let us explain the procedure is some more detail. For each irreducible representation $R$ of $su(N)$, we make an ansatz for the operator $G_{R}$. The ansatz is a sum over the weights $\lambda$ of the representation $R$,
\be
G_R \cdot \cI( a_i ) = \sum_{\lambda} C_{R,\lambda}(p,q,t,a_j) \, \cI ( q^{-(\lambda,h_i)} a_i ) 
\ee
with some unknown functions $C_{R,\lambda}(p,q,t,a_j)$. Here, the
bracket $(\, , \,)$ denotes the standard inner product on the Cartan
subalgebra normalized so that $(e_i,e_i)=2$ for all simple
roots. Furthermore,  
$h_i$ are the weights of the fundamental
representation. They obey $(h_i,h_j)=\delta_{i,j}-1/N$. 

The weights of an irreducible representation $R$ of $su(N)$ can be represented
by semi-standard Young tableaux, that are obtained by placing a number
$1, \ldots, N$ in each box of the Young diagram (as we review in
Appendix~\ref{MacPolys}). Each weight can be written as a sum
\be
\lambda = \sum_{j=1}^N m_j \, h_j \, ,
\ee 
where $m_j$ are the filling numbers of the corresponding semi-standard
Young tableau. In particular,
the weights of the $r$-th symmetric tensor representation are given by  
\be
\lambda = \sum_{j=1}^N m_j \, h_j \, ,
\ee 
where the numbers $m_i$ are such that $\sum_j m_j =
r$. Since $(\lambda,h_i) = m_i - \frac{r}{N}$, the chosen 
ansatz is compatible with the symmetric tensor operators $G_r$ that we already know.

Now we substitute the coefficients $C_{R,\lambda}(p,q,t,a_j)$ for the
symmetric tensor operators, as well as our ansatz for the remaining
representations, into the algebra relations
\be
G_{R_1} \circ G_{R_2} = \sum_{R_3} {\cN_{R_1,R_2\,}}^{R_3}~G_{R_3} \, .
\ee
We first solve these relations for the $su(2)$ coefficients
${\cN_{r_1,r_2\,}}^{r_3}(p,q,t)$, and propose a generalization
for the $su(N)$ coefficients ${\cN_{R_1,R_2\,}}^{R_3}(p,q,t)$. Then we
find that the remaining coefficients
$C_{R,\lambda}(p,q,t,a_j)$ are determined uniquely. The fact that 
this procedure works requires intricate theta-function identities, providing a
strong self-consistency check of our ansatz. 

As a preliminary step, we  introduce a small normalization of the
operators $G_{r}$ labeled by $r$-th symmetric tensor
representations. We redefine the operators by multiplying them by the factor
\be
\cN_r  = t^{-r(N-1)/2}\prod_{i=0}^{r-1}\frac{\theta(q^{-1-i},p)}{\theta(t q^i,p)}\, .
\ee
The purpose of the normalization is to render the leading algebra
coefficient \label{footnote:orderYoungdiagram} equal to one. In the Schur limit $\{p,q,t\} \to \{p,q,q\}$ this normalization factor reduces to $\cN_r \to (-1)^rq^{-\frac{1}{2}r(r+N)}$, in agreement with
the normalization factor in \cite{Alday:2013kda}. 

\subsubsection{Rank 1}\label{subsubsec:SU2}

A good starting point is $su(2)$, since its irreducible representations are exhausted by
$r$-fold symmetric products of the fundamental representation. Thus, the algebra of difference operators should close without introducing any new operators. In particular, we
expect that the product $ G_{r_1}  \circ  G_{r_2}$ can be decomposed according to the tensor
product of the corresponding irreducible representations 
\begin{align}
G_{r_1}  \circ G_{r_2} = \sum_{r=|r_1-r_2|}^{r_1+r_2}
{\cN_{r_1,r_2}}^{ r_3} \, G_{r_3} \, ,
\end{align}
where we can compute the OPE coefficients
${\cN_{r_1,r_2}}^{r_3}(p,q,t)$. Consistency of this structure demands that
the coefficients ${\cN_{r_1,r_2}}^{r_3}$ constructed in this way are independent of the
fugacity parameter $a$. 

For simplicity, let us first consider the Macdonald limit $p\to0$. In this limit, the ratios of theta-functions in the
operators are replaced by rational functions of the remaining
variables $q$ and $t$. The operators $G_{r}$ become
\be
G_{r} \cdot \cI(a_i) = \cN_r \sum_{m_1 + m_2 = r}  \;
\prod_{i,j=1}^2 \left[  \prod_{m=0}^{m_j-1} \frac{\left( 1-q^{m+m_j-m_i} \frac{t a_i} { a_j} \right)}{\left(1- q^{m-m_j} \frac{a_j}{a_i} \right)} \right]  \; \cI\left(a_i \mapsto q^{\frac{r}{N}-m_i} \, a_i \right),
\ee
where $a_1 = a$ and $a_2 = a^{-1}$. 

When composing any two such rational operators $G_{r_1}$ and
$G_{r_2}$, we indeed find that the product
$G_{r_1} \circ G_{r_2}$ 
decomposes according to the tensor product of the corresponding
irreducible representations, in such a way that the structure
constants ${\cN_{r_1,r_2}}^{r_3}(q,t)$ are rational functions of $q$
and $t$. 

As mentioned above, we have
 normalized the difference operators such that the
structure constant for the leading OPE coefficient ${\cN_{r_1,r_2}}^{r_1+r_2}=1$. The remaining structure constants
can be computed straightforwardly in each case. For example,
$G_1 \circ G_1 =
G_2 + {\cN_{1,1}}^{0} \, G_0$, where 
\be
{\cN_{1,1}}^{0}(q,t)  = \frac{(1+t)(1-q)}{(1-q t)}\, .
\ee
This is a particular case of the more general decomposition
\be
G_1 \circ G_r = G_{r+1} + {\cN_{1,r}}^{r-1} \, G_{r-1},
\ee 
where
\be
{\cN_{1,r}}^{r-1}(q,t) = \frac{(1-t^2q^{r-1})(1-q^{r})}{(1-tq^{r-1})(1-tq^r)}\, .
\ee
Similar formulae can be derived for any other example. 

Remarkably,
we observe that the structure constants ${\cN_{r_1,r_2}}^r(q,t)$ are equal
to the $(q,t)$-deformed Littlewood-Richardson coefficients. In other
words, the operators $G_r$ in the limit $p\to0$ obey the same algebra as the Macdonald
polynomials $P_r(a,q,t)$ for $su(2)$.  (We refer to
Appendix~\ref{MacPolys} for more details regarding Macdonald polynomials and
$(q,t)$-deformed Littlewood-Richardson coefficients.)

It turns out that the structure constants of the general elliptic operator algebra can
be obtained in a canonical way by ``lifting'' the structure constants
${\cN_{r_1,r_2}}^{r_3}(q,t)$  of
the Macdonald algebra. This works as follows.  First we express the $(q,t)$-deformed
Littlewood-Richardson coefficients as rational
functions consisting of factors of the form $(1-x)$, where $x$ is a
monomial of the form $q^{\alpha}t^{\beta}$. Then we ``lift'' each
factor to an elliptic function $\theta(x,p)$ whose second argument is the
additional parameter $p$. The original coefficients are obtained in
the limit $p\to0$. 

Note that even though there are ambiguities in writing the $(q,t)$-deformed Littlewood-Richardson coefficients as rational functions of the form $(1-x)$, such as for example in 
\begin{align}
{\cN_{1,1}}^0(q,t) = \frac{(1-t^2) (1-q)}{ (1-t)  (1-qt)} =  \frac{(1-\frac{1}{t^2}) (1-\frac{1}{q}) }{(1-\frac{1}{t}) (1-\frac{1}{qt})} \, ,
\end{align}
the elliptic lift 
\bea
{\cN_{1,1}}^0(p,q,t) &= \frac{\theta(t^2,p) \,
  \theta(q,p)}{\theta(t,p) \, \theta(qt,p)} \, .
\eea 
is uniquely defined because of the theta-function identity
\begin{align}
\theta(z^{-1};p) = - \frac{1}{z} \, \theta(z;p) \, . 
\end{align}

Verifying the composition rules for the elliptic difference operators $G_r$ now requires numerous theta-function identities. For instance,  
checking that $G_1 \circ G_1 =G_2 + {\cN_{1,1}}^{0}(p,q,t) \, G_0$ 
requires
\bea
\frac{\theta(t^2,p)\theta(q,p)}{\theta(t,p)\theta(qt,p)} =&+\frac{\theta \left(q^{-2},p\right) \theta \left(t^{-1},p\right)
  \theta \left(ta^{-2},p\right) \theta \left(t a^2,p\right)}{\theta
  \left(q^{-1},p\right) \theta \left(q^{-1}a^{-2},p\right) \theta
  \left(q^{-1}a^2,p\right) \theta (q t,p)} \\
  &-\frac{\theta \left(t^{-1},p\right) \theta \left(ta^{-2},p\right)
  \theta \left(t q^{-1}a^2,p\right)}{\theta \left( a^{-2},p\right)
  \theta (t,p) \theta \left(q^{-1}a^2,p\right)} \\ 
  & - \frac{\theta
  \left(t^{-1},p\right) \theta \left(t a^2,p\right) \theta \left(t
    q^{-1}a^{-2},p\right)}{\theta \left(a^2,p\right) \theta (t,p)
  \theta \left(q^{-1}a^{-2},p\right)}  \, ,
\eea
which can be checked for instance by expanding around $p=0$. 

Similarly, when composing the fundamental operator $G_1$ with the
operator $G_r$ for any other
irreducible representation of $su(2)$, we find that another elliptic
theta-function identity brings the non-trivial structure constant into the form 
\be
{\cN_{1,r}}^{r-1}(p,q,t) = \frac{\theta(t^2
  q^{r-1},p)\, \theta(q^{r},p)}{\theta(t q^{r-1},p) \, \theta(t q^r ,p)}\, .
\ee
In fact, for any other check we did, we find that the structure constants
${\cN_{r_1,r_2}}^{r_3}$ are independent of the fugacity parameter $a$ and can be expressed as 
ratios of theta-functions. Even better, we find that they are elliptic
(lifts of $(q,t)$-deformed) Littlewood-Richardson coefficients, in the sense explained above. 

The elliptic
operators $G_r$ thus obey an elliptic version of the Macdonald
polynomial algebra. In particular, this provides evidence for the
conjecture that the surface defects labeled by $r\in
\mathbb{Z}_{\geq0}$ are to be identified with irreducible
representations of $su(2)$.

\subsubsection{Higher rank}

For $su(N)$, with $N>2$, the algebra of the difference operators $G_r$ is not
closed. We introduce a new set of operators $G_R$ labeled by
irreducible representations of $su(N)$, and identify the difference operators
$G_r$ with the operators $G_{(r)}$ labeled by the rank
$r$ symmetric tensor representation\footnote{We label by $(\ell_1, \ldots, \ell_{N-1})$ the representation associated to the Young diagram whose $j$-th row has length $\ell_j$. }. We systematically find 
expressions for the novel operators by imposing the algebra
\begin{align}
G_{R_1}  \circ G_{R_2} = \sum_{R_3}
{\mathcal{N}_{R_1,R_2}}^{R_3}\, G_{R_3}\, ,
\end{align}
where we assume that the coefficients $\mathcal{N}_{R_1,R_2}^{\qquad
  R_3}(p,q,t)$ are given by the elliptic (lifts of $(q,t)$-refined)
Littlewood-Richardson coefficients, which can be found uniquely for any
triple of representations $R_1$, $R_2$ and $R_3$. 

In the rank 2 and 3 cases, we have explicitly computed a large set of elliptic
difference operators $G_R$, and performed ample consistency
checks amongst them. These computations reveal several structures
amongst the difference operators, and we
are to make some proposals for general $N$. Let us give a few
examples here. 

First, consider the composition
of two operators each
labeled by the fundamental representation, $G_{(1)} \circ G_{(1)}$.
 This representation $(1) \otimes (1)$ decomposes into the
symmetric tensor $(2)$ and the antisymmetric tensor
$(1,1)$ representations. 
The coefficient of the operator $G_{(2)}$ labeled by the symmetric
tensor representation
is one, following from our choice of normalization. Choose the
coefficient 
\begin{equation}
{\mathcal{N}_{(1),(1)}}^{(1,1)}(p,q,t) = \frac{\theta(q,p)\,
  \theta(t^2,p)}{\theta(t,p) \,\theta(qt,p)}
\end{equation} 
to be the uplift of the corresponding $(q,t)$-deformed Littlewood-Richardson coefficient.
The difference operator ${ G}_{(1,1)}$ labeled by the
rank-two antisymmetric tensor representation of $su(N)$ can then be
determined from the equation  
\begin{align}
G_{(1)}  \circ
  G_{(1)}  = 
 G_{(2)} + 
\frac{\theta(q,p) \, \theta(t^2,p)}{\theta(t,p) \, \theta(qt,p)} \, 
 G_{(1,1)}\, .
\end{align}
By this method, we find that the elliptic difference
operator $G_{(1,1)}$ for the antisymmetric tensor representation is
given by 
\be
  G_{(1,1)}  \cdot \, \cI(a_i) = t^{-1} \sum_{j_1<j_2} \prod_{k \neq
   \{j_1,j_2 \}}
 \frac{ \theta \left( \frac{t }{q } a_{j_1}/a_k, p \right) \theta \left( \frac{t }{q } a_{j_2}/a_k, p \right) }{ \theta \left( a_k / a_{j_1} , p \right) \theta \left( a_k / a_{j_2} , p \right)} \, \cI \left( q^{\frac{2}{N} - \delta_{i,\{j_1,j_2\}} } a_i \right)   .
\ee
The term in the sum labeled by $j_1 < j_2$ corresponds to 
the weight $\lambda = h_{j_1} + h_{j_2}$ in the antisymmetric tensor
representation $(1,1)$.

Next, we determine the difference operator
$G_{(2,1)}$ from the equation
\begin{align}
 \left( G_{(2)}  \circ
  G_{(1)}  \right) \cdot \cI = 
G_{(3)}\, \cdot \cI + 
\frac{\theta(q^2,p) \, \theta(q t^2,p)}{\theta(q t,p) \, \theta(q^2
  t,p)} \, 
G_{(2,1)}\, \cdot \cI \, , 
\end{align}
where 
\begin{equation}
{\mathcal{N}_{(2),(1)}}^{(2,1)}(p,q,t) = \frac{\theta(q^2,p) \, \theta(q t^2,p)}{\theta(q t,p) \, \theta(q^2
  t,p)} 
\end{equation} 
is the elliptic lift of the $(q,t)$-deformed Littlewood-Richardson
coefficient ${\mathcal{N}_{(2),(1)}}^{(2,1)}(q,t)$.

We verify that the difference operator $G_{(2,1)}$ can
indeed be written as a sum over the weights $\lambda = \sum_ i m_i
h_i$ with $\sum_i m_i = 3$, i.e.\ as a sum over  the weights in the
representation labeled by the Young diagram $(2,1)$. These weights can
be divided into two groups. The weights $\{m_{i_1}=m_{i_2} = m_{i_3} =1\}$
occur with multiplicity two, whereas the weights $\{m_{j_1}=2,\,
m_{j_2}=1\}$ occur with multiplicity one. 

We then expand the resulting operator to lowest order in $p$, read off its elliptic
lift and check this in an expansion in $p$. For instance, for $su(3)$
we find that  
\bea 
 G_{(2,1)}  \cdot  \cI(a_1, a_2, a_3) =  & \sum_{\sigma \in S_3}
C_{210}(a_{\sigma(1)},a_{\sigma(2)},a_{\sigma(3)}) \,
\cI\left(\frac{a_{\sigma(1)}}{q}, a_{\sigma(2)}, q
  a_{\sigma(3)}\right) \\ & +
C_{111}(a_1,a_2,a_3) \, \cI\big(a_1, a_2, a_3\big) \, .
\eea
The first group of terms in this sum correspond to weights $\lambda = 2
h_{\sigma(1)} + h_{\sigma(2)}$ that occur with multiplicity one. These terms
are given by a single product over ratios of theta-functions:
\begin{align}
C_{210}(a_{\sigma(1)},a_{\sigma(2)},a_{\sigma(3)})=\,
 t^{-2} \,\frac{
\theta \left(\frac{t a_{\sigma(1)}}{q a_{\sigma(2)}},p\right)
\theta \left(\frac{t   a_{\sigma(1)}}{q a_{\sigma(3)}},p\right)
\theta \left(\frac{t a_{\sigma(2)}}{q   a_{\sigma(3)}},p\right)
\theta \left(\frac{t  a_{\sigma(1)}}{q^2 a_{\sigma(3)}},p\right)}
{ \theta  \left( \frac{a_{\sigma(2)}}{a_{\sigma(1)}},p\right)
\theta  \left( \frac{a_{\sigma(3)}}{a_{\sigma(1)}},p\right)
\theta  \left( \frac{a_{\sigma(3)}}{a_{\sigma(2)}},p\right)
\theta \left( \frac{q   a_{\sigma(3)}}{a_{\sigma(1)}},p\right) } \, .
\end{align}
The last term corresponds to the weight $\lambda = h_1 + h_2
+ h_3$, which occurs with multiplicity two. Its contribution is given
by
\bea
& C_{111}(a_1,a_2,a_3) = - t^{-3} \frac{  \theta \left(t ,p\right) \theta
  \left(q^2 t ,p\right) }{ \theta \left(q^{-1},p\right) \theta \left(q
    t^2,p\right) } \\
&\times \left( \, \sum_{\sigma \in S_3}  \frac{  \theta \left(  \frac{t a_{\sigma(1)}}{a_{\sigma(2)}} ,p\right) \theta
  \left(\frac{t a_{\sigma(2)}}{a_{\sigma(1)}} ,p\right)  \theta \left(
    \frac{t a_{\sigma(1)}}{a_{\sigma(3)}} ,p\right) \theta
  \left( \frac{t a_{\sigma(2)}}{a_{\sigma(3)}} ,p\right)  \theta
  \left( \frac{t a_{\sigma(3)}}{ q a_{\sigma(1)}} ,p\right) \theta
  \left( \frac{t a_{\sigma(3)}}{q a_{\sigma(2)}} ,p\right) }{ \theta
  \left( \frac{a_{\sigma(1)}}{q a_{\sigma(2)}} ,p\right) \theta
  \left(\frac{a_{\sigma(2)}}{q a_{\sigma(1)}} ,p\right)  \theta \left( \frac{a_{\sigma(3)}}{a_{\sigma(1)}} ,p\right) \theta
  \left( \frac{q a_{\sigma(1)}}{a_{\sigma(3)}} ,p\right)  \theta \left( \frac{a_{\sigma(3)}}{  a_{\sigma(2)}} ,p\right) \theta
  \left( \frac{q a_{\sigma(2)}}{a_{\sigma(3)}} ,p\right)  } \right.  \\
&  \qquad \left. + \frac{ \theta(\frac{1}{t},p)^2 \, \theta(q^3,p)}{\theta(q ,p)^2\,
  \theta(\frac{1}{q t^2},p)} 
 \frac{
\theta(\frac{t a_1}{a_2},p) \,  
 \theta(\frac{t a_2}{a_1},p) \, 
 \theta(\frac{t a_1}{a_3},p) \, 
\theta(\frac{t a_2}{a_3},p) \, 
 \theta( \frac{t a_3}{a_1},p) \, 
\theta(\frac{t a_3}{a_2},p) 
}{
\theta(\frac{a_1}{q a_2} ,p) \,
\theta(\frac{a_2}{q a_1} ,p) \, 
 \theta(\frac{q a_3}{a_1} ,p) \,
   \theta(\frac{q a_1}{a_3},p) \, 
 \theta(\frac{q a_3}{a_2} ,p) \,
    \theta(\frac{q a_2}{a_3},p)}    \,  \right) .
\eea
The last term in this expression is invariant itself under permutations of $a_1$,
$a_2$ and $a_3$, the first six terms permute into each other. 

Continuing this strategy, one can systematically find the elliptic
difference operators for any given representation $R$ and perform
consistency checks on it. We have explicitly computed 
all $su(3)$ and $su(4)$ difference operators labeled by Young
diagrams with up to four boxes. From these results we infer that 
the difference operator $ G_{(1^r)} $, corresponding to the
rank $r$ antisymmetric representation of $su(N)$, is given by
\be
  G_{(1^r)} \cdot \cI(a_i) = t^{r(r-N)/2} \sum_{|I|=r}
 \prod_{\substack{j \in I \\ k\notin I}} \frac{ \theta ( \frac{t}{q}
   a_j / a_k , p ) }{ \theta ( a_k / a_j , p ) } \, \cI\left(
   q^{\frac{r}{N} - \delta_{i,I} } a_i \right)  , 
 \label{antisymmetric4d}
\ee
where the summation is over subsets $I \subset \{1,\ldots, N\}$ of
length $|I|=r$ and where the symbol $\delta_{i,I}$ is one if $i\in I$
and zero if $i\notin I$. As we will show in more detail in the next section, these operators are related by conjugation to the
Hamiltonians of the elliptic Ruijsenaars-Schneider model. 

\subsection{Properties of difference operators}

Let us summarize a few properties of the resulting difference
operators $G_R$:
\begin{itemize}
\item They are formally self-adjoint respect to the vectormultiplet measure $\Delta(a) \, \cI_V(a)$ on the unit circle $|a|=1$.
\item The composition $G_{R_1} \circ G_{R_2}$ is commutative. 
\item The difference operators $G_R$ obey the algebra
\begin{align}
G_{R_1} \circ G_{R_2} = \sum_{R_3} {\cN_{R_1,R_2}}^{R_3} \, G_{R_3} \, ,
\end{align}
where the coefficients ${\cN_{R_1,R_2}}^{R_3}$ are elliptic lifts of the $(q,t)$-deformed
Littlewood-Richardson coefficients.
\item They can be expanded as 
\begin{align*}
G_R \cdot \cI( a_i ) = \sum_{\lambda} C_{R,\lambda}(p,q,t,a_j) \, \cI ( q^{-(\lambda,h_i)} a_i ),
\end{align*}
where the summation is over weights $\lambda$ in the representation $R$.
\end{itemize}
While we have not found a closed expression for the coefficients $ C_{R,\lambda}(p,q,t,a_j)$, it may be useful to point out the following structures:
\begin{itemize}
\item For general values of the fugacities, $C_{R, \lambda} \neq
  C_{R', \lambda}$ when $\lambda$ is a weight of two different
  representations $R$ and $R'$. Only in the Schur limit
  $q=t$ do the coefficients $C_{R, \lambda} $ depend only on the weight $\lambda$.
\item The coefficients $C_{R, \lambda}$ are given by a single ratio of
  theta-functions when the weight $\lambda$ occurs in the
  representation $R$ with multiplicity one. If $\lambda$ occurs with
  higher multiplicity the coefficient $C_{R, \lambda}$ is a sum of
  ratios of theta-functions.
\item If $\lambda$ and
  $\lambda'$ are in the same Weyl orbit, the coefficients $C_{R, \lambda}$ and $C_{R, \lambda'}$ are related by a permutation of fugacities $a_i$.
\item If $\lambda$ is in the Weyl orbit of the highest weight in the
  representation $R$, the coefficient $C_{R, \lambda}$ does not
  contain theta-functions that are independent of the fugacities
  $a_j$. 
\item If $\lambda$ is not in the Weyl orbit of the highest weight in
  the representation $R$, the coefficient
  $C_{R, \lambda}$ does contain a ratio of such theta-functions that are independent of $a_j$. This ratio can be obtained as an elliptic lift of the corresponding
  coefficient for the Macdonald polynomial $P_{R}(q,t)$. 
\end{itemize}

We also note that the elliptic lift of the $(q,t)$-deformed Littlewood-Richardson coefficients
${\mathcal{N}_{R_1,R_2}}^{R_3}(p,q,t) $ have the same number of
terms in the numerator and denominator. Moreover, when $q=t$ these 
terms all cancel each other. This implies that the elliptic algebra
reduces to the Schur algebra when $q=t$. In this limit all
coefficients $C_{R, \lambda}(p,q,t)$ reduce to a single product
$C_{\lambda}(q)$ depending only on the weight $\lambda$ as found previously in~\cite{Alday:2013kda}.

\subsection{Two-dimensional worldvolume theory}

So far, we have constructed an algebra of operators $G_R$ that
compute the superconformal index in the presence of a set of surface
defects labeled by irreducible representations $R$ of $su(N)$. The
operators $G_{(r)}$, labeled by symmetric tensor representations,
were found in an infinite tension limit of two-dimensional half-BPS
vortices. The operators for the remaining representations were obtained in
a canonical way by completing the algebra.   

An alternative and more direct way to define surface
defects is by coupling to two-dimensional degrees of freedom on
the supported surface. For instance, consider a
two-dimensional $\cN=(2,2)$ gauge theory with flavor symmetry group
$SU(N)$. This two-dimensional theory can be coupled to the four-dimensional theory by gauging the
2d flavor symmetry using the 
restriction of a 4d dynamical or background $SU(N)$
vectormultiplet to the surface $S$. A large class of half-BPS surface
defects defined in this way have been studied in~\cite{Gaiotto:2009fs,Gaiotto:2013sma}. 

The superconformal index of four-dimensional linear $\cN=2$ quiver
theories in the presence of such surface defects can be found by
combining the two-dimensional elliptic genus with the four-dimensional
superconformal index~\cite{Gadde:2013dda}. Let us consider this combined
index in a few
examples of surface defects in $\cN=2$ superconformal QCD,
i.e.~a four-dimensional $SU(N)$ gauge theory coupled to $2N$
hypermultiplets.  

\begin{figure}[t]
\centering
\includegraphics[width=0.45\textwidth]{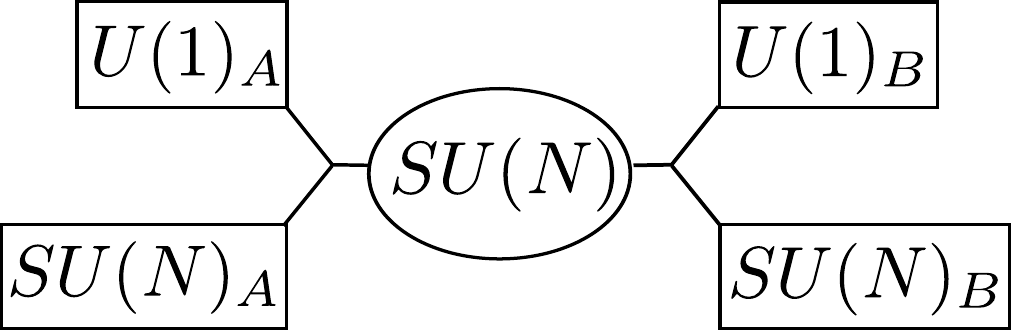}
\caption{Linear quiver description of $\cN=2$ superconformal QCD. The flavor symmetry group of each set of $N$ hypers is enhanced to $U(N)$. This splits into an $SU(N)$ plus a diagonal $U(1)$ flavor symmetry group.}
\label{fig:QCDquiver}
\end{figure}

Before introducing surface defects, let us remind ourselves that
$\cN=2$ superconformal QCD has a dual description as a degeneration 
limit of a Riemann surface with two simple and two full
punctures. Equivalently, its matter content
can be read off from a linear quiver, see Figure~\ref{fig:QCDquiver}. The manifest global
symmetry in this presentation is $$SU(N)_A \times SU(N)_B \times
U(1)_A \times U(1)_B \, .$$ If we denote the 
corresponding fugacities by $(a_i,b_i,x,y)$, the superconformal index
of superconformal QCD is 
\be
\int \Delta(z_j) \, \cI_V(z_j) \, \cI_H(z^{-1}_j,a_i,x) \, \cI_H(z_j,b_i,y) \, .
\ee
Notice that we could have equivalently considered the same theory with
a $U(N)$ gauge group, since the center of mass $U(1)$ decouples
in the IR.

\begin{figure}[t]
\centering
\includegraphics[width=.65\textwidth]{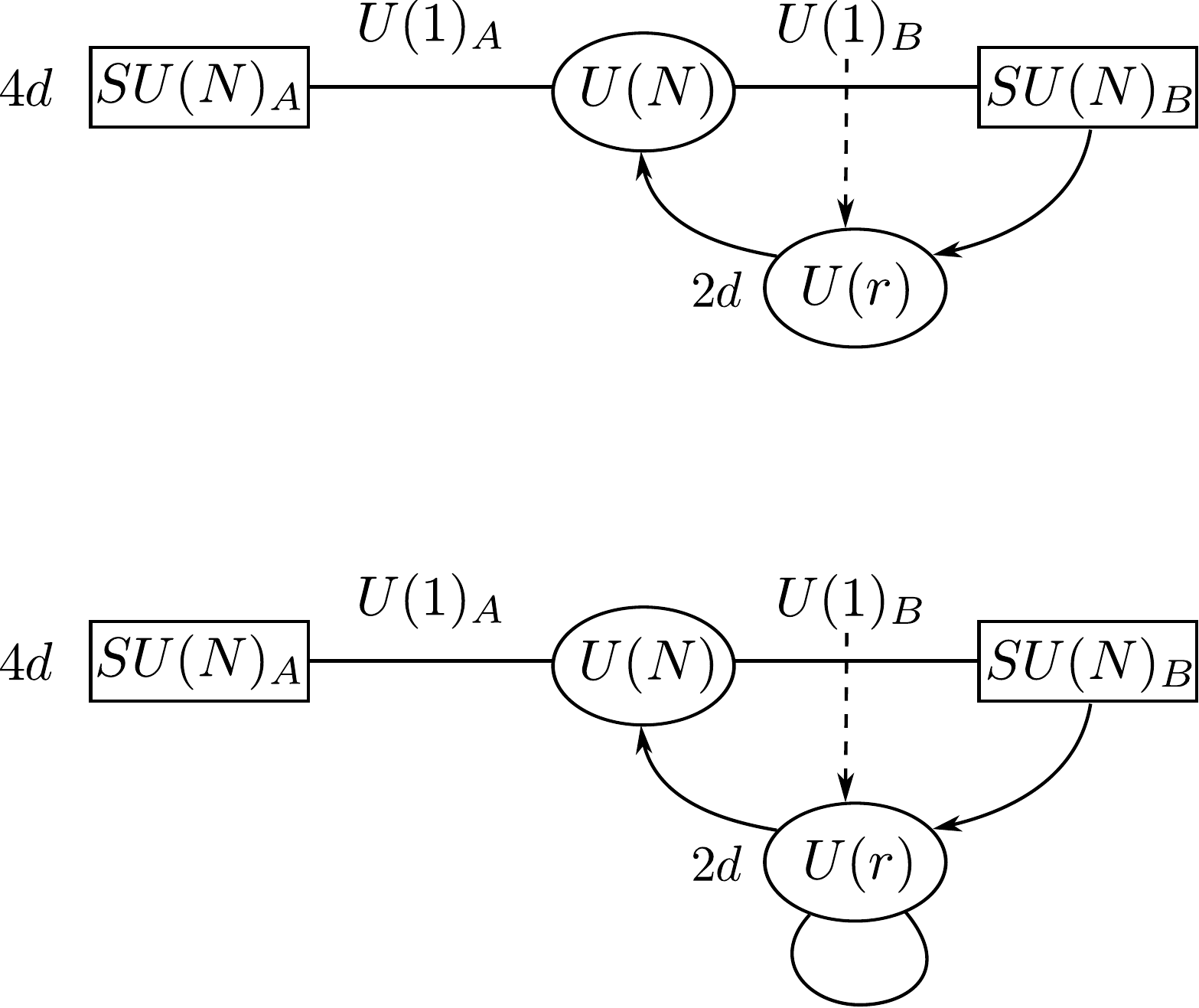}
\caption{On top (bottom): 2d-4d quiver description of a fully antisymmetric (symmetric) surface defect in $\cN=2$ superconformal QCD.} 
\label{fig:QCDquiver+defect}
\end{figure}

Let us now add two-dimensional degrees of freedom to the four-dimensional
superconformal QCD theory with gauge group $U(N)$.  We give two
examples whose 2d-4d quiver descriptions are shown in 
Figure~\ref{fig:QCDquiver+defect}.   

As a first example, we consider a two-dimensional $\cN=(2,2)$ gauge theory with
gauge group $U(r)$ coupled to $N$ fundamental and $N$ anti-fundamental
chiral fields. The two-dimensional flavor
symmetry group is thus $U(N)_f 
\times U(N)_a$. We couple the $N$ fundamental chirals to the $U(N)$ gauge
symmetry, and the $N$
anti-fundamental chirals to the $SU(N)_B \times U(1)_B$ global  
symmetry, as described in \cite{Gadde:2013dda}. The resulting quiver is illustrated on top in
Figure~\ref{fig:QCDquiver+defect}. The superconformal index
of the resulting 2d-4d system is  
\be
\int \Delta(z_j) \, \cI_V(z_j) \, \cI_H(z^{-1}_j,a_i,x) \left( \cO_r \cdot \cI_H(z_j,b_i,y)  \right)  ,
\ee
where the operator $\cO_r$ acts as
\be \label{eq:2d4dindexanti}
\cO_r \cdot \cI(z_i) = \sum_{|I|=r}
 \prod_{\substack{j \in I \\ k\notin I}} \frac{ \theta ( 
   \frac{t}{q} z_j/z_k , p ) }{ \theta ( z_k/z_j, p ) } \, \cI \left( q^{
     - \delta_{i,I} } z_i \right) 
\ee
on the fugacities $z_i$ of the $U(N)$ gauge symmetry.

The terms in the above expression are in one-to-one  correspondence with the
$N \choose r$ Higgs branch vacua of the two-dimensional theory, in
which certain components of the chiral fields get a vacuum expectation
value. Each term in equation~(\ref{eq:2d4dindex}) can be
interpreted as computing the index of the 2d-4d system in one of these
vacua. 

The operator $\cO_r$ agrees with the elliptic difference operator $G_{(1^r)}$ labeled by the antisymmetric tensor representation $(1^r)$ of rank $r$ up to an overall fractional shift by $q^{\frac{r}{N}}$. Since the shifts $z_i \mapsto q^{-\delta_{i,I}} z_i$ do not preserve the condition $\prod_j z_j =1$,  the fugacities $z_j$ should really be interpreted as $U(N)$ (instead of
$SU(N)$) fugacities. 
To find the exact operators $G_{(1^r)}$, however, we
would need to find a system that couples the same two-dimensional
degrees of freedom to a 4d theory with genuine $SU(N)$ symmetry groups. 
This is for instance required to understand surface defects
in the four-dimensional $\mathcal{T}_N$ theory, whose flavor symmetry groups cannot be enlarged to $U(N)$.   

As a second example, we add a chiral field in the adjoint representation to the
two-dimensional $\cN=(2,2)$ theory that we considered before. The
quiver description can be found on the bottom of
Figure~\ref{fig:QCDquiver+defect}. The
presence of the adjoint field drastically changes the vacuum
structure, which is mirrored in the expression for the superconformal
index.  The index of the 2d-4d system is the same as before, except
that the operator $\cO_r$ now acts as
\be \label{eq:2d4dindex}
\cO_r \cdot \cI(z_i) = \sum_{\sum_{j=1}^N m_j = r}  \;
\prod_{i,j=1}^N \left[ \, \prod_{m=0}^{m_j-1} \frac{\theta\left( q^{m+m_j-m_i} t z_i / z_j , p\right)}{\theta\left( q^{m-m_j} z_j/z_i , p \right)} \right]  \; \cI\left(q^{-m_i} \, z_i \right) .
\ee
This expression coincides with the operator $G_{(r)}$
associated to the symmetric representation $(r)$ of rank $r$, as was
already noted in \cite{Gadde:2013dda},  except that the
fractional shift $q^{\frac{r}{N}}$ is again missing.

For the symmetric as well as the antisymmetric tensor representations the
two-dimensional degrees of freedom on the surface defect introduced by 
the operators $G_R$, can thus be identified with certain two-dimensional
$\cN=(2,2)$ gauge theories, up to some shifts. 

It would  be interesting to observe whether the $S^2$ partition
function of these $\cN=(2,2)$ theories can be obtained from Toda
correlators with degenerate vertex operators labeled by highest weights
of the symmetric and antisymmetric tensor representations. This has
been demonstrated for the fundamental representation
in~\cite{Doroud:2012xw} (see also \cite{Dimofte:2010tz, Benini:2012ui}). 


\section{Two-dimensional TQFT and Verlinde algebra} \label{sec:2dTQFT}

In this section we identify the difference operators $G_R$ with local operators in a topological quantum field theory (TQFT) of the Riemann surface $C$. In the case $p=0$, this can be identified with an analytic continuation of refined Chern-Simons theory on $S^1 \times C$ and the relevant local operators arise from Wilson loops in the representation $R$ and wrapping the $S^1$.  

\subsection{TQFT structure of the superconformal index}
\label{TQFTstructure}

Recall that for any superconformal field theory of class $\cS$ the superconformal
index is independent of marginal couplings and hence of the complex structure of the Riemann surface $C$. This suggests that the superconformal index of these theories has a
dual description as a two-dimensional TQFT
on the Riemann surface $C$~\cite{Gadde:2009kb}. In the Schur limit
(when $p\to 0$ and $q=t$), the TQFT has been
identified as $q$-deformed Yang-Mills theory on $C$ in the zero area limit~\cite{Gadde:2011ik},
or equivalently as an analytic continuation of Chern-Simons theory on
$C \times S^1$. This picture can be extended to the Macdonald limit
$(p \to 0 )$ when the superconformal index has a dual description as
an analytic continuation of refined Chern-Simons 
theory on $C \times S^1$~\cite{Aganagic:2011sg}. 

In order to verify the above relation, it is necessary to extract a
certain function $K(a)$ from the superconformal index for each $SU(N)$
flavor puncture. In what follows, 
we define the normalized index $\cI^{(n)}$ through the equation
\be
\cI(a,b,\ldots) = (K(a) K(b) \cdots )\, \cI^{(n)}(a,b,\ldots ) \, ,
\ee
where 
\be
K(a) = \prod_{i \neq j}^N \Gamma(t a_i/a_j ,p,q) \, .
\ee
The normalized index $\cI^{(n)}$ is now gauged using the measure
\be
\Delta^{(n)}(a) = K(a) \Delta(a)  =  \frac{1}{N!} \left( \frac{(p,p)(q,q)}{\Gamma(t,p,q)} \right)^{N-1} \; \prod_{i \neq j} \frac{\Gamma(t a_i / a_j,p,q)}{\Gamma(a_i/a_j, p,q)}\, .
\ee

The difference operators $\bar G_R$ acting on the normalized index
are thus obtained by conjugation
\be
\bar G_R = \frac{1}{K(a)}   (G_R \cdot K(a)) \, .
\ee
This conjugation leaves the algebra of difference operators unchanged. 
After a long, yet straightforward, computation we find that the
conjugated operators for the fully symmetric representations
$R=(r)$ are given by
\be
\bar G_{(r)} \cdot \cI^{(n)}(a_i)= \cN_r \, \sum_{\sum_{j=1}^N m_j = r}  \; \prod_{i,j=1}^N \prod_{m=0}^{m_j-1} \frac{\theta\left( t q^m a_i /a_j ,p \right)}{\theta\left( q^{m-m_i}a_i / a_j ,p \right)}\; \cI^{(n)}\left(q^{\frac{r}{N}-m_i} a_i \right)
\ee
while those for the fully antisymmetric representations
$R=(1^r)$ are
\be
\bar G_{(1^r)} \cdot \cI^{(n)}(a_i) = t^{r(r-N)/2} \sum_{|I|=r} \prod_{\substack{k\in I \\ j\notin I}} \frac{\theta\left( t a_j /a_k ,p \right)}{\theta\left( a_j / a_k , p \right) } \cI^{(n)}(q^{\frac{r}{N} - \delta_{i,I}}a_i) \, ,
\ee
where the summation is over all subsets $I\subset\{1,2,\ldots,N\}$
of length $r$. Comparing with~\eqref{eqn:symmelldiffop} and~\eqref{antisymmetric4d} noting the reflection property $\theta(z,p) = \theta(p/z,p)$ we see that the effect of the conjugation is simply to interchange $t \leftrightarrow p q /t $. In summary, we have found that
\be
\bar G_R(p,q,t,a) = G_R\left(p,q,\frac{pq}{t},a\right)\, .
\ee
Remarkably, the conjugated antisymmetric tensor operators $\bar G_{(1^r)}$ are
precisely the Hamiltonians of the elliptic Ruijsenaars-Schneider model,
extending the observation made in~\cite{Gaiotto:2012xa}.

We will assume that the difference operators $\bar G_{R}$ admit
a complete set of eigenfunctions $\{\psi_{S}(a_i)\}$, indexed by 
irreducible representations
$S$ of $su(N)$, which are orthogonal with respect to the measure $\Delta^{(n)}(a)$
and have non-degenerate eigenvalues $E^{S}_{(R)}$. In fact, the eigenfunctions are determined by the fully antisymmetric operators $\bar G_{(1^r)}$. With the help of these eigenfunctions $\{\psi_{S}(a_i)\}$ the TQFT structure of the superconformal index can be made very explicit~\cite{Gaiotto:2012xa}. 

Consider for instance the sphere with three maximal
punctures. The corresponding four-dimensional conformal field theory is known
as $\cT_N$. It has at least $SU(N)^3$ flavor
symmetry. Write the normalized superconformal
index of the $\cT_N$ theory as $$\cI^{(n)}(a_i,b_i,c_i) \, ,$$
where the parameters $a_i$, $b_i$ and
$c_i$ are three sets of fugacities dual to the maximal torus of the
$SU(N)$ flavor symmetries.  This superconformal index can be expanded in terms of the set of eigenfunctions
$\{ \psi_S(a_i) \}$ as 
\begin{align}
\cI^{(n)}(a_i,b_i,c_i) = \sum_{S_1,S_2,S_3}
C_{S_1,S_2,S_3} \,\psi_{S_1}(a_i) \psi_{S_2}(b_i) \psi_{S_3}(c_i)\, ,
\end{align}
where $C_{S_1,S_2,S_3}$ are the structure constants of the two-dimensional
TQFT. If we impose that acting with any one of the
operators $\bar G_{(1^r)}$ gives the same result, and assume that the eigenvalues are non-degenerate then the superconformal index is in fact diagonal in this basis
\begin{align}
\cI^{(n)}(a_i,b_i,c_i) = \sum_{S}
C_{S} \, \psi_{S}(a_i) \psi_{S}(b_i) \psi_{S}(c_i)\, .
\end{align}
This is illustrated in Figure~\ref{fig:indexTQFT} for the case of a sphere with three punctures. 
As explained in~\cite{Gaiotto:2012xa}, the constants $C_S$ can be found by comparing two degeneration limits of the $N+1$ punctured sphere with $N-1$ maximal and two minimal punctures.

\begin{figure}[t]
\centering
\includegraphics[width=0.8\textwidth]{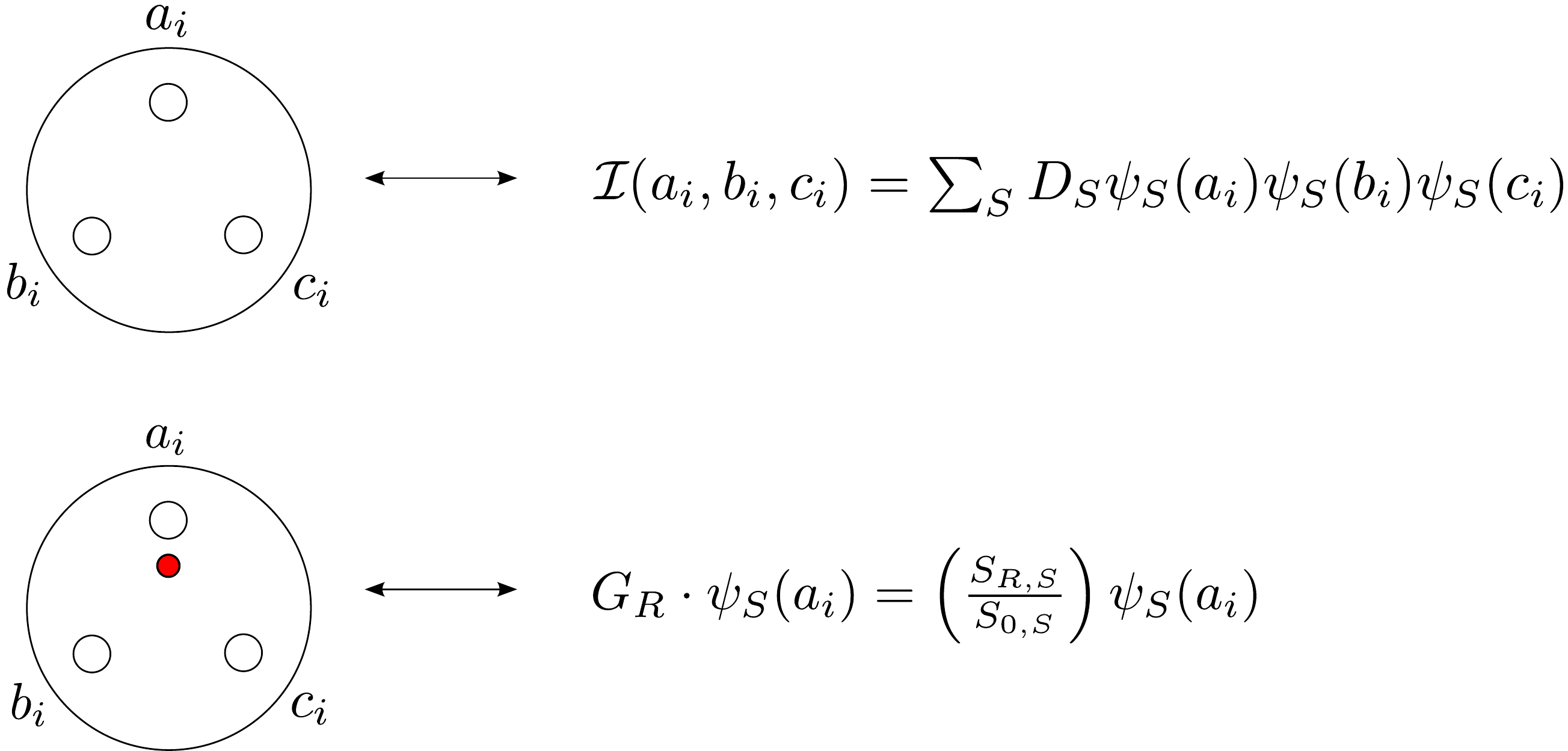}
\caption{The superconformal index can be written as a TQFT
  correlator. This correlator is diagonal in the eigenfunctions
  $\psi_S(a_i)$ of the difference operators $G_R$. } 
\label{fig:indexTQFT}
\end{figure}

In the remainder of this section we will restrict ourselves to the
Macdonald slice $(p,q,t)=(0,q,t)$. In this limit, the antisymmetric
difference operators $\bar{G}_{(1^r)}$ turn into the Macdonald operators 
\be
\bar{G}_{(1^r)} \cdot \cI^{(n)}(a_i) = t^{r(r-N)/2}\sum_{|I|=r} \prod_{\substack{k\in I \\ j\notin I}} \frac{\left( 1-q a_j /a_k ,p \right)}{\left(1- a_j / a_k ,p\right) } \cI^{(n)}(q^{\frac{r}{N} - \delta_{i,I}}a_i)
\ee
while the normalized vectormultiplet measure becomes
\be
\Delta^{(n)}(a) = \prod_{i \neq j} \frac{(1-t a_i / a_j,q)}{(1-a_i/a_j,q)}
\ee
and coincides with the standard Macdonald measure. 

The operators $\bar{G}_{(1^r)}$ are self-adjoint with respect to this measure on the unit circle $|a|=1$ and their common eigenfunctions are the Macdonald polynomials $P_S(a_i;q,t)$, which are labeled by irreducible representations of
$su(N)$. They are by construction orthogonal with respect to the measure
$\Delta^{(n)}(a)$ and are normalized such that  
\be
P_S(a_i;q,t) = \chi_S(a_i) + \sum_{T<S} c_{S,T}(q,t) \, \chi_{T}(a_i)\, .
\ee
In this equation, the $c_{S,T}$ are rational functions of $q$ and $t$ that are
uniquely fixed by ensuring that $P_S(a_i;q,t)$ is an eigenfunction of
the Macdonald operators $\bar{G}_{(1^r)}$ for $r=1,\ldots,N-1$.  In this limit the
structure constants $C_S$ are given by  
\begin{align}
C_S = \frac{1}{S_{0,S}}\, ,
\end{align}
where $S_{R,S}$ is an analytic continuation of the modular S-matrix of
refined Chern-Simons theory.

\subsubsection{Operator algebra from Macdonald polynomials}

The Macdonald polynomials obey
\be
P_{S_1} \cdot P_{S_2} = \sum_S {\cN_{S_1,S_2}}^{S_3} P_{S_3} \, ,
\ee
where ${\cN_{S_1,S_2}}^{S_3}$ are the $(q,t)$-deformed
Littlewood-Richardson coefficients. Remarkably, we have found that the
conjugated difference operators $\bar{G}_{R}$ obey the
same algebra. Let us try to understand this fact. 

Consider for instance the case $N=2$. The eigenvalues of the difference operators
$\bar{G}_{r}$ can be computed from explicit formulae that we have found. By experimentation, we find that they are given by 
\be
\bar{G}_{r_1} \cdot P_{r_2}(a)  = \frac{S_{r_1,r_2}}{S_{0,r_2}} \, P_{r_2} \, ,
\ee
where $S_{r_1,r_2}$ is an analytic continuation of the modular
S-matrix of refined Chern-Simons theory  (see Appendix \ref{MacPolys} for the
construction of this quantity). This formula can be proven by a lengthy computation using the residue construction, in which the eigenvalue is given by
\be
\frac{
  \mathrm{Res}_{a=t^{\frac{1}{2}} q^{\frac{r_1}{2}}} \frac{1}{a}
  P_{r_2}(a)} { \mathrm{Res}_{a=t^{\frac{1}{2}}} \frac{1}{a}
  P_{r_2}(a)} 
  =\frac{S_{r_1,r_2}}{S_{0,r_2}} \, .
  \ee 
This S-matrix is known to obey the
$(q,t)$-deformed Verlinde formula
\be
S_{r_1,s} \cdot S_{r_2,s} = S_{0,s} \sum_r {\cN_{r_1,r_2}}^r S_{r,s} \, .
\ee

Let us now act with the composition of the operators
$\bar{G}_{r_1}$ and $\bar{G}_{r_2} $ on the Macdonald polynomial $P_{s}$ and apply the refined Verlinde formula
\bea
( \bar{G}_{r_1} \circ \bar{G}_{r_2} ) \cdot P_s & = \frac{S_{r_1,s}}{S_{0,s}} \frac{S_{r_2,s}}{S_{0,s}} \, P_s \\
& = \sum_r {\cN_{r_1,r_2}}^r \frac{S_{r,s}}{S_{0,s}} \, P_s \\
& = \sum_r {\cN_{r_1,r_2}}^r \bar{G}_r \cdot P_s\, .
\eea
Since the Macdonald polynomials form a complete basis of symmetric
functions, we find that the structure constants of the difference operator
algebra are indeed the $(q,t)$-deformed Littlewood-Richardson
coefficients. 

Similarly, we have verified that the generalized difference operators $\bar{G}_R$,
labeled by irreducible representations $R$ of $su(N)$, satisfy the eigenvalue equation
\be\label{eqn:eigenvaluesSmatrix}
\bar{G}_{R_1} \cdot P_{R_2}  = \frac{S_{R_1,R_2}}{S_{0,R_2}} P_{R_2}
\ee
in the Macdonald slice.

\subsection{Wilson loops in refined Chern-Simons theory}

In the Macdonald slice the superconformal index is dual to an analytic
continuation of the refined Chern-Simons theory on $S^1 \times
C$. Similar to the discussion in \cite{Alday:2013kda} we can
identify the surface defect operators $\bar{G}_R$ in this refined
Chern-Simons theory as the Wilson loop operator 
\begin{align}
 \cO_R = \tr_R\, \mathrm{P \, exp} \left( i \oint_{S^1} A \right)
\end{align}
in the representation $R$ wrapping the $S^1$. This is of course a local operator from the perspective of the two-dimensional TQFT on $C$, in accordance with our expectations from six-dimensional engineering. 

Correlation functions of this operator are independent of its position on $C$ and simply insert a ratio $S_{R,S} / S_{0,S}$ in the sum over representations $S$ is any correlator. For example, inserting the operator $\cO_R$ in a correlator on the three-punctured sphere is computed as 
\begin{align}
\langle \cO_R \rangle_{0,3} = \sum_S \frac{P_S(a_1) P_{S}(a_2)
  P_S(a_3)}{S_{S,0}} \, \frac{S_{R,S}}{S_{0,S}} \, , 
\end{align}
where $S_{S,R}$ is an analytic continuation of the modular S-matrix of refined Chern-Simons theory. Hence inserting the local operator $\cO_R$
in a TQFT correlation function is equivalent to acting on any of the punctures with the difference operator $\bar{G}_R$.

Moreover, from the $(q,t)$-deformed Verlinde formula
\be
S_{R_1,S} \cdot S_{R_2,S} = S_{0,S} \sum_R {\cN_{R_1,R_2}}^R S_{R,S}
\ee
we derive the operator product expansion
\begin{align}
\cO_{R_1} \cdot \cO_{R_2} = \sum_R {\cN_{R_1,R_2}}^{R} \, \cO_R \, , 
\end{align}
where ${\cN_{R_1,R_2}}^{R}$ are the analytically continued
$(q,t)$-deformed Littlewood-Richardson coefficients. Thus in the Macdonald limit, the algebra of the difference operators $G_{R}$ is equivalent to the Verlinde algebra in refined Chern-Simons theory on $S^1 \times C$. 

The general superconformal index could be taken to define a $(p,q,t)$-deformed Yang-Mills theory on $C$, whose structure constants are given in
terms of the eigenfunctions $\psi_R(a_i;p,q,t)$ of the elliptic difference operators. The difference operators satisfy an algebra whose structure constants
${\cN_{R_1,R_2}}^{R_3}$ are elliptic functions. It would be fascinating to understand this theory.









\section{Algebra of three-dimensional line defects} \label{section_Reduction_to_3d}

In section~\S\ref{section:index}, we constructed the superconformal
index of $\cN=2$ theories on $S^1 \times S^3$ in the presence of
certain surface defects supported on $S^1 \times S^1$. These
surface defects were labeled by an irreducible representation $R$
of $su(N)$ and could be added to any superconformal theory with an $SU(N)$ flavor
symmetry. In this section, we consider the reduction of the four-dimensional
superconformal index to a partition function on a squashed three-sphere $S^3$, following~\cite{Gadde:2011ia,Dolan:2011rp,Imamura:2011uw}. In this limit, the surface defects become codimension-two defects in the three-dimensional
theory wrapping an $S^1 \subset S^3$.

For four-dimensional theories of class $\mathcal{S}$, upon dimensionally reducing on $S^1$ the theory flows to an $\cN=4$ superconformal field theory in three-dimensions. Moreover, this has a mirror description in terms of a star-shaped
quiver theory~\cite{Benini:2010uu}. It is expected
that the surface defects introduced by the difference operators $G_{R}$ become supersymmetric
Wilson loops in representation $R$ for the central node of this
star-shaped quiver upon dimensional reduction. We demonstrate this explicitly for antisymmetric tensor
representations $R=(1^r)$ and the case of a round three-sphere. 
For non-minuscule representations
$R$, however, we find that the difference operators $G_{R}$ introduce a linear
combination of Wilson loops in irreducible representations $S$ with
$|S| \leq |R|$.

\subsection{From superconformal index to 3d partition function}

The four-dimensional superconformal index on $S^1\times S^3$ can be reduced to
a partition function on the squashed three-sphere $S^3$, as demonstrated in~\cite{Gadde:2011ia,Dolan:2011rp,Imamura:2011uw}. This limit is taken by parametrizing the fugacities by 
\be
p = e^{-\beta b^{-1}} \, , \qquad q = e^{-\beta b} \, , \qquad t = e^{-\beta \epsilon } \, , \qquad a_j = e^{ - i \beta x_j} \, , 
\label{3dlimit}
\ee
with $\beta >0$ and then taking the limit $\beta \to 0^{+}$. Here we have introduced the convenient notation $\epsilon = \frac{q}{2} + i m$ where $q=b+b^{-1}$.

The real
parameter $b>0$ encodes the geometry of the three-sphere, defined by
the embedding
\be
b^{-2} |z|^2 + b^2 |w|^2 = 1 \, 
\ee
into $\mathbb{C}^2$ with complex coordinates $(z,w)$. The parameters $x_i$ with $\sum_{i=1}^N x_i=0$ are real mass parameters for the global
$SU(N)$ symmetry that is inherited by the three-dimensional theory.
It is convenient to repackage the components $x_j$ into a
vector $x$ such that $x_j = (x,h_j)$.
In addition, the real parameter $m$ gives a mass to
the adjoint chiral multiplet inside the background $\cN=4$
vectormultiplet, breaking the supersymmetry to $\cN=2$ in three dimensions.

Let us consider two important examples. Firstly, the three-dimensional limit~\eqref{3dlimit} of the superconformal index of a free hypermultiplet is given by
\bea
\cZ_H(x) = S_b\left( \frac{\epsilon}{2} \pm i x \right) \, .
\label{3dhyper}
\eea
Secondly, the superconformal index of an $SU(N)$ vectormultiplet combined with the Haar measure becomes the partition function of a three-dimensional $\cN=4$ vectormultiplet
\be
\cZ_V(x) = \prod_{i<j}^N 2 \sin \left( i \pi b^{\pm}  x_{ij} \right) \, \cK(x)
\label{4dvmlimit}
\ee
where
\be
\cK ( x )  = \frac{1}{S_b(\epsilon^*)}  \, \prod_{i, j = 1 }^N \, S_b(\epsilon^* + i x_{ij} )
\label{Kdef}
\ee
where we have denoted $\epsilon^* = \frac{q}{2}-im$. Note that $\cK(x)$ is the contribution from the $\cN=2$ adjoint chiral multiplet inside the three-dimensional $\cN=4$ vectormultiplet and cancel in pairs in the limit $m \to 0$. We use the double sine function that obeys the difference equation $S_b(x+b^{\pm}) = 2 \sin (\pi b^{\pm}x)S_b(x)$ and the reflection property $S_b(x)S_b(q-x)=1$. Further properties of this special function can be found in appendix~\ref{appendix:special}.

Let us now consider the three-dimensional limit of the difference operators $G_R$ that
introduce surface defects into the four-dimensional $\cN=2$ theory. The three-dimensional limit can be evaluated using the fact that the ratio of theta-functions with a common second argument reduces to a ratio of sine-functions, 
\begin{equation}
 \frac{\theta\left( e^{\alpha \rho}, e^{\beta \rho} \right)}{\theta
   \left( e^{\gamma \rho}, e^{\beta \rho} \right)} \xrightarrow{\rho
   \rightarrow 0} \, \frac{\sin \left( \pi \alpha/\beta \right)}{\sin \left(\pi \gamma/\beta\right)} \, .
\end{equation}

Let us first consider the difference operator $G_{(1)}$ labeled by the
fundamental representation of $su(N)$. In four dimensions this
operator is given by
\be
 G_{(1)} \cdot \cI(a_i, \ldots) = \sum_{j=1}^N \prod_{k \neq
  j}^N \frac{ \theta ( \frac{t}{q} a_j / a_k , p ) }{ \theta ( a_k /
  a_j , p ) } \, \cI\left(  q^{\frac{1}{N} - \delta_{k,i} } a_i, \ldots \right)
\label{fundamental4d}
\ee
up to some overall $t$-dependent factor. Applying the above formula to each term, we find that the three-dimensional limit of the fundamental difference operator $G_{(1)}$ acts on the three-dimensional
partition function $\cZ(x, \ldots)$ as
\be
G^{(\mathrm{3d})}_{(1)} \cdot \cZ(x,\ldots) =\sum_{j=1}^N \prod_{k \neq j}^N
\frac{ \sin \pi b \left( \epsilon^*- i x_{jk} \right)}{ \sin \pi
  b \left( -i x_{jk} \right)} \cZ\left( x+i b h_j, \ldots \right) ,
\label{fundamental3d}
\ee
where we use the shorthand $x_{jk} = x_j -x_k$. We also recall that the weights $h_i$ obey $(h_i,h_j) = \delta_{ij}-1/N$. 

Let us now extend this computation to the rank $r$ antisymmetric tensor representation
$(1^r)$ of $su(N)$. In section~\S\ref{section:index} we found that up to a power of $t$ the
corresponding difference operator is
\be
  G_{(1^r)} \cdot \cI(a_i) = \sum_{|I|=r}
 \prod_{\substack{j \in I \\ k\notin I}} \frac{ \theta ( \frac{t}{q}
   a_j / a_k , p ) }{ \theta ( a_k / a_j , p ) } \, \cI\left(
   q^{\frac{r}{N} - \delta_{i,I} } a_i \right)  , 
\ee
where the summation is over subsets $I \subset \{1,\ldots, N\}$ of
length $|I|=r$ and where the symbol $\delta_{i,I}$ is one if $i\in I$
and zero if $i\notin I$. In the three-dimensional limit, we obtain the operator
\begin{align}
  G^{(\mathrm{3d})}_{(1^r)} \cdot \cZ(x) = \sum_{|I|=r} \prod_{
   \substack{j\in I \\ k \notin I}} \frac{ \sin \pi b \left(\epsilon^*
     - i x_{jk}\right)}{ \sin \pi b \left( - i x_{jk} \right)} \,
 \cZ\left(x + i b \sum_{j\in I} h_j  \right),
 \label{antisymmetric3d}
\end{align}
acting on the squashed three-sphere partition function $\cZ(x)$. Similar computations can be performed for the four-dimensional difference operators $G_{R}$
corresponding to any irreducible representation $R$ of $su(N)$.

The dimensionally reduced operators $G^{(\mathrm{3d})}_R$ have similar properties to
their four-dimensional ancestors. The adjoint operator of $G^{(\mathrm{3d})}_R$ with respect to the three-dimensional $\cN=4$ vectormultiplet measure~\eqref{4dvmlimit} on $\mathbb{R}^{N-1}$,
\be \label{3dopsself-adjoint}
 \langle  f_1 , f_2 \rangle = \int \left[ \frac{\rd ^{N-1}x}{N!}  \prod_{i<j}^N 2 \sin \left( i \pi b^{\pm}  x_{ij} \right) \cK(x) \right] \cdot f_1 (x) f_2 (x)  \, ,
\ee
is given by the operator $G^{(\mathrm{3d})}_{R^*}$ for the conjugate representation $R^{*}$. Furthermore, the operators $G^{(\mathrm{3d})}_R$ generate a commutative algebra that can be derived by applying the limit~\eqref{3dlimit} to the general structure constants ${\cN_{R_1,R_2}}^{R_3}(p,q,t)$ that we found in the four-dimensional case. 

Finally, in section~\ref{TQFTstructure} we pointed out that the four-dimensional difference operators with the replacement $ t \to pq / t $ were related by a similarity transformation. In the three-dimensional limit, this corresponds to $m \to -m$ or quivalently $\epsilon \to \epsilon^*$. Thus, either by direct computation or by taking the three-dimensional limit of the the result in section~\ref{TQFTstructure} we find that
\be
G_R^{(3d)}(x)=  \frac{1}{\cK(x)} \cdot  \bar{G}^{(3d)}_R(x) \cdot  \cK(x)
\label{switch} 
\ee
where the operator $\bar{G}_R^{(3d)}(x)$ is related to $G_R^{(3d)}(x)$ by the replacement $\epsilon \to \epsilon^*$ and $\cK(x)$ is the partition function of an adjoint $\cN=2$ chiral multiplet inside an $\cN=4$ vectormultiplet. A consequence is that the eigenfunctions of the two sets of operators are proportional.




\subsection{Wilson loops in 3d star-shaped quivers}

Since the dimensional reduction is performed along a circle on which
the surface defect is supported, we expect that the difference operators~\eqref{fundamental3d} and~\eqref{antisymmetric3d}
introduce defects in the three-dimensional theory supported on the circle $|z|^2 = b^2$.  In the following we will perform indirect checks of this prediction by exploiting three-dimensional mirror symmetry to relate these defects to supersymmetric Wilson loops.

\begin{figure}[htp]
\centering
\includegraphics[width=.7\textwidth]{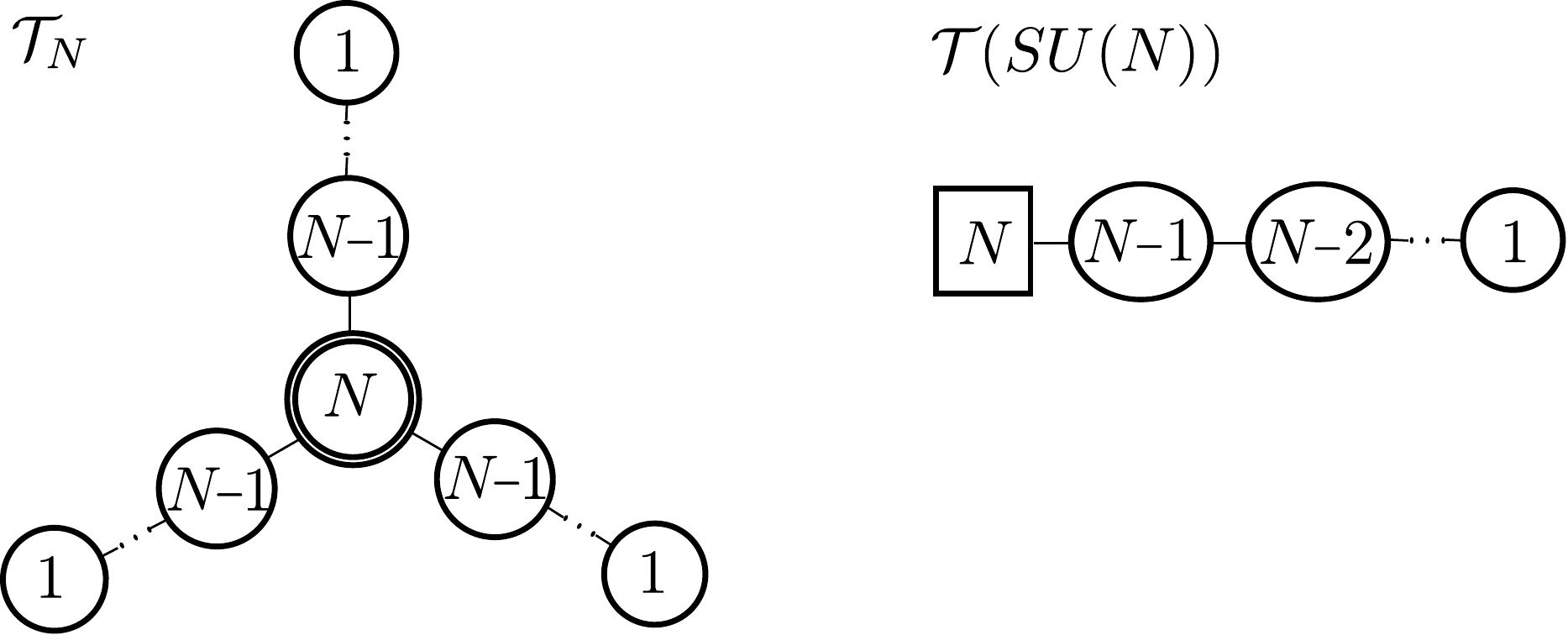}
\caption{Left: Star-shaped quiver corresponding to the mirror of the
  three-dimensional $\cT_N$ theory. Right: Linear quiver
  description of the three-dimensional $\cT(SU(N))$ theory.}
\label{fig:starquiver}
\end{figure}

Upon dimensional reduction, a four-dimensional $\cN=2$ theory of class $\cS$ flows to an $\cN=4$ superconformal field theory in three dimensions, which is related by mirror symmetry to a
star-shaped quiver gauge theory~\cite{Benini:2010uu}. For
example, the mirror of the three-dimensional reduction of the $\mathcal{T}_N$ theory is
given by a star-shaped quiver with three legs, shown in
Figure~\ref{fig:starquiver}. More generally, each full puncture in four dimensions gives rise to one copy of the linear quiver illustrated on the right in Figure~~\ref{fig:starquiver}. The corresponding field theory is known as the $\cT(SU(N))$ theory~\cite{Gaiotto:2008sa,Gaiotto:2008ak}. This theory contains $\cN=4$ vectormultiplets for the
gauge groups $U(1), \ldots, U(N-1)$. These gauge groups
are coupled linearly through $U(k) - U(k+1)$ bifundamental hypermultiplets. Lastly, there are
$N$ hypermultiplets in the fundamental representation of 
the largest gauge group $U(N-1)$. 

The $\cT(SU(N))$ theory has manifest $SU(N)$ Higgs branch symmetry
acting on the $N$ hypermultiplets whose mass parameters are denoted by the vector $ x$. In addition, we can introduce $N-1$ Fayet-Illiopoulos parameters
$t_1, \ldots, t_{N-1}$, which are mass parameters for the
topological $U(1)$ symmetries associated to each node of the
quiver. Let us express these parameters in terms of a new vector $y$ such that $t_ k =
y_k - y_{k+1}$. Let us denote the partition function of $\cT(SU(N))$ by $\cZ_{\epsilon}(x,y)$. This is
expected to be invariant under exchanging $x \leftrightarrow
y$ and $\epsilon \leftrightarrow \epsilon^*$. This symmetry reflects an enhancement of the Coulomb branch symmetry to ${}^LSU(N)$ in the infrared, as well as the presence of mirror symmetry 
exchanging $SU(N)\leftrightarrow {}^LSU(N)$.  

The partition function of any three-dimensional theory of class $\cS$
with full punctures can then be constructed according to the
star-shaped quiver description. For example, the partition function of
the three-dimensional mass-deformed $\cT_N$ theory is given by
\be
\cZ_{\cT_N} (x,y,z ) = \int \frac{d^{N-1} w}{N!} \cZ_V(w) \; \cZ_\epsilon( w, x ) \, \cZ_\epsilon ( w, y ) \, \cZ_\epsilon ( w, z ) \, ,  
\ee
where $\cZ_V(w)$ is the partition function of an $\cN=4$
vectormultiplet. This multiplet is used to gauge the diagonal
combination of $SU(N)$ Higgs branch symmetries. 

Let us consider the action of the three-dimensional operators on the partition
function $\cZ (x,y,z)$. Similar as in four
dimensions, the result should be independent of which puncture we act
on. This condition is guaranteed if the partition function of each
quiver tail $\cT(SU(N))$ is an eigenfunction of the operator
$G^{(\mathrm{3d})}_{(1^r)}$. We will now show that in fact we have
\be \label{eq:monodromy-wilson}
G^{(\mathrm{3d})}_{(1^r)}(y)  \cdot \cZ_\epsilon ( x ,y ) = W_{(1^r)}(x) \, \cZ_\epsilon
(x,y ) \, , 
\ee
where
\be
W_{(1^r)}(x)  =  \sum_{j_1<\ldots < j_r} e^{-2 \pi b (x_{j_1} + \cdots + x_{j_r} ) } 
\ee
is the expectation value of a supersymmetric Wilson loop in the rank $r$ antisymmetric
representation of the $SU(N)$ flavor symmetry\footnote{Recall that Wilson loops are labeled by irreducible representations of the gauge group. Here we use that each irreducible representation $R$ of $su(N)$ corresponds to an irreducible representation $R$ of $SU(N)$, and vice versa.} with mass parameter $x$.

But let us first remark that equation~(\ref{eq:monodromy-wilson}) tells us that introducing a  
background defect in the $\cT(SU(N))$ theory for the Coulomb branch
symmetry, through the operator $G^{(\mathrm{3d})}_{(1^r)}(y)$, is equivalent to
introducing a background Wilson loop $W_{(1^r)}(x)$ for the
Higgs branch symmetry.  
This means
that mirror symmetry interchanges the defects introduced by the operators $G^{(\mathrm{3d})}_{(1^r)}$ and
supersymmetric Wilson loops in the $r$-th antisymmetric representation of $SU(N)$.
In the context of the mirror description of the three-dimensional
$\cT_N$ theory, the operators $G^{(\mathrm{3d})}_{(1^r)}$ therefore introduce a dynamical Wilson loop
for the central node of the star-shaped quiver theory.

\subsubsection{$\cT(SU(2))$}

We first show equation~(\ref{eq:monodromy-wilson}) in complete generality for the partition function of the mass deformed $\cT(SU(2))$ theory on a squashed three-sphere. This partition function is given by
\be
\cZ_{\epsilon}(x,y)
= \frac{1}{S_b(\epsilon^*)} \int dz \, \cQ(x,z)\, e^{4\pi i y z} \, ,
\label{TSU2pf}
\ee
where
\be
\cQ(x,z) =  \frac{S_b\left(\frac{\epsilon^*}{2} \pm ix - iz \right)}{S_b\left(q - \frac{\epsilon^*}{2} \pm ix - iz \right)} \, ,
\label{Qdef}
\ee
$x$ is the $SU(2)$ mass parameter and $y$ the FI parameter and the contour is given by $z \in \mathbb{R}+i\epsilon$ with $\epsilon>0$. Note that the $\cN=2$ mass deformation $m$ in the hypermultiplet contribution appears with the opposite sign compared to equation~\eqref{3dhyper}. The reason is that after dimensional reduction, there is a mirror symmetry required to reach the star-shaped quiver description.

It is expected that the partition function has the following properties
\bea
\cZ_\epsilon(x,y) & = \cZ_\epsilon(-x,y) = \cZ_\epsilon(x,-y) \\
\cZ_\epsilon(x,y) & = \cZ_{\epsilon^*}(y,x) \\
G^{(3d)}_{(1)}(y)  \, \cZ_\epsilon(x,y) & = W_{(1)}(x) \, \cZ_\epsilon(x,y),
\label{Zproperties}
\eea
where 
\be
G^{(3d)}_{(1)}(x)  = \frac{\sin \pi b \left( \epsilon^* - 2ix \right)}{ \sin \pi b(-2ix)} e^{\frac{ib}{2}\del_x} + \frac{\sin \pi b \left( \epsilon^* + 2ix \right)}{ \sin \pi b(2ix)} e^{-\frac{ib}{2}\del_x}
\ee
is the fundamental difference operator~\eqref{fundamental3d} when $N=2$, and $W_{(1)}(x) = e^{2\pi b x}+e^{-2\pi b x}$ is the fundamental Wilson loop expectation value. 

The first line of equation~\eqref{Zproperties} represents the enhancement of the Higgs and Coulomb branch symmetry to $SU(2) \times {}^LSU(2)$ in the infrared. The second line illustrates the mirror symmetry of the mass-deformed $\cT(SU(2))$ theory. These properties were demonstrated in~\cite{Hosomichi:2010vh}. Here we would like to prove the final line of equation~\eqref{Zproperties}. Using mirror symmetry this line is equivalent to
\be
\bar{G}_{(1)}^{(\mathrm{3d})}(x) \, \cZ_\epsilon(x,y) = W_{(1)}(y) \, \cZ_\epsilon(x,y) \, ,
\ee
where $\bar{G}_{(1)}^{(\mathrm{3d})}(x)$ is obtained from $G_{(1)}^{(\mathrm{3d})}(x)$ by the replacement $m \to -m$. Let us prove the intertwining property in this equivalent form.

As a preliminary step, we derive a few properties of the function $\cQ(x,z)$ defined in equation~\eqref{Qdef}. From the difference equation and the reflection property obeyed by the double sine function $S_b(x)$, it is straightforward to show that 
\be
e^{ib\del_z} \cQ(x,z) = \frac{\sin\pi b(\frac{\\epsilon^*}{2} \pm i x - i z )}{\sin\pi b(q-\frac{\epsilon^*}{2}\pm ix -iz)} \cQ(x,z)
\ee
and
\be
e^{\frac{ib}{2} \del_x} \cQ(x,z) = e^{-\frac{ib}{2}\del_z } \left[ \, \frac{\sin\pi b(\frac{\epsilon^*}{2} - i x-iz)}{\sin\pi b(q-\frac{\epsilon^*}{2} - i x-iz)} \, \cQ(x,z) \, \right]
\ee
\be
e^{-\frac{ib}{2} \del_x} \cQ(x,z) = e^{-\frac{ib}{2}\del_z } \left[ \, \frac{\sin\pi b(\frac{\epsilon^*}{2} + i x-iz)}{\sin\pi b(q-\frac{\epsilon^*}{2} + i x-iz)} \, \cQ(x,z) \, \right],
\ee
where we have used the notation $\epsilon = \frac{q}{2} +im$. Using these results we can now compute the action of the difference operator $\bar{G}_{(1)}^{(\mathrm{3d})}(x)$ on this function,
\bea
\bar{G}_{(1)}^{(\mathrm{3d})}(x) \cdot \cQ(x,z)  
%
& = e^{-\frac{ib}{2} \del_z} \left[ \, \frac{\sin \pi b \left( \epsilon - 2ix \right)}{ \sin \pi b(-2ix)} \frac{\sin\pi b(\frac{\epsilon^*}{2} - i x-iz)}{\sin\pi b(q-\frac{\epsilon^*}{2} - i x-iz)}\right. \\
& \hspace{1.5cm} \left. + \frac{\sin \pi b \left(\epsilon + 2ix \right)}{ \sin \pi b(2ix)} \frac{\sin\pi b(\frac{\epsilon^*}{2} + i x-iz)}{\sin\pi b(q-\frac{\epsilon^*}{2} + i x-iz)} \, \right] \cQ(x,z)
\\ & = e^{-\frac{ib}{2} \del_z} \left[ 1+\frac{\sin\pi b(\frac{\epsilon^*}{2} \pm i x-iz)}{\sin\pi b(q-\frac{\epsilon^*}{2} \pm i x-iz)} \right] \cQ(x,z)
\\ & = ( e^{\frac{ib}{2} \del_z}+e^{-\frac{ib}{2} \del_{z}}) \cQ(z,x) \, .
\label{GonQ}
\eea
In going from the first to the second line we have applied a simple trigonometric identity.

Armed with this result, we now consider the action of the difference operator $\bar{G}_{(1)}^{(\mathrm{3d})}(x)$ on the full partition function~\eqref{TSU2pf} of the $\cT(SU(2))$ theory. The difference operator can be brought inside the integral to act on $\cQ(x,z)$ as in equation~\eqref{GonQ}. By shifting the contour of integration by $z \to z\pm \frac{ib}{2}$, we find 
\bea
 \bar{G}_{(1)}^{(\mathrm{3d})}(x) \cdot \cZ_\epsilon(x,y) & = \frac{1}{S_b(\frac{q}{2}-im)} \int dz\left[ ( e^{\frac{ib}{2} \del_z}+e^{-\frac{ib}{2} \del_z})\, \cQ(x,z) \right] e^{4\pi i y z} 
\\ & = \frac{1}{S_b(\frac{q}{2}-im)} \int \cQ(x,z) \, \left[ \, ( e^{\frac{ib}{2} \del_z}+e^{-\frac{ib}{2} \del_z}) \, e^{4\pi i y z} \right] 
\\ & = W_{(1)}(y) \, \cZ_{\epsilon}(x,y) \, .
\eea
Using the analytic structure of the double sine function $S_b(x)$, it is straightforward to check that no poles are crossed in shifting the contours provided the mass are real. Now, applying mirror symmetry we have
\be
G_{(1)}^{(\mathrm{3d})}(y) \, \cZ_\epsilon(x,y) = W_{(1)}(x) \cZ_{\epsilon}(x,y) \, ,
\ee
which is the required result.

An important consequence of this result, together with the similarity transformation~\eqref{switch} relating the operators $G^{(3d)}_{(1)}$ and $\bar{G}^{(3d)}_{(1)}$, is that the partition function of mass deformed $\cT(SU(2))$ theory obeys
\be
\cZ_{\epsilon^*}(x,y) =  \cK(y) \, \cZ_{\epsilon}(x,y) 
\ee
which is rather non-obvious from the integral representation.

\subsubsection{$\cT(SU(N))$}

Let us now consider equation~(\ref{eq:monodromy-wilson}) for the general $\cT(SU(N))$ theory. In this case, we simplify the problem and prove a weaker result by taking the limit of $\cN=4$ supersymmetry ($m=0$) and a round three-sphere $(b=0)$. 

In this limit, the operators for the fully antisymmetric
representations are given by
\be
G^{(\mathrm{3d})}_{(1^r)} \cdot \cZ ( x ) = (-1)^{r(N-r)} \sum_{j_1<\cdots < j_r} \cZ ( x+ i(h_{j_1} + \cdots + h_{j_r}) ) \, .
\ee 
Hence, up to a sign, the operators are simply a sum of shift operators with coefficient 1 with the shifts determined by the weights of the representation $(1^r)$. 

Furthermore, the zero mode integrals in the partition function of the $\cT(SU(N))$ theory can be performed explicitly~\cite{Benvenuti:2011ga,Nishioka:2011dq}. The result is\footnote{Here we drop the subscript $m$ on the partition function of $\cT(SU(N))$ because we have set $m=0$.}
\be
\cZ ( x, y ) = \frac{\sum_{\rho \in S_N} (-1)^\rho e^{2\pi i \sum_{j=1}^N x_{\rho(j)} y_j}}{\prod_{i<j} 2\sinh \pi (x_{i}-x_j ) \, 2\sinh\pi (y_{i}-y_j ) }\, ,
\label{eq:tsun}
\ee
where the summation in the numerator is over the Weyl group $S_N$ of
permutations of $\{1,\ldots,N\}$. Mirror symmetry $x\leftrightarrow y$
in this case follows from the identity
\be
\sum_{\rho \in S^N} (-1)^\rho e^{2\pi i \sum_{j=1}^N x_{\rho(j)} y_j}
= \sum_{\rho \in S^N} (-1)^\rho e^{2\pi i \sum_{j=1}^N x_{j}
  y_{\rho(j)}} \, . 
\ee
Thus there is a vast simplification in the limit $b=1$ and $m=0$.

Let us now act with the operator $G^{(\mathrm{3d})}_{(1^r)}$ on the partition function $\cZ(x,y)$. First, note that each term in the operator $G^{(\mathrm{3d})}_{(1^r)}$ leaves the
denominator invariant up to a factor $(-1)^{r(N-r)}$ which cancels the overall sign in the operator. Thus we can concentrate on the
numerator of $\cZ (x,y)$ and find
\begin{align}
G^{(\mathrm{3d})}_{(1^r)}(y) & \cdot \sum_{\rho \in S^N} (-1)^\rho e^{2\pi i \sum_{j=1}^N
  y_{\rho(j)} x_j} =
 \sum_{|I| = r} \left[ \sum_{\rho \in S^{N}} (-1)^{\rho} e^{2 \pi i \sum_{k=1}^{N} y_{\rho (k)} x_{k} } e^{-2 \pi
     \sum_{\rho(i)\in I} x_{i}} \right] .  
\end{align}
Imposing the condition $\sum_{i=1}^{N} x_{i} = 0$ leads to vanishing
overall background shifts. 

Label the subsets $I \subset \{1,\ldots,N\}$ with $|I|=r$ by
$I_\ell$ for $1\leq \ell \leq N_r={N \choose r}$. Furthermore split
$S^{N}$ into $\mathbb{Z}_{N_r} \otimes (S^{N-r} \otimes S^{r})$, where
$S^{(N-r)} \otimes S^{r}$ gives all the permutations satisfying $\rho
(I_\ell) = I$ for fixed $I$ and $\ell$, and $\mathbb{Z}_{N_r}$ gives
the different choices of $\ell$ (or $I$). Since this splitting is an isomorphism\footnote{Any $\rho\in S^{N}$ can be uniquely characterized by $\rho (\sigma (I_\ell)) = \pi(I)$ for a unique $I_\ell$ or $I$ and $\sigma \in S^{r}$, $\pi \in S^{N-r}$ permutations of $I$, $\mathbb{Z}_N \setminus I$ respectively.}, we can write the sum over all $\rho \in S^{N}$ as a double sum over sets $I$ with $|I|=r$ and permutations in $S^{N-r} \otimes S^{r}$ preserving $\rho (I_\ell) = I$. With that in mind, we can rewrite the above as
\begin{align}
G^{(\mathrm{3d})}_{(1^r)}(y) & \cdot \sum_{\rho \in S^N} (-1)^\rho e^{2\pi i \sum_{j=1}^N
  y_{\rho(j)} x_j} =\sum_{\ell=1}^{N_r} e^{-2 \pi \sum_{i \in I_\ell} x_{i}} \sum_{|I|=r}
\sum_{\substack{\rho \in S^{N} \\ \rho (I_\ell) = I}}
(-1)^{ \rho } e^{2 \pi i \sum_{k=1}^{N} y_{\rho (k)}
  x_{k} }\, ,
\end{align}
which is equal to the Wilson loop vacuum expectation value $W_{(1^r)}(x)$
times the numerator of $\cZ(x,y)$. The eigenvalue of the operator
$G^{(\mathrm{3d})}_{(1^r)}$ acting on the full partition function $\cZ(x,y)$ is thus precisely
the localization expression for a supersymmetric Wilson loop in the antisymmetric
representation $(1^r)$ of $SU(N)$.

\subsection{Three-dimensional algebra}

In the above, we have shown that the defect operator $G^{(\mathrm{3d})}_{(1^r)}$ is dual
to a Wilson loop in the rank $r$ antisymmetric representation of $SU(N)$ under
mirror symmetry. This turns out $\emph{not}$ to be the case for 
non-minuscule representations. 

One immediate way to see this is the following. Denote the operator
that \emph{is} exactly dual to a Wilson loop in the representation $R$ by
$\tilde{G}^{(\mathrm{3d})}_{R}$. The operators 
$\tilde{G}^{(\mathrm{3d})}_{R}$ must obey the algebra
\be\label{eq:Wilsonalgebra}
\tilde{G}^{(\mathrm{3d})}_{R_1} \circ \tilde{G}^{(\mathrm{3d})}_{R_2} = \sum_{R_1} {N_{R_1,R_2}}^{R_3} \, \tilde G^{(\mathrm{3d})}_{R_3}\, ,
\ee
where ${N_{R_1,R_2}}^{R_3}$ are the (plain) Littlewood-Richardson
coefficients. Indeed, 
the supersymmetric Wilson loops are characters and hence obey this
algebra. Instead, the elliptic Littlewood-Richardson 
coefficients ${\cN_{R_1,R_2}}^{R_3}(p,q,t)$ reduce in general to
non-integer coefficients in three dimensions. 

For example, let us consider $SU(2)$ and the composition of
two operators in the fundamental representation. In the
three-dimensional limit we find that 
\be
G^{(\mathrm{3d})}_{(1)} \circ G^{(\mathrm{3d})}_{(1)} = G^{(\mathrm{3d})}_{(2)} + \left(\frac{1}{2 \cos \left(\pi  b^2\right)-1}-1\right) G^{(\mathrm{3d})}_{(0)}\, .
\ee
Since $\tilde{G}^{(\mathrm{3d})}_{(1)} = G^{(\mathrm{3d})}_{(1)}$ and $\tilde{G}^{(\mathrm{3d})}_{(0)} =
G^{(\mathrm{3d})}_{(0)} =1$, we read off from
equation~(\ref{eq:Wilsonalgebra}) that
\be
\tilde{G}^{(\mathrm{3d})}_{(2)} = G^{(\mathrm{3d})}_{(2)} + \left(\frac{1}{2 \cos \left(\pi
      b^2\right)-1}-2 \right) G^{(\mathrm{3d})}_{(0)}\, .
\ee
The operator $\tilde{G}^{(\mathrm{3d})}_{(2)}$ that \emph{is} dual to a Wilson loop thus
differs from the difference operator $G^{(\mathrm{3d})}_{(2)}$ by lower order contributions.

In general, the relation between the operators $G^{(\mathrm{3d})}_{R}$ appearing in the vortex construction and the operators $\tilde{G}^{(\mathrm{3d})}_R$ that are exactly dual to Wilson loops in the three-dimensional limit is given by
\be
\tilde{G}^{(\mathrm{3d})}_{R} = G^{(\mathrm{3d})}_{R} + \sum_{|S|<|R|} c_S \, G^{(\mathrm{3d})}_S \, .
\ee 
Even though the difference operators $G^{(\mathrm{3d})}_R$ are thus not exactly dual
to Wilson loops, this is merely an invertible linear transformation on the algebra that these operators obey.  

The original basis of operators $G_{R}$ appears to be more fundamental from a four-dimensional perspective, since in the limit $p\to0$ they are precisely dual to Wilson 
loop operators in refined Chern-Simons theory on $S^1 \times C$. On the other hand, in the three-dimensional limit, the
basis $\tilde{G}^{(\mathrm{3d})}_{R}$ seems  more fundamental since it is dual to a basis of Wilson loop operators in the star-shaped quiver theories. 


\section{'t Hooft loops in the four-dimensional $\cN=2^*$ theory}\label{sec:thooft}

In this section, we realize the mass-deformed theory $\cT(SU(N))$ on a squashed three-sphere as an S-duality domain wall in four-dimensional $\cN=2^*$ theory on an ellipsoid, as described in~\cite{Hosomichi:2010vh,Terashima:2011qi}. We then use this observation to interpret the three-dimensional difference operators $G^{(\mathrm{3d})}_R$ as operators that introduce supersymmetric 't Hooft loops in the $\cN=2^*$ theory partition function on a four-sphere. 

The four-dimensional $\cN=2^*$ theory can also be obtained by compactifying the
six-dimensional $(2,0)$ theory of type $A_{N-1}$ on a torus with a
simple puncture. A consequence of this construction is that via the AGT correspondence~\cite{Alday:2009aq,Wyllard:2009hg}, the four-sphere partition function of the $\cN=2^*$
theory can also be computed as a correlation function in 
Liouville or Toda CFT on the punctured torus. The difference operators
$G^{(\mathrm{3d})}_{R}$ can then be interpreted as Verlinde loop operators that act
on suitably normalized Virasoro or $W_N$-algebra conformal blocks on a punctured torus.

\subsection{Four-sphere partition function}

The exact partition function of $\cN=2$ supersymmetric gauge
theories on an ellipsoid has been computed by
supersymmetric localization in~\cite{Hama:2012bg}, extending the
computation of Pestun for the round four-sphere
$S^4$~\cite{Pestun:2007rz}. The ellipsoid geometry can be embedded
into five-dimensional Euclidean space as 
\be
x_0^2 + \frac{1}{b^2} ( x_1^2 + x_2^2 ) + b^2 ( x_3^2+x_4^2 ) = 1\, ,
\ee
where $b\in \mathbb{R}_{\geq 0}$ is a real parameter. The
equator $\{x_0=0\}$ is identified with the squashed three-sphere
geometry considered in the previous section by setting $z = x_1+ix_2$ and $w = x_3+ix_4$. 

Let us concentrate on the $\cN=2^*$ theory and denote the real
hypermultiplet mass parameter by $m$ and the complexified gauge
coupling by $\tau$. The result of the localization computation can be
written as a matrix integral
\be\label{eqn:foursphere3d}
\cZ_{S^4_b}(m,\tau) = \int da \, \big|
\,\cZ(a,m;\tau) \, \big|^2
\ee
over a real slice of the Coulomb branch. In this integral $\cZ(a,m;\tau)$
is the Nekrasov partition function for the 
four-dimensional $\cN=2^*$ theory in the Omega-background
$\mathbb{R}^4_{\epsilon_1, \epsilon_2}$, with equivariant 
parameters $\epsilon_1 = b$ and $\epsilon_2 = b^{-1}$~\cite{Nekrasov:2002qd, Nekrasov:2003rj}. It can be split into 
a classical, 1-loop and instanton piece as
\be \label{Nekpartasdf}
\cZ(a,m;\tau) := \cZ_{\mathrm{cl}}(a;\tau)\,
\cZ_{1-\mathrm{loop}}(a,m) \, \cZ_{\mathrm{inst}}(a,m;\tau) \, .
\ee

In this paper we advertise an alternative factorization of the
ellipsoid partition function $\cZ_{S^4_b}$. We find it insightful
to rewrite the matrix integral \ref{eqn:foursphere3d} in the form (for a derivation of this representation see appendix~\ref{appendix:special})
\be\label{eqn:foursphere3dalt}
\cZ_{S^4_b}(m;\tau) = \int da \, \mu(a) \, \big|
\,\cG(a,m;\tau) \, \big|^2 \, , 
\ee
where
\be
\mu (a)  =\prod\limits_{e>0} \, 2 \sinh\left( \pi b (e,a) \right) \, 2\sinh\left( \pi b^{-1} (e,a) \right)
\ee
is the Haar measure times the partition function of a three-dimensional $\cN=2$
vectormultiplet on the squashed three-sphere at the equator $\{ x_0=0\}$. 

We expect that the factorization (\ref{eqn:foursphere3dalt}) has the
following interpretation~\cite{Drukker:2010jp}. We can cut the ellipsoid into two
half-spheres $\{x_0>0\}$ and $\{x_0<0\}$ and impose Dirichlet boundary
conditions on the fields in the $\cN=2^*$ theory at the boundary $\{x_0=0\}$. This decouples the
dynamics on both half-spheres. Restricting the gauge transformations
to the identity on the boundary, leaves a flavor symmetry
group $SU(N)$ acting on the values of the fields at $x_0=0$. We can reconstruct the partition function of an ellipsoid by inserting a three-dimensional $\cN=2$ $SU(N)$ vectormultiplet on the boundary $\{x_0=0\}$ and gauging the diagonal $SU(N)$ symmetry. We thus claim that $\cG(a,m;\tau)$ in the matrix integral \eqref{eqn:foursphere3dalt} is the partition function of $\cN=2^*$ theory on the upper half of the ellipsoid $\{x_0>0\}$ with Dirichlet
boundary conditions, and similarly for $\overline{\cG(a,m;\tau)}$
on the lower half $\{x_0<0\}$. 

Note that $\cG(a,m;\tau)$ can be split into classical, one-loop and
instanton contributions just like the Nekrasov partition function in \eqref{Nekpartasdf}. 
Whereas we take its classical and instanton contributions to be the
same as those of $\cZ(a,m;\tau)$, i.e. $\cG_{\mathrm{cl}}(a;\tau) \equiv \cZ_{\mathrm{cl}}(a;\tau)$ and $\cG_{\mathrm{inst}}(a,m;\tau) \equiv \cZ_{\mathrm{inst}}(a,m;\tau)$, the one-loop factor $\cG_{\rm 
  1-loop}$ is not canonically determined. We claim that it is
fixed by imposing Dirichlet boundary conditions on the half-sphere, in
such a way that 
\bea
\label{eqn:norm-1-loop}
\cG_{1-\mathrm{loop}}(a,m) 
& = \frac{\prod\limits_{w \in \mathrm{adj}} \Gamma_b\left( \frac{q}{2}+ i(a,w)+im\right) }{ \prod\limits_{e>0}^{\,} \Gamma_b(q+i(a,e)) \Gamma_b(q-i(a,e)) } \, ,
\eea
where $q =b+b^{-1}$ and $\Gamma_b(x)$ is the Barnes' double gamma
function. The numerator contains the contribution from the
vectormultiplet and the denominator that from the adjoint hypermultiplet with mass $m$ in the $\cN=2^*$ theory. 

Let us mention that via the AGT correspondence, in the case $N=2$, this choice is equivalent to a commonly used normalization of Virasoro conformal blocks in Liouville theory, as described for example in~\cite{Drukker:2009id}. For this choice of normalization, we will show that the expectation values of 't Hooft loop operators in the $\cN=2^*$ theory are given by acting on $\cG(a,m;\tau)$ with the three-dimensional difference operators $G^{(\mathrm{3d})}_{R}$, constructed in~\S\ref{section_Reduction_to_3d}. 

\subsection{S-duality domain wall}

The three-dimensional theory $\cT(SU(N))$ appears as an S-duality domain wall between two four-dimensional $\cN=4$ SYM theories with gauge groups $SU(N)$ and ${}^LSU(N)$ respectively and equal holomorphic gauge coupling $\tau$~\cite{Gaiotto:2008sa,Gaiotto:2008ak}. Furthermore, the mass deformation $m$ of the domain wall theory can be identified with the canonical mass deformation of the bulk theory to $\cN=2^*$ by giving a mass to the adjoint $\cN=2$ hypermultiplet. 

On the ellipsoid $S^4_b$, one can introduce the S-duality domain wall at the equator $\{x_0=0\}$ in a way that preserves half of the supersymmetries of the bulk~\cite{Drukker:2010jp}. As above, let us assume that the normalized function  $\overline{\cG(a,m;\tau)}$ corresponds to the partition function of the $\cN=2^*$ theory with gauge group $SU(N)$ on $\{x_0<0\}$ with Dirichlet boundary conditions for the vectormultiplet, and similarly that $\cG(a',m;\tau)$ corresponds to the partition function of the $\cN=2^*$ theory with gauge group ${}^LSU(N)$  on $\{x_0>0\}$. Let us also denote the partition function of the $\cT(SU(N))$ theory on the squashed three-sphere at the equator $\{ x_0=0\}$ by $\cZ(a,a',m)$, where $a$ and $a'$ are mass parameters for the $SU(N)\times {}^L SU(N)$ global symmetry as in~\S\ref{section_Reduction_to_3d}. Then the combined partition function in the presence of the S-duality domain wall is
\be
\int da\,da'\, \mu(a)\, \mu(a') \,  \overline{\cG(a,m;\tau)} \, \cZ(a,a',m) \, \cG(a',m;\tau) \, ,
\ee
where $\mu(a)\, \mu(a')$ is the partition function of three-dimensional $\cN=2$ vectormultiplets on the equator $\{x_0=0\}$ that gauge the symmetry $ SU(N)\times {}^L SU(N)$ (see Figure~\ref{fig:foursphere-S-wall}).

\begin{figure}[t]
\centering
\includegraphics[width=0.75\textwidth]{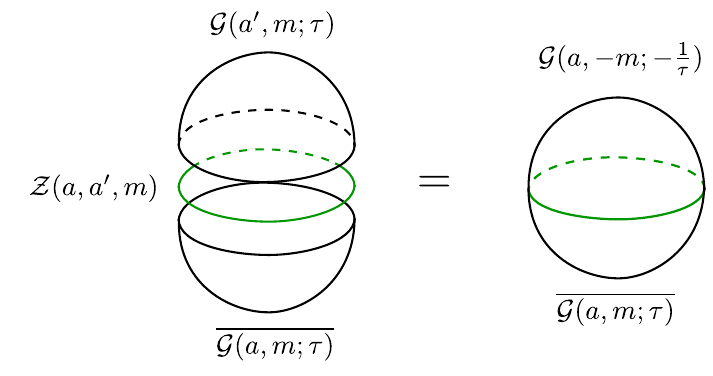}
\caption{Left: The ellipsoid partition function in the presence of an S-duality domain wall can be constructed by gluing in the domain wall partition function $\cZ(a,a',m)$ in between the half-sphere partition functions $\overline{\cG(a,m;\tau)}$ and $\cG(a',m;\tau)$, while gauging their flavor symmetries. Right: The same ellipsoid partition function can be constructed by gluing the half-sphere partition functions $\overline{\cG(a,m;\tau)}$ and $\cG(a,-m;-\frac{1}{\tau})$. 
\label{fig:foursphere-S-wall}}
\end{figure}

Another interpretation of the same domain wall is as a Janus domain wall interpolating between holomorphic gauge coupling $\tau$ for $\{x_0<0\}$ and $-1/\tau$ for $\{x_0>0\}$. The two pictures are related by an S-duality transformation of the theory on $\{x_0>0\}$. Another way of saying this is that the partition function $\cZ(a,a',m)$ should be an S-duality kernel relating the functions $\cG(a,m;\tau)$ and $\cG(a,-m;-1/\tau)$ through the measure $d\mu(a)$. This statement is rather hard to check in field theory because there is in general no closed expression for the $\cG(a,m;\tau)$. In Appendix~\ref{appendix:sdualitykernel}, however, we demonstrate this explicitly in the limit $m=0$ and $b=1$ and explain some of the subtleties involved in making this statement precise. 

On the other hand, in the context of the AGT correspondence, it has been checked in~\cite{Hosomichi:2010vh,Terashima:2011qi} that the partition function $\cZ(a,a',m)$ on a squashed three-sphere of the $\cT(SU(2))$ theory is precisely equal to the S-duality kernel of a normalized conformal block in Liouville theory~\cite{Teschner:2003at} under the relevant identification of parameters. We are not aware of a similar computation for $N>2$ and Toda theory.

\subsection{Supersymmetric loop operators}

Since we can embed the mass-deformed $\cT(SU(N))$ theory as a domain
wall in the four-dimensional $\cN=2^*$ theory on an ellipsoid, it is natural to think that supersymmetric loop operators
in the two theories on $\{x<0\}$ and $\{x>0\}$ are related. In particular, one can introduce a
loop operator on one hemisphere and push it through the domain wall
to find another loop operator on the other hemisphere. For an
S-duality wall one expects that this process turns a Wilson loop
operator in the four-dimensional $\cN=2^*$ theory into a 't Hooft loop
operator.  

Let us briefly summarize a few facts that are known about
supersymmetric loop operators on the four-sphere. The four-sphere
partition function can for instance be enriched with Wilson and 't Hooft loop
operators. To preserve half of the supersymmetries such loop operators should be
supported on the circle   
\bea
 x_0 &= \cos \rho \\
 x_1 &= b \sin \rho \cos \varphi \\
 x_2 &= b \sin \rho \sin \varphi \\
 x_3 &= x_4 = 0\, ,
\eea
where $0<\varphi<2\pi$ and $0<\rho<\pi$, or alternatively supported on the circle obtained by
interchanging $b \leftrightarrow b^{-1}$ and $\{ x_1,x_2\}
\leftrightarrow \{x_3,x_4\}$. The support of the loop operator lies in the squashed three-sphere at the equator $\{x_0=0\}$ when $\rho=\pi/2$. However, the expectation value of the loop operator is independent of $\rho$.

\subsubsection*{Wilson loops}

Supersymmetric Wilson loops in the four-dimensional $\cN=2^*$ theory are labeled by
irreducible representations of the gauge group $G$. The expectation
values of supersymmetric Wilson loops on the ellipsoid have been
computed in~\cite{Hama:2012bg}. 

The expectation value for a supersymmetric Wilson loop in the
irreducible representation $R$ around a circle in the $(x_1, x_2)$-plane is obtained by inserting the factor 
\be
W_R(a) = \sum_{w \in R} e^{-2\pi b (w,a)}
\ee
into the matrix integral. For example, for a rank $r$ antisymmetric
tensor representation of $SU(N)$ we insert the factor
\be
W_{(1^r)}(a) = \sum_{\{j_1<\ldots<j_r\}} e^{-2\pi b \,(a_{j_1}+ \cdots + a_{j_r})} \, .
\ee
 The expectation value for
supersymmetric Wilson loops in the $(x_3,x_4)$-plane is obtained by
replacing $b\to b^{-1}$. 

Sometimes it is convenient to normalize the
above expression by dividing by the quantum dimension $\dim_{q}R$ of
the representation, where $q=e^{i\pi b^2}$, but we will not do this here.  

\subsubsection*{'t Hooft loops}

A supersymmetric 't Hooft loop is defined by computing the path
integral in the presence of a singular boundary condition along a
circle that preserves half of the supersymmetries. The boundary
condition is specified by the image of an abelian 't Hooft monopole
under a homomorphism $\rho : U(1) \to G$, with gauge transformations
acting by conjugation on $\rho$. These configurations are classified
by irreducible representations $R$ of the Langlands dual ${}^LG$~\cite{Kapustin:2005py}. 

The expectation values of supersymmetric 't Hooft loop operators in
the $\cN=2^*$ theory on the round four-sphere have been computed in~\cite{Gomis:2011pf}. It was found that the expectation value can be expressed as (where now $b=1$)
 \be
 \int da \, \overline{\cZ(a,m;\tau)} \left( T_{R} \cdot
   \cZ(a,m;\tau) \right),
 \ee
 where $T_{R}$ is a difference operator that acts on the Coulomb branch parameters $a$. The difference operator
 takes the general form  
 \be
 T_R \cdot \cZ(a) = \sum_{\nu} C_{\nu}(a,m) \, \cZ(a+ i \nu)\, ,
 \ee
 where the sum is taken over the weights $\nu$ of the representation
 $R$. 

For the antisymmetric tensor representations $R = (1^r)$ the coefficients
$C_{\nu}(a,m)$ only receive one-loop 
 contributions. In this case
 \be
 C_{\nu}(a,m) = \prod_{\substack{j \in I \\ k \notin I }} \left[ \frac{\sinh \pi (a_{jk}-m) \sinh\pi(-a_{jk}-m) }{ \sinh\pi(a_{jk}) \sinh\pi(-a_{jk} ) } \right]^{1/2}\, ,
 \ee
 where we have denoted the weights of the $r$-th antisymmetric tensor
 representation by $\nu = \sum_{j \in I} h_j $ for $I = \{ j_1 <
 \ldots <
 j_r \}$. For general representations $R$ there are additional
non-perturbative monopole bubbling contributions to the coefficients
 $C_{\nu}(a,m)$.

Here, we want to re-express the expectation value of the 't Hooft operator in terms of a difference
operator $\widetilde{T}_{R}$ acting on the half-sphere partition 
function $\cG(a,m;\tau)$ in the case $b=1$. In other words, the expectation value of the 't Hooft loop is given by 
\be
\int da \, \mu(a) \,  \overline{\cG(a,m;\tau)} \,  \left( \widetilde{T}_{R} \cdot \cG(a,m;\tau) \right) .
\ee
 The difference operator $\widetilde{T}_{R}$ is related to
 $T_{R}$ by conjugating with the one-loop factor that relates the
 Nekrasov partition function $\cZ(a,m;\tau)$ to the half-sphere
 partition function $\cG(a,m;\tau)$.  Later it will be important that $\widetilde{T}_R$ is self-adjoint with respect to the measure $\mu(a) da$.

In appendix~\ref{appendix:special}, we perform this conjugation explicitly for the antisymmetric tensor representations to find
\be \label{eq:hooftb=1}
\widetilde{T}_{(1^r)} \cdot \cG(a) = \sum_{|I|=r} \prod_{\substack{ j \in I \\ k
    \notin I }}  \frac{\sin\pi(-ia_{jk}-im)}{\sin\pi(ia_{jk}) }
\cG \left( a+i \sum_{j \in I} h_j \right).
\ee
 Remarkably, this difference operator is in agreement with the difference operators $G^{(\mathrm{3d})}_{(1^r)}\, (= \tilde{G}^{(\mathrm{3d})}_{(1^r)})$ that introduce codimension-two defects in the $\cT(SU(N))$ theory by acting on the three-dimensional partition function $\cZ(a,a',m)$, in the limit $b\to1$. 

\subsection{Intertwining Wilson and 't Hooft loops}

Let us now explain why the difference operators $G_{R}^{(\mathrm{3d})}$ are related to 't Hooft operators $\widetilde{T}$. We consider the four-sphere partition function in the presence of both an S-duality wall and a supersymmetric loop operator. Recall that on the lower half-sphere $\{x_0<0\}$ we have the gauge group $SU(N)$ for which the Wilson loops are labelled by irreducible representations of $SU(N)$. On the upper half-sphere $\{x_0>0\}$, we have the gauge group ${}^L SU(N)$ for which the 't Hooft operators are labelled by irreducible representations of $SU(N)$.

Thus, let us now consider an 't Hooft loop labelled by an irreducible representation of $SU(N)$ inserted at some point $\rho>\pi/2$ in the upper half-sphere $\{x_0>0\}$. The expectation value of this system takes the form
\be
\int da\,da'\, \mu(a)\, \mu(a') \,  \overline{ \cG(a,m;\tau)} \, \cZ(a,a',m) \, \left[\, \widetilde{T}_{(1^r)}(a') \cdot \cG(a',m;\tau) \right]\, ,
\ee
where for simplicity we focus on antisymmetric tensor representations. 

Now, the expectation value is independent of the position $\rho$. Thus we can imagine moving the 't Hooft loop through the S-duality domain wall to some point $\rho<\pi/2$ in the region $\{x_0<0\}$. According to the transformation of loop operators under S-duality, it should become a Wilson loop in the antisymmetric tensor representation $R=(1^r)$. At the level of the partition function, since the operator $\widetilde{T}_{(1^r)}(a')$ is self-adjoint with respect to the measure $\mu(a')$, we can bring it to act on $\cZ(a,a',m)$. Provided that $\cZ(a,a',m)$ is an eigenfunction such that
\be
\widetilde{T}_{(1^r)}(a') \cdot \cZ(a,a',m) = W_{(1^r)}(a)\, \cZ(a,a',m)\, ,
\label{eq:intertwine}
\ee
we find
\bea
\int da\,da'\, &\mu(a)\, \mu(a') \, \overline{\cG(a,m;\tau)} \, \cZ(a,a',m) \, \left[\, \widetilde{T}_{(1^r)}(a') \cdot \cG(a',m;\tau) \right]\\
&=\int da\,da'\, \mu(a)\, \mu(a') \,  \overline{\cG(a,m;\tau)} \, \left[ \, \widetilde{T}_{(1^r)}(a')  \cdot \cZ(a,a',m) \, \right] \, \cG(a',m;\tau) \\
&=\int da\,da'\, \mu(a)\, \mu(a') \,  \left[ \, \overline{ W_{(1^r)}(a)\, \cG(a,m;\tau)}\, \right] \, \cZ(a,a',m)\, \cG(a',m;\tau) ,
\eea
which is the expectation value of an S-duality domain wall together with a Wilson loop in the representation $(1^r)$ at some point $\rho<\pi/2$. Thus compatibility with S-duality demands that $\cZ(a,a',m)$ intertwines 't Hooft loops and Wilson loops according to equation~\eqref{eq:intertwine}. See Figure~\ref{fig:foursphere-S-wall-Hooft}. In section \S\ref{section_Reduction_to_3d} we have argued that three-dimensional mirror symmetry requires $\cZ(a,a',m)$ to obey the same relation with respect to the three-dimensional limit of the surface defect operators $G^{(\mathrm{3d})}_{(1^r)}$. Thus the corresponding operators should agree. Above we checked that this is indeed the case for a round four-sphere.

\begin{figure}[t]
\centering
\includegraphics[width=1.0\textwidth]{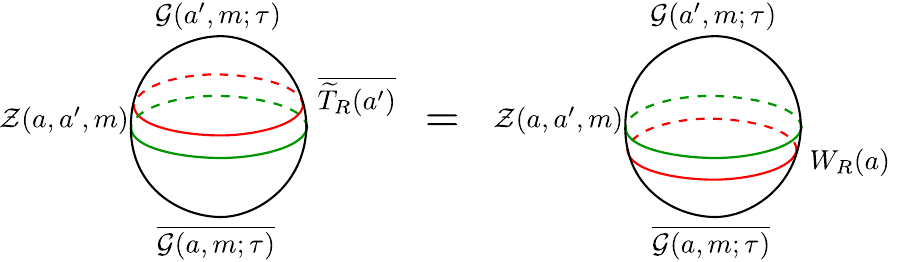}
\caption{A 't Hooft loop operator $T_R$ can be moved through the S-duality domain wall to obtain a Wilson loop operator $W_R$. 
\label{fig:foursphere-S-wall-Hooft}}
\end{figure}

Let us now make some comments on non-minuscule representations $R$. Since Wilson loop operators labeled by $R$ are defined by a trace over the representation $R$, they obey the character algebra
\be
W_{R_1} \circ W_{R_2}= \sum_{R_3} {N_{R_1,R_2}}^{R_3} \, W_{R_3} \, ,
\ee
where ${N_{R_1,R_2}}^{R_3}$ are the standard Littlewood-Richardson coefficients. In particular, they can be generated from Wilson loops labeled by fully antisymmetric tensor representations by composition and addition/subtraction. 

Therefore, we can define a new set of operators $\hat{T}_{R}$ by taking $\hat{T}_{(1^{r})}\equiv \widetilde{T}_{(1^{r})}$, or equivalently $\hat{T}_{(1^{r})}\equiv G^{(\mathrm{3d})}_{(1^{r})}$, for antisymmetric representations and imposing the character algebra
\be
\hat{T}_{R_1}\circ \hat{T}_{R_2} = \sum_{R_3} {N_{R_1,R_2}}^{R_3} \, \hat{T}_{R_3} \, .
\ee
The resulting operators $\hat{T}_{R}$ automatically transform in the expected way under S-duality, and it is natural to expect that these operators encode the expectation value of 't Hooft loops for general representations. 

However, we emphasize that the $\hat{T}_{R}$ do not seem to correspond to the expectation value of a 't Hooft loop with magnetic weight given by the highest weight of the representation $R$, when the representation is non-minuscule. For example, for $SU(2)$ the 't Hooft loop whose magnetic weight is double that of the 't Hooft loop of minimal charge is given by $T_1 \circ T_1$ rather than $T_1 \circ T_1 - T_0$. This is again an invertible linear transformation on the algebra of the operators. In this case, the origin of the basis transformation is a natural resolution of the Bogomolnyi moduli space that arises for representations with non-perturbative monopole bubbling effects~\cite{Gomis:2011pf}. Once again, we emphasize that the simplest and unambiguous operators are those in antisymmetric tensor representations.

\subsection{Verlinde operators in Toda CFT}

All we have discussed so far in this section can also be framed in the
language of Liouville or Toda conformal field theory. This approach has the benefit 
that, at least for the 't Hooft loop in the fundamental representation, we can
compute the required operator for general squashing parameter
$b$.

\begin{figure}[t]
\centering
\includegraphics[height=3cm]{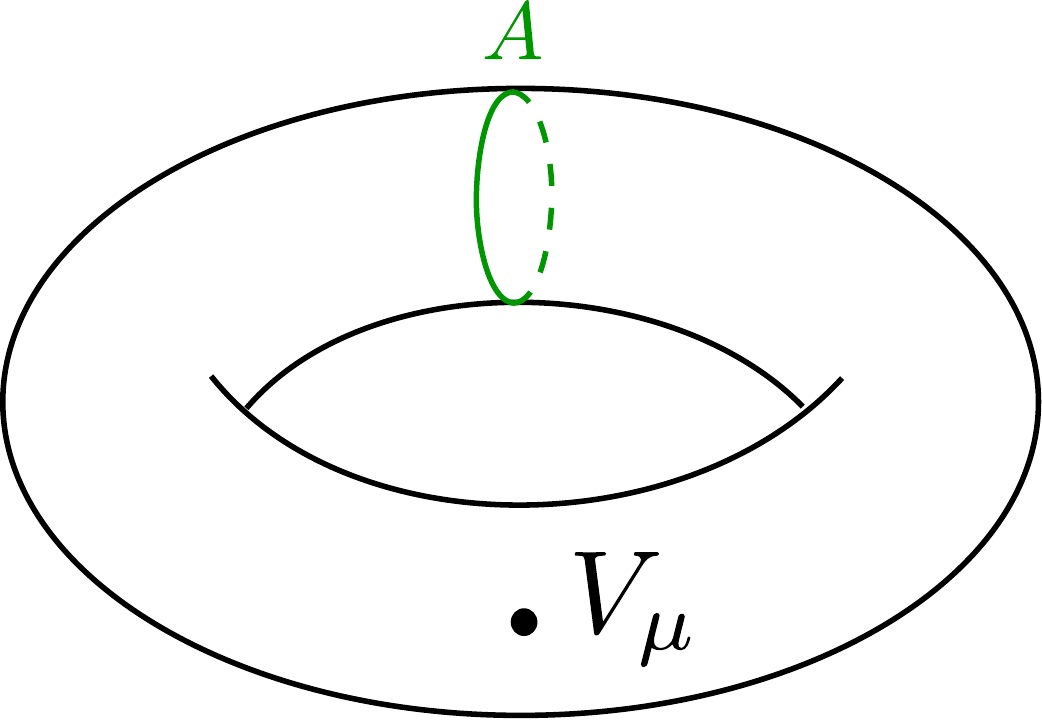}
\caption{The four-sphere partition function of the $\cN=2^*$ theory
    is equal to a Toda correlation function on the punctured torus with
      a semi-degenerate vertex operator $V_{\mu}$, with momentum $\mu$, inserted  at the puncture.} \label{fig:N=2star}
\end{figure}

Let us briefly review aspects of this
correspondence. For the $\cN=2^*$ theory with gauge group $SU(N)$, the
ellipsoid partition function is related to a Liouville
or type $A_{N-1}$ Toda
correlator on the punctured torus with an insertion of a
semi-degenerate primary field. The parameters on both sides of the
correspondence are related as follows:
\begin{enumerate}
\item The geometric parameter $b$ is a
  dimensionless coupling in the conformal field theory and gives the central charge $c=(N-1)(1+N(N+1) q^2)$, where $q=b+b^{-1}$.
\item The holomorphic gauge coupling $\tau$ is the complex structure
  parameter of the punctured torus.
\item The mass $m$ of the adjoint hypermultiplet is encoded in the momentum of the semi-degenerate primary field,
\be
\mu = N \left( \frac{q}{2}+im \right) \omega_{N-1}\, .
\ee
\end{enumerate}

Choosing a pants decomposition, the correlation function of the primary field on the punctured torus can be written as an expansion in Liouville or $W_N$--algebra conformal blocks
\be
\int d\al \, C(\mu,\al,2Q-\al) \, \overline{\cF(\al,\mu;\tau)} \cF(\al,\mu;\tau) \, ,
\ee
where the integral is over non-degenerate momenta $\al = Q +i a$, with $a\in \mathbb{R}^{N-1}$ and $Q= q \rho$, where $\rho$ is the Weyl vector of $A_{N-1}$. 

The conformal blocks $\cF(a,\mu;\tau)$ are normalized to
contain the classical and instanton contributions to the Nekrasov
partition function. The three-point function $C(\mu,\al,2Q-\al)$ is
proportional (up to an $m$-dependent piece that can be absorbed in the
normalization of the primary field) to the modulus squared of the
1-loop contribution $| \, \cG_{1-\mathrm{loop}} \, |^2$ times the measure
$\mu(a)$. The meromorphic function $\cG(a,m;\tau)$ that we introduced
earlier corresponds to a convenient normalization of the
conformal block $\cF(\al,\mu;\tau)$ that absorbs the three-point functions. This
is an extension to higher rank of a frequently used normalization in Liouville theory~\cite{Drukker:2009id}. 

Loop operators in the four-dimensional gauge theory are realized as
Verlinde operators in the dual conformal field
theory~\cite{Alday:2009fs,Drukker:2009id}. The Verlinde operators act on the space of Virasoro or $W_N$-algebra
conformal blocks by transporting a chiral primary field around a
simple closed curve $C$ on the Riemann surface. The operators constructed in this way depend only on the homotopy class of
the curve $C$ up to a choice of `framing' that will not be important here.

If we choose the pants decomposition of the punctured torus determined by the A-cycle in Figure~\ref{fig:N=2star}, a
supersymmetric Wilson loop in $\cN=2^*$ theory in the rank $r$
antisymmetric tensor representation corresponds to transporting a
degenerate chiral primary with momentum $\eta = -b \omega_j$ around
that A-cycle. The resulting expression changes from the original conformal
block by the factor
\be
W_{(1^{r})}^{\mathrm{CFT}}=\sum_{\{j_1<\ldots<j_r \}} e^{-2\pi b\, (a_{j_1}+\cdots+j_{r})}\, ,
\ee
which is in agreement with the localization computation. 

An 't Hooft loop in the $r$-th fundamental representation corresponds
to transporting the same chiral primary around the B-cycle of the punctured 
torus. This Verlinde operator has been computed directly in Toda theory
for the fundamental representation in~\cite{Gomis:2010kv}. Acting on the 
conformal blocks $\cF(\al,\mu;\tau)$, the operator is given by 
\be
T^{\mathrm{CFT}}_{(1)} \cdot \cF(\al) = \sum_{j=1}^N \prod_{k \neq j}^N \frac{\Gamma(iba_{jk}) \Gamma\left(bq+iba_{jk}\right)}{\Gamma\left( \frac{bq}{2}+iba_{jk}-ibm \right)\Gamma\left( \frac{bq}{2}+iba_{jk}+ibm\right) } \cF(\al-bh_j) \, ,
\ee
where $\al = Q+ia$ is the momentum around the loop that defines the
pants decomposition. 

To construct an operator that acts on the normalized conformal
blocks $\cG(a,m;\tau)$, we have to conjugate by the one-loop
contribution~\eqref{eqn:norm-1-loop}. In Appendix~\ref{appendix:special} we perform this conjugation to find
\begin{align}
  \widetilde{T}^{\mathrm{CFT}}_{(1)} \cdot \cG(a) = \sum_{j=1}^N \prod^N_{
   \substack{k \neq j}} \frac{ \sin \pi b \left( \frac{q}{2}
     - i a_{jk} -i m \right)}{ \sin \pi b \left( -i a_{jk} \right)} \,
 \cG \left(a + i b h_j  \right),
 \label{}
\end{align}
which is precisely equal to the three-dimensional operator $G^{\mathrm{(3d)}}_{(1)}$ for any real $b$ (see equation~\eqref{antisymmetric3d}). This provides another check on the relation of the difference operators
$G^{(\mathrm{3d})}_R$ to the 't Hooft loop operators for the fundamental representation.


\section{Discussion}\label{section_Conclusions}

In this paper we generated an algebra of difference operators $G_R$ acting on the $\cN=2$ superconformal index,
labeled by irreducible representations $R$ of $SU(N)$. Generalizing the
arguments of \cite{Gaiotto:2012xa}, we claim that these difference operators
represent half-BPS surface defects in four-dimensional $\cN=2$ theories
of class $\cS$. We discussed several arguments in favour of this
claim. Most importantly, we emphasized that it is highly non-trivial
that we indeed managed to consistently 
close the algebra, and that the difference operators have a natural interpretation in various dual frames. Let us mention a few open questions and interesting links.

\medskip

A microscopic gauge theory understanding of these defects is
unfortunately still lacking, either in terms of a defect description
or alternatively as a description of the two-dimensional degrees of
freedom living on the support of the defect. We did find a
two-dimensional field theory description in two extreme
cases: fully antisymmetric and fully symmetric representations. 
It is however not at all clear that there exists a Lagrangian
description for the two-dimensional degrees of freedom living on the
support of the surface defect for a generic representation $R$.

The
operators $G_R$ can be written as a sum of weights in the
representation $R$, which in the field theory description of the
defect should have the interpretation as a sum over vacua. In case
a weight $\lb$ appears with multiplicity one in the representation $R$, the
contribution to $G_R$ is a single ratio of theta-function and seems
likely to have an interpretation as the contribution to the superconformal index in a vacuum
characterized by $\lambda$. When the weight $\lambda$ appears with higher  
multiplicity, however, the contribution to $G_R$ is a sum of such ratios and
is less likely to have such an interpretation. 

A similar structure can be observed from the perspective of the AGT correspondence. In particular, reference~\cite{Doroud:2012xw} has demonstrated that a ratio of Toda correlation functions involving a degenerate momentum $\mu = -b h_1$ captures the two-sphere partition function of the $\cN=(2,2)$ theory that we have associated to the surface defect labelled by the fundamental representation. However, for representations with multiple weight
contributions, the Toda three-point function with degenerate insertions (see~\cite{Fateev:2005gs,Fateev:2007ab}) do not appear to have the structure of one-loop contributions to the two-sphere partition function of an $\cN=(2,2)$ theory.  
   
\begin{figure}[t]
\centering
\includegraphics[width=0.75\textwidth]{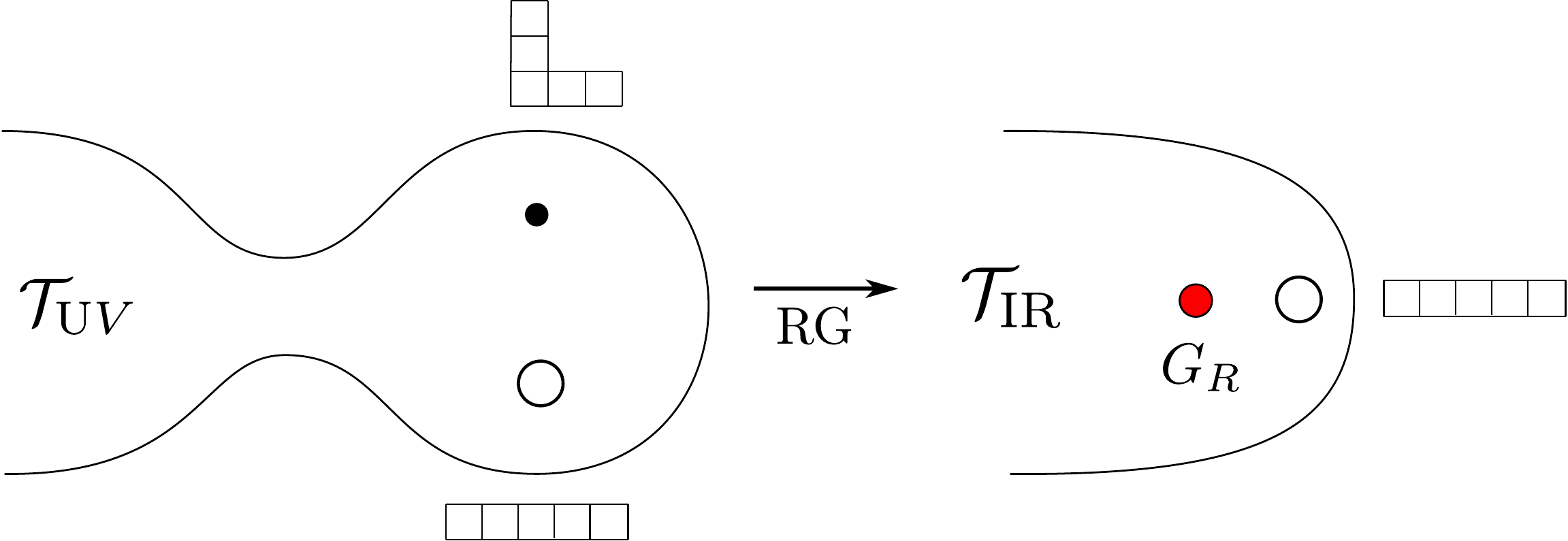}
\caption{Left: UV curve of the theory $\cT_{\rm UV}$, obtained by gluing a three-punctured
  sphere with two full punctures and one hook-shaped puncture to the
  UV curve of the theory $\cT_{\rm IR}$. Right: A renormalization group flow connects
  $\cT_{\rm UV}$ to $\cT_{\rm IR}$ with a possibly general surface defect
  $G_{R}$. }
\label{fig:UVIR3}
\end{figure}

\medskip

As briefly mentioned in the introduction and main text, we expect that there is an alternative method to find the difference operators $G_R$. Instead of coupling the theory
$\cT_{\rm IR}$ to a bifundamental hypermultiplet, corresponding to adding a puncture with $U(1)$ symmetry, one could add a puncture with a larger flavor symmetry group. This generically involves coupling $\cT_{IR}$ to a non-Lagrangian theory corresponding to a sphere with two full punctures and one intermediate puncture. An example is illustrated in
Figure~\ref{fig:UVIR3}. The superconformal index for $\cT_{\rm UV}$ should then contain a larger spectrum of residues. One might expect any difference operator $G_R$ to originate from such a residue computation.

Again, this is analogous to the Toda perspective, where non-maximal flavor punctures correspond to insertions of semi-degenerate vertex operators. By analytic continuation correlation functions of such operators have poles, whose residues correspond to reductions to a completely degenerate vertex operator. If we insert a semi-degenerate vertex operator that corresponds to a simple $U(1)$ puncture, we can only access completely degenerate vertex operators with momentum $\al = -b \lb_1-b^{-1}\lb_2$, where $\lb_1 = r_1 \omega_1$ and $\lb_2 = r_2 \omega_2$ are the highest weights of two symmetric tensor representations. To find completely degenerate vertex operators with generic weights $\lb_1$ and $\lb_2$, one must insert a generic semi-degenerate vertex operator, corresponding  to a generic flavor puncture.

\medskip

The difference operators $G_R$ are elliptic generalizations of the Macdonald operators. Although these operators have not been constructed mathematically for all representations $R$ (as far as we know), the elliptic Ruijsenaars-Schneider difference operators have been related to exterior powers of the vector representation of the elliptic quantum group $E_{\tau,\eta}(gl_N)$ \cite{FelderVarchenko}. It would be very interesting to interpret this connection to elliptic quantum groups physically. This relation could appear naturally when interpreting the difference operators $G_R$ in terms of a three-dimensional topological field theory on $C \times S^1$. 

In the Macdonald limit $p=0$ we have found that the difference operators correspond to Wilson loops in an analytic continuation of refined Chern-Simons theory on $C \times S^1$ (see \S\ref{sec:2dTQFT}). The particular ratio of modular S-matrices that appears in the operator product expansion of the $G_R$, suggests that when we take a Wilson loop operator close to a puncture on $C$, it can be interpreted as a Verlinde loop operator on the boundary torus. For example, taking the Schur limit $t=q$, and replacing $q \to \exp\left( 2\pi i / (k+N) \right)$ in the modular S-matrix, we would recover the modular S-matrix elements for characters of integrable representations of the affine current algebra $\widehat{\mathfrak{su}}(N)_k$. However, for the superconformal index it is important that we have an analytic continuation of this statement to $|q|<1$.

\medskip

In the dimensional reduction of the four-dimensional superconformal index to a three-dimensional partition function, we found that the difference operators $G_R$ are related to operators that introduce line defects. For theories of class $\cS$, there is a mirror description as a star-shaped quiver and we showed that the difference operators introduce Wilson loops for the central node of the quiver, at least in the case of antisymmetric tensor representations. For non-minuscule representations, we found that there is some mixing. 

It would be interesting and important to understand these line defects in three-dimensions from first principles by localization. For the fully symmetric and anti-symmetric tensor representations, we expect this could be done by coupling to a supersymmetric quantum mechanics on a circle, in a similar spirit to~\cite{Gadde:2013dda} but in one dimension lower. On the other hand, we expect that the rank-$r$ anti-symmetric tensor operator has another description as a monodromy defect breaking the gauge group to $S(U(r) \times U(N-r))$, which might also be used to perform an exact localization computation by extending the computations of~\cite{Kapustin:2012iw,Drukker:2012sr} for abelian monodromy defects.


\acknowledgments

We would very much like to thank Fernando Alday for collaboration on part of this project. The work of M.B. is supported by the Perimeter Institute for Theoretical Physics. Research at the Perimeter Institute is supported by the Government of Canada through Industry Canada and by the Province of Ontario through the Ministry of Research and Innovation. The work of M.F. and P.R. is supported by ERC STG grant 306260. The work of L.H. is supported by a Royal Society Dorothy Hodgkin fellowship.


\appendix


\section{Macdonald polynomials and the refined S-matrix}\label{MacPolys}

\subsection{Group theory}

The finite dimensional irreducible representations of $A_{N}$ are in one-to-one correspondence with dominant integral weights,
$$ \lambda = \sum_{i=1}^{N-1} \lambda_i \, \omega_i $$
whose Dynkin labels $(\lambda_1,\lambda_2,...,\lambda_{N-1})$ are nonnegative integers. Equivalently, irreducible representations are labeled by partitions $(\ell_1,\ell_2,\ldots,\ell_{N})$ where $\ell_1 \geq \ell_2 \geq... \geq \ell_{N}=0$, such that
\begin{equation}
\ell_i = \lambda_i + \lambda_{i+1}+...+\lambda_{N-1} \, .
\end{equation}
Each partition is associated to a Young diagram whose $i$-th row has length $\ell_i$. For instance,  the following diagram
$$\yng(4,2,2)$$
corresponds to the partition $( 4,2,2,0)$. The partition labels $(\ell_1,\ell_2,\ldots,\ell_{N})$ are related to the components of the weight in the orthogonal basis
\be
 \omega_i = \epsilon_1+\cdots+\epsilon_i -\frac{i}{N}\sum_{j=1}^N \epsilon_j
\ee
where
\be
\lambda = \sum_{i=1}^N \kappa_i \epsilon_i \, , \qquad \kappa_i = \ell_i -\frac{1}{N} \sum_{j=1}^{N-1}j (\ell_j - \ell_{j+1})\, .
\ee

The states in a given irreducible representation are in one-to-one correspondence with {\it semi-standard Young tableaux}. They are obtained by filling the boxes of a Young diagram with the numbers $\{1,\ldots,N\}$, such that the numbers are non-decreasing from left to right and strictly increasing from top to bottom. Finally, to each semi-standard Young tableau, we attach the labels $(n_1,...,n_N)$, where $n_i$ denotes the number of times that $i$ appears in the semi-standard tableau. As an example below we include a few semi-standard tableaux  for the adjoint ${\bf 8}$ representation of $SU(3)$ with their corresponding labels.

\begin{center}
\begin{tabular}{c c c c}
 $\young(11,2)$ & $\young(12,3)$ &$\young(13,2)$ &$\young(22,3)$\\
 $(2,1,0)$  &$(1,1,1)$ &$(1,1,1)$ & $(0,2,1)$
\end{tabular}
\end{center}

\subsection{Schur polynomials and the modular S-matrix}

Introduce coordinates $a_j$, for $j=1,...,N$, obeying $\prod_{i=1}^N a_i=1$. For $a_j=e^{i \theta_j}$ they are coordinates on the maximal torus of $SU(N)$. The Schur polynomials form a basis of symmetric functions in the variables $\{a_1,\ldots,a_N\}$ labeled by irreducible representations. The Schur polynomial of the irreducible representation $\lambda$ labeled by the partition $(\ell_1,\ldots,\ell_{N})$ is given by the determinant formula
\begin{equation}
\chi_{\lambda}(a) = \frac{\det a_j^{\ell_i+N-i}}{\det a_j^{N-i}}\, .
\end{equation}
An important property of the Schur polynomials is that they are orthonormal
with respect to the inner product on the space of symmetric functions
\be
\la f , g\ra = \int \Delta(a) f(a) g(a^{-1}) 
\ee
where
\be
\Delta(a) = \frac{1}{N!} \prod\limits_{i=1}^{N-1} \frac{da_i}{2 \pi i a_i} \prod\limits_{i \neq j} \left(1-\frac{a_i}{a_j} \right) .
\ee
is the Haar measure and the integration is over the maximal torus of $SU(N)$. Products of Schur polynomials decompose according to the tensor product of the irreducible representations
\be
\chi_{\lambda_1} (a) \, \chi_{\lambda_1}(a)  = \sum_{\mu} {N_{\lambda_1,\lambda_2}}^{\mu}\,  \chi_{\mu}(a)
\ee
where ${N_{\lambda_1,\lambda_2}}^{\mu}$ are the Littlewood-Richardson numbers.

In order to construct the modular S-matrix we introduce the Weyl weight $\rho$, which is the highest weight of the adjoint representation of $SU(N)$. Its components in the Dynkin basis are $\rho=(1,1,\ldots,1)$. In the orthogonal basis mentioned above, 
\begin{equation}
\rho = \left( \frac{N-1}{2},\frac{N-3}{2},...,\frac{1-N}{2} \right) ,
\end{equation}
and we will denote these components by $\rho_j = (N-2j+1)/2$. Now consider two irreducible representations $\lambda$ and $\lambda'$ with components $\kappa_i$ and $\kappa'_i$ in the orthogonal basis. Then the modular S-matrix is given by
\begin{equation}
S_{\lambda \lambda'} = S_{00} \, \chi_{\lambda}( q^{\rho_1},...,q^{\rho_N})\chi_{\bar\lambda'}( q^{\rho_1+\kappa_1},...,q^{\rho_N+\kappa_N}) \, ,
\end{equation}
where $\bar\lambda'$ denotes the complex conjugate representation of $\lambda'$. We will not need the overall normalization $S_{00}$.

\subsection{Macdonald polynomials and the refined S-matrix}

The Macdonald polynomials are symmetric polynomials in the variables $\{a_1,\ldots,a_N\}$ that depend on two additional complex parameters $q$ and $t$. The Macdonald polynomials are labeled by irreducible representations $\lambda$ of $SU(N)$ and reduce to the corresponding Schur polynomials when $q=t$. 

The Macdonald polynomial labeled by the irreducible representation $\lambda$ is 
\be
P_{\lambda}(a,q,t) = \chi_{\lambda}(a) + \sum_{\mu<\lambda} c_{\lb,\mu}(q,t) \, \chi_{\mu}(a)
\ee
where $c_{\lambda,\mu}(q,t)$ are rational functions of $q$ and $t$ that are uniquely determined by ensuring $P_\lambda(a,q,t)$ is a simultaneous eigenfunctions of the difference operators 
\be
G_r = t^{r(1-N)} \sum\limits_{ \substack{ I\subset\{1,\ldots,N\} \\ |I|=r } } \, \prod_{i\in I ,j\notin I} \frac{t a_i - a_j}{a_i - a_j} \, T_{I} \, ,\quad\qquad r=1,\ldots,N-1
\ee
where
\be
T_I : a_i \to \begin{cases} 
q^{1-1/N}a_i & \mbox{if}\, i\in I \\
q^{-1/N}a_i & \mbox{if}\, i \notin I \, .
\end{cases}
\ee
Here we have included a background shift by $q^{-1/N}$ compared to the standard Macdonald difference operators in order to preserve the condition $\prod_i^N a_i=1$ relevant for $SU(N)$. For example, the first few Macdonald polynomials for $SU(2)$ are
\bea
P_0(a,q,t) & =1 \\
P_1(a,q,t) & = \chi_1(a) \\
P_2(a,q,t) & = \chi_2(a) + \frac{q-t}{1-q t} \\
P_3(a,q,t) & = \chi_3(a) + \frac{(q-t)(1+q)}{1-t q^2} \chi_1(a) \\
P_4(a,q,t) & = \chi_4(a) + \frac{(q-t)(1-q^3)}{(1-q)(1-q^3 t)} \chi_2(a) + \frac{q(q-t)(1+q^2)(1-t)}{(1-q^2 t)(1-q^3 t)}\, .
\eea

The difference operators are self-adjoint with respect to the inner product
\be
\la f,g \ra = \int \Delta_{q,t} (a) f(a)g(a^{-1}) \, , \quad\qquad  \Delta_{q,t}(a) = \frac{1}{N!} \prod\limits_{i=1}^{N-1} \frac{da_i}{2 \pi i a_i} \prod\limits_{i \neq j} \frac{(a_i/a_j;\,q)}{(t a_i/a_j;\,q)}
\ee
where $(a;\,q) = \prod_{i=0}^{\infty} (1-q^i a)$ is the q-Pochhammer symbol, and consequently, the Macdonald polynomials are non-degenerate and orthogonal with respect to the same measure. In order to obtain functions orthonormal with respect to the measure, a normalization factor must be included.

The product of Macdonald polynomials decomposes according to the tensor product of irreducible representations
\be
P_{\lambda_1}(a,q,t) P_{\lambda_2}(a,q,t) = \sum_{\mu} {N_{\lambda_1,\lambda_2}}^{\mu} (q,t) P_{\mu}(a,q,t)
\ee
where the ${N_{\lambda_1,\lambda_2}}^{\mu} (q,t)$ are rational functions in $q$ and $t$.

Analogous to the modular S-matrix, the refined S-matrix is given by

\be
S_{\lambda \lambda'} = S_{00} \, P_{\lambda}( t^{\rho_1},...,t^{\rho_N}) \, P_{\bar\lambda'}( t^{\rho_1}q^{\kappa_1},...,t^{\rho_N}q^{\kappa_N}) \, .
\ee
It is then an easy exercise to check that the ratios $S_{R,S}/S_{0,S}$ are indeed the eigenvalues of the difference operators $G_R$ in the Macdonald limit, namely

\be
 G_R \cdot P_S(a_i,q,t) = \frac{S_{R,S}}{S_{0,S}} \, P_S(a_i,q,t) \, .
\ee

\section{S-duality kernel} \label{appendix:sdualitykernel}

Instead of merely inserting a three-dimensional $\cN=2$ vectormultiplet on the
three-dimensional boundary $\{ x_0 = 0\}$, we
could also glue in the three-dimensional mass-deformed linear quiver
theory $\cT(SU(N))$. In fact, its $\cN=4$ variant was introduced
as an S-dual of the Dirichlet boundary condition in the four-dimensional $\cN=4$ theory with gauge group $SU(N)$ \cite{Gaiotto:2008ak,Gaiotto:2008sd}. 

It is thus
natural to expect that the mass-deformed $\cT(SU(N))$ theory encodes the field theory degrees of freedom on a so-called S-duality domain wall in the
$\cN=2^*$ theory. Such a domain wall is defined so that the four-dimensional
 theories on either side are related by
the transformation $S:(\tau,m) \to (-1/\tau,-m)$. 
In this Appendix we will verify that this is indeed the case if we
assume that $\cG_b$ is the partition function on the
half-sphere with Dirichlet boundary conditions.

Before introducing the S-duality domain wall, let us briefly consider the
ellipsoid partition function $\cZ_{S^4_b}$ of the  $\cN=2^*$ theory with gauge
group $SU(N)$. The AGT correspondence relates this to a Toda
correlator on the once-punctured torus. We thus expect that the ellipsoid
partition function transforms as a modular form. More precisely, it
should transform as \cite{Hadasz:2009sw}
\be
\cZ_{S^4_b}(-m;-1/\tau) = |\tau|^{2 \Delta(m)} \cZ_{S^4_b}(m;\tau)\, ,
\ee
with modular weight 
\be
\Delta(m) = \frac{N(N-1)}{2} \left( \frac{Q^2}{4} +m^2 \right).
\ee

This modular property of the ellipsoid partition function is
guaranteed if the half-sphere partition function $\cG_b$
transforms as 
\be
\cG_b(-m,a;-1/\tau) = (-i\tau)^{\Delta(m)} \int da' \, \mu_b(a') \, \cZ_b(a_i,a',m)\, \cG_b(m,a';\tau)\, ,
\label{modularblocks}
\ee
where we integrate over a real slice of the Coulomb branch (just like
in all matrix integrals in the remainder of this section). The
integration kernel $\cZ_b(a,a',m)$ must obey two important 
properties. 
First, it must obey the symmetry 
$$\cZ_b(a,a',m)=\cZ_b(a',a,-m)\, .$$ 
Second, it must be a unitary with respect to the
measure $\mu_b(a) \, d a$, in the sense that 
\be\label{eqn:unitarytransfo}
 \int da \, \mu_b(a) \,  \overline{\cZ_b(a',a,m)} \, \cZ_b(a,a'',-m)  = \mu_b(a') \,
\delta(a',a'') \, . 
\ee

Now consider the ellipsoid partition function 
with the insertion of an S-duality domain wall. Assuming that
$\cG_b$ is the half-sphere partition function of the
$\cN=2^*$ theory with Dirichlet boundary conditions, the S-duality
partition function on the squashed four-sphere should be  given by  
\begin{multline}\label{eqn:Swallpartfunction}
\int da \, \mu(a) \, \overline{\cG_b(m,a;\tau)} \,
\cG_b(-m,a;-1/\tau) \\ = \int d a \, \mu(a) \, \int da'\, \mu(a')  \,
\overline{\cG_b(m,a;\tau)} \, \cZ_b(a,a',m) \,
\cG_b(m,a';\tau) \, . 
\end{multline}
Consequently, $\cZ_b(a,a',m)$ should encode the gauge degrees of freedom localized
on the domain wall. 
Specifically, we expect that $\cZ_b(a,a',m)$ is the partition function of the mass-deformed $\cT(SU(N))$ theory on a squashed three-sphere. 

In this context, the
symmetry $\cZ_b(a,a',m) = \cZ_b(a',a,-m)$ is equivalent to
three-dimensional mirror symmetry. The unitary property
(\ref{eqn:unitarytransfo}) follows
because the partition function $\cZ_b(a,a',m)$ is 
an eigenfunction of the self-adjoint operator $G^{(\mathrm{3d})}_R$ with respect to the
measure $\mu_b(a)\, da$ (see equation \eqref{3dopsself-adjoint}). 

Indeed, let us denote the integral (\ref{eqn:unitarytransfo}) by 
\be
\cI(a',a'') = \int da \, \mu_b(a) \,
\overline{\cZ_b(a',a,m)}  \, \cZ_b(a,a'',-m)\, . 
\ee
The self-adjoint operator
$G^{(\mathrm{3d})}_R(a')$ can act inside this integrand in either direction, which
must lead to the same answer. Consequently we find 
\be
\left(W_R(a) - W_R(a'') \right) \, \cI(a,a'') = 0\, ,
\ee
where $W_R(a)$ is the expectation value of a Wilson loop in the
representation $R$. This implies that the integral vanishes if $a \neq
a''$ modulo Weyl transformations.  

\subsection{Example}

Let us check the above transformation properties of the half-sphere
partition function $\cG_{b}$ on the round four-sphere, when $b=1$, and in the $\cN=4$
limit, when $m \to 0$.

First, we compute the explicit expression for $\cG_{b=1}(\tau,m,a_i)$ for gauge group $SU(N)$.
Its one-loop contribution~\eqref{eqn:norm-1-loop} simplifies to the formula
\be
\cG_{\rm 1-loop}(m,a_i) = \frac{1}{\sqrt{2\pi}} \prod_{i<j} \frac{\pi a_{ij} }{\sinh(\pi a_{ij})}\, ,
\ee
where $a_{ij}=a_i-a_j$ with the constraint that $\sum_{i=1}^N a_i=0$. 
Its classical contribution times its instanton contribution is given by 
\be
\cG_{\rm cl}(a_i;\tau) \, \cG_{\rm inst}(a_i;\tau) =  e^{-\pi
  i \tau \left( \sum_{i=1}^N a_i^2\right)}\, m(\tau)^{1-N}\, .
\ee
This can be argued as follows. If the gauge group would be $U(N)$, the instanton contribution would be $\cG_{\rm inst}=1$ \cite{Okuda:2010ke}. For gauge group $SU(N)$, however, one must first
divide by the $U(1)$ factor. We can find this $U(1)$ factor by comparing with the $q= \exp(\tau)$-expansion of the Toda conformal block
\be
\cF (a_i;\tau) = q^{\Delta(a_i)-\frac{c}{24}} \sum_k q^k F_k \, .
\ee
In particular, using the known expressions for the Toda central charge $c$ and the momentum $\Delta(a_i)$, we can verify the classical contribution to $\cG_{\rm cl} \, \cG_{\rm inst}$ for any $N$. Furthermore, we can match the full expressions in an expansion of the instanton parameter~$q$ for $N=2,3$.

Putting the pieces together, we have 
\be
\cG_{b=1}(a_i;\tau) = e^{- \pi i \tau \left( \sum_{i=1}^N a_i^2 \right)}
\frac{1}{\sqrt{2\pi}} \prod_{i<j} \frac{\pi a_{ij} }{\sinh(\pi
  a_{ij})} m(\tau)^{1-N} \, .
\ee
After performing $(N-1)$ Gaussian integrals we expect to find the partition function
\be
\cZ_{S^4_{b=1}}(\tau) \sim \frac{1}{|m(\tau)|^{2(N-1)} \, \mathrm{Im}(\tau)^{(N^2-1)/2}}\, ,
\ee
which has the expected transformation 
\be
\cZ_{S^4_{b=1}}\left(-\frac{1}{\tau}\right) = |\tau|^{N(N-1)}\,
\cZ_{S^4_{b=1}}(\tau)
\ee
 under S-duality. We have indeed verified this for $N=2,3$. In the above, we have used $\mu_{b=1}(a) =
 \prod_{i<j} 4
 \sinh(\pi a_{ij})^2$ and $\Delta(0)=\frac{N(N-1)}{2}$.

We can also check that the three-sphere partition function
\be
\cZ_{b=1}(a_i,a'_i)= \frac{\sum_{\rho \in S_N} (-1)^\rho e^{2\pi \sum_{j=1}^N a_{\rho(j)} a'_j}}{\prod_{i<j} 2\sinh \pi (a_{ij} ) \, 2\sinh\pi (a'_{ij} ) }\, ,
\ee
is the S-duality kernel for the half-sphere partition
function $\cG_{b=1}(\tau,a_i)$. This is again a matter of performing Gaussian integrals and using the modular property of the $\eta$-function. In particular, for $N=2,3$ we explicitly verified that 
\begin{align}
\int da'_i \, \mu_{b=1}(a'_i) \, \cZ_{b=1}(a_i,a'_i) \, \cG_{b=1}(a'_i;\tau) &  \sim (-i\tau)^{\frac{-N(N-1)}{2}} \cG_{b=1}(a_i;-1/\tau)\, .\notag
\end{align}
This completes the argument and gives some evidence that 
$\cG_b$ is indeed the half-sphere partition function  with Dirichlet boundary conditions.  

\section{Factorization of Toda 3-point function}
\label{appendix:special}

Let us briefly review some properties of special functions we need in order to manipulate one-loop contributions. As in the main text, $b\in \mathbb{R}_{>0}$ is a real parameter and we define $q\equiv b+b^{-1}$. 

The double gamma function $\Gamma_b(x)$ is a meromorphic function of $x$ characterized by the functional equation
\be
\Gamma_b(x+b) = \sqrt{2\pi}\, b^{bx-\frac{1}{2}}  
\Gamma_b(x) / \Gamma(bx)
\ee
where $\Gamma(x)$ is the Euler gamma function and its value $\Gamma_b(q/2)=1$. We will also need the double sine function, which is a meromorphic function that can be defined in terms of the double gamma function by the formula $S_b(x) \equiv \Gamma_b(x) / \Gamma_b(q-x)$. The double sine function is characterized by the functional equation
\be
S_b(x+b)=2\sin(\pi b x)S_b(x)\, .
\ee
We will furthermore need the function $\Upsilon_b(x)^{-1}=\Gamma_b(x)\Gamma_b(q-x)$ which is entire analytic. A more complete discussion of the properties of these functions can be found, for example, in~\cite{Vartanov:2013ima}.

Let us begin by considering the three-point function $C(\al,2Q-\al,\nu)$ in $A_{N-1}$ Toda theory corresponding to the trivalent vertex in the pants decomposition of a torus with a simple puncture. The momentum in the internal channel $\al=Q+ia$, with $a\in \mathbb{R}$, is non-degenerate and describes a delta-function normalizable state, while the momentum $\nu=N(q/2+im)\omega_{N-1}$, with $m\in \mathbb{R}$, is semi-degenerate. Substituting these momenta into the more general result of~\cite{Fateev:2005gs,Fateev:2007ab} we find that
\bea
C(\al,2Q-\al,\nu) & = f(m) \frac{ \prod\limits_{i<j}^N \Upsilon_b\left( ia_{ij} \right)\Upsilon_b\left( -ia_{ij} \right)}{\prod\limits_{i,j=1}^N \Upsilon_b\left( \frac{q}{2} + ia_{ij} + im \right)} 
\eea
where $a_{ij} = a_i-a_j$. The proportionality factor $f(m)$ is independent of the internal parameter $a$. Since we will be concerned with difference operators acting only on the internal parameter $a$, we will not need to know the details of $f(m)$ and it will be omitted whenever convenient in what follows.

The complete correlation function on a torus with simple puncture is
\be
\int da \, C(\al,2Q-\al,\mu) \overline{\cF(\al,\mu;\tau)} \, \cF(\al,\mu;\tau)
\ee
where $\cF(\al,\mu;\tau)$ are the $W_N$-algebra conformal blocks. This correlation function computes the ellipsoid partition function of the four-dimensional $\cN=2^*$ theory on an ellipsoid, with the parameters identified as in the main text.

We now consider two different ways of factorizing the three-point function and absorbing it into the $W_N$-algebra conformal blocks. The first way is chosen to maximally simplify the expressions for the Verlinde operators and we expect that this corresponds to a half-sphere partition function of $\cN=2^*$ theory with Dirichlet boundary conditions for the vectormultiplet. The second way corresponds to computing the Nekrasov partition function of the $\cN=2^*$ theory with deformation parameters $\epsilon_1 = b$ and $\epsilon_2=b^{-1}$.

\subsection*{Renormalized Conformal Blocks}

Let us express the Toda three-point function in terms of double gamma functions and manipulate the answer into a convenient factorized form. For the hypermultiplet contribution, we have
\bea
\prod_{i,j=1}^N \Upsilon_b\left( \frac{q}{2} + ia_{ij} + im \right)^{-1} 
& = \prod_{i,j=1}^N \Gamma_b\left( \frac{q}{2} +ia_{ij}+im \right) \Gamma_b\left( \frac{q}{2} - ia_{ij} -im \right)\\
& = \left| \, \prod_{i,j=1}^N \Gamma_b\left( \frac{q}{2} +ia_{ij}+im \right) \, \right|^2 \, .
\eea
For the vectormultiplet contribution
\bea
\prod_{i < j}^N \Upsilon_b\left( ia_{ij} \right) \Upsilon_b\left( - ia_{ij} \right) 
& = \prod_{i\neq j}^N \frac{1 }{ \Gamma_b\left( ia_{ij} \right) \Gamma_b\left( q-ia_{ij} \right) } \\
& =  \prod_{i\neq j}^N \frac{\Gamma_b(q+ia_{ij})}{\Gamma_b(ia_{ij})}  \, \prod_{i\neq j}^N \frac{1}{ \Gamma_b\left( q + ia_{ij} \right)\Gamma_b\left( q - ia_{ij} \right)  } \\
& = \prod_{i<j} \frac{S_b(q+ia_{ij})}{S_b(ia_{ij})} \left| \, \prod_{i\neq j}^N \frac{1}{ \Gamma_b\left( q+ ia_{ij} \right) } \, \right|^2 \\
&= \mu(a) \, \left| \, \prod_{i < j}^N \frac{1}{ \Gamma_b\left( q + ia_{ij} \right) \Gamma_b(q-ia_{ij} )} \, \right|^2 ,
\eea
where
\be
\mu(a) = \prod_{i<j} \, 2\sinh\left( \pi ba_{ij} \right) \,  2\sinh\left( \pi b^{-1} a_{ij} \right)
\ee
is the $3d$ partition function of an $\cN=2$ vectormultiplet on a squashed three-sphere~\cite{Hama:2011ea}, which is identified here with the equator $\{x_0=0\}$.

As described in the main text, we can now absorb the three-point function into the $W_N$-algebra conformal blocks, by defining new renormalized blocks
\be
\cG(a,m;\tau) = \frac{ \prod\limits_{i,j=1}^N \Gamma_b\left( \frac{q}{2} +ia_{ij}+im \right)  }{ \prod\limits_{i<j}^N \Gamma_b\left( q + ia_{ij} \right) \Gamma_b(q-ia_{ij} ) } \, \cF(a,m;\tau) 
\label{FtoG}
\ee
such that the correlation function becomes
\be
\int da \, \mu(a) | \, \cG(a,m;\tau) \, |^2
\ee
We believe that the renormalized conformal block $\cG(a,m;\tau) $ correspond to the partition function on the upper half-sphere $\{x_0>0\}$ with Dirichlet boundary conditions for the vectormultiplet at the equator. Thus, in order to transform between Verlinde operators acting on $\cF(a,m;\tau)$ and those acting on $\cG(a,m;\tau)$ we have to conjugate by the factor in equation~\eqref{FtoG}. 

Let us concentrate on the Verlinde operator corresponding to the fundamental 't Hooft loop. Acting on the unnormalized conformal blocks, the difference operator has been computed in~\cite{Gomis:2010kv}. The result is given by
\be
\sum_{j=1}^N \left[ \, \prod_{k\neq j}^N \frac{\Gamma\left(i ba_{kj}\right)}{\Gamma\left( \frac{bq}{2}+iba_{kj}-i b m \right)} \frac{\Gamma\left(b q+i ba_{kj} \right)}{\Gamma\left( \frac{bq}{2}+iba_{kj}+i b m \right)} \,  \right] \Delta_j
\label{unnorm}
\ee
where we have introduced the notation $\Delta_j:a \to a+ibh_j$. Now, by patient and repeated application of the functional equation for the double gamma function, we find
\begin{multline}
\Bigg[ \, \frac{ \prod\limits_{i,j=1}^N \Gamma_b\left( \frac{q}{2} +ia_{ij}+im \right)  }{ \prod\limits_{i<j}^N \Gamma_b\left( q + ia_{ij} \right) \Gamma_b(q-ia_{ij} ) }  \, \Bigg] \Delta_j \, \Bigg[ \, \frac{ \prod\limits_{i,j=1}^N \Gamma_b\left( \frac{q}{2} +ia_{ij}+im \right)  }{ \prod\limits_{i<j}^N \Gamma_b\left( q + ia_{ij} \right) \Gamma_b(q-ia_{ij} ) }  \, \Bigg]^{-1}  \\
 = \prod_{k\neq j}^N \frac{\Gamma\left( \frac{qb}{2}+iba_{kj}+ibm \right)}{ \Gamma\left( 1-\frac{qb}{2}-iba_{kj} +ibm\right)} \frac{\Gamma\left(1-iba_{kj} \right)}{\Gamma\left( bq+iba_{kj}\right)} \, .
\end{multline}

We can immediately see that two sets of gamma functions the second line will cancel against the same gamma functions in the unnormalized operator in~\eqref{unnorm}. The remaining gamma functions combine to give only trigonometric functions for the renormalized operator,
\begin{multline}
\sum_{j=1}^N \left[ \, \prod_{k\neq j}^N \frac{\Gamma\left(i ba_{kj}\right)\Gamma\left( 1-iba_{kj} \right)}{\Gamma\left( \frac{bq}{2}+iba_{kj}-i b m \right)\Gamma\left( 1-\frac{qb}{2}-iba_{kj} +ibm\right)} \right]  \Delta_j \\
 = \, \sum_{j=1}^N \left[ \,  \prod_{k\neq j}^N \frac{\sin\pi b\left( \frac{q}{2} + ia_{kj}-im \right)}{\sin \pi b \left( i a_{kj}  \right)}  \right] \Delta_j
 \label{Gop}
\end{multline}
as claimed in the main text. With patient bookkeeping, the same computation can be performed for the difference operators in any other completely antisymmetric tensor representation.

\subsection*{Nekrasov Partition Function}

For comparison with the exact computation of an 't Hooft loop on the four-sphere in~\cite{Gomis:2011pf}, it is necessary to consider another factorization of the Toda three-point function. In this factorization the difference operators act on the Nekrasov partition function $\cZ(a,m;\tau)$, with $\epsilon_1 = b$ and $\epsilon_2 = b^{-1}$, which we named as in the main text. 

Thus we now express the three-point function as
\be
C(\al,2Q-\al,\mu) = f(m)\,  |\, \cZ^{1-\mathrm{loop}}(a,m;\tau) \, |^2, 
\ee
where 
\be
\cZ^{1-\mathrm{loop}}(a,m;\tau) = \left[ \, \frac{ \prod\limits_{i<j}^N \Upsilon_b\left( ia_{ij} \right) \Upsilon_b\left( -ia_{ij} \right) }{ \prod\limits_{i,j=1}^N \Upsilon_b\left( \frac{q}{2}+ia_{ij}+im \right)} \, \right]^{1/2}
\ee
are the one-loop contributions to the Nekrasov partition function and $f(m)$ is independent of the internal momentum $a$ as before. The classical and instanton contributions to the Nekrasov partition function are encoded in the $W_N$-algebra conformal blocks. Thus, up to the factor $f(m)$, the complete Toda correlator can be expressed
\be
\int da \, | \, \cZ(a,m;\tau) \, |^2
\ee
in agreement with the exact computation of the partition function of the $\cN=2^*$ theory on an ellipsoid in~\cite{Hama:2012bg}.

To obtain difference operators acting on the Nekrasov partition function, it is easier at this stage to   start from the relationship to the renormalized $W_N$-algebra conformal blocks. In fact, from the relationship between the double gamma, the double sine functions and upsilon functions, we find that
\be
\cZ(a,m;\tau) = \left[ \, \frac{\prod\limits_{i<j}^N S_b(q+ia_{ij}) S_b(q-ia_{ij} ) } { \prod\limits_{i,j=1}^N S_b(\frac{q}{2}+ia_{ij}+im )} \, \right]^{1/2} \cG(a,m;\tau) \, .
\ee
Now, using the functional equation for the double sine function, we compute
\bea
\left[ \, \frac{\prod\limits_{i<j}^N S_b(q+ia_{ij}) S_b(q-ia_{ij} ) } { \prod\limits_{i,j=1}^N S_b(\frac{q}{2}+ia_{ij}+im )} \, \right]^{1/2} \, \Delta_j \, \left[ \, \frac{\prod\limits_{i<j}^N S_b(q+ia_{ij}) S_b(q-ia_{ij} ) } { \prod\limits_{i,j=1}^N S_b(\frac{q}{2}+ia_{ij}+im )} \, \right]^{-1/2} \\
= \left[ \; \prod_{k \neq j}^N \frac{\sin\pi b \left( \frac{q}{2} +ia_{kj}+im \right) \sin\pi b(ia_{kj})}{ \sin\pi b\left( \frac{q}{2} +ia_{kj}-im \right) \sin\pi b( q+ia_{kj} )}  \, \right]^{1/2} \, .
\eea

Thus, combining with equation~\eqref{Gop}, we conjecture that the fundamental 't Hooft loop operator acting on the Nekrasov partition function with $\epsilon_1=b$ and $\epsilon_2=b^{-1}$ has the general form
\be
\sum_{j=1}^k \left[ \; \prod_{k \neq j}^N \frac{\sin\pi b \left( \frac{q}{2} +ia_{kj}+im \right) \sin\pi b\left( \frac{q}{2} + ia_{kj} -im\right)}{ \sin\pi b\left(ia_{kj}\right) \sin\pi b( q+ia_{kj} )}  \, \right]^{1/2}  \Delta_j \, .
\ee
This agrees with the exact computation of the fundamental 't Hooft loop operator in the case of a round four-sphere $b = 1$~\cite{Gomis:2011pf}. Again, with patient bookkeeping the same conclusion can be reached for 't Hooft loops labeled by any antisymmetric tensor representation.

\bibliography{SurfaceOperators}

\end{document}